\documentclass[conference]{IEEEtran}
\makeatletter
\def\ps@headings{%
\def\@oddhead{\mbox{}\scriptsize\rightmark \hfil \thepage}%
\def\@evenhead{\scriptsize\thepage \hfil \leftmark\mbox{}}%
\def\@oddfoot{}%
\def\@evenfoot{}}
\makeatother
\IEEEoverridecommandlockouts
\usepackage[dvips]{graphicx}
\usepackage[cmex10]{amsmath}
\usepackage{amsfonts,amssymb,bm}
\usepackage[tight,footnotesize]{subfigure}
\usepackage{fixltx2e}
\usepackage{dblfloatfix}
\usepackage{rotating}
\usepackage{multirow}
\usepackage{url}
\usepackage{cite}
\usepackage{algorithmic,algorithm}
\usepackage{slashbox}
\usepackage{cite}
\usepackage{setspace}
\usepackage{footnote}
\usepackage[T1]{fontenc}
\usepackage{ae,aecompl}
\usepackage{epsfig}
\usepackage{multicol}
\usepackage{wrapfig}
\usepackage[normalem]{ulem}

\usepackage{subfigure}
\usepackage{times}

\usepackage[10pt]{moresize}

\usepackage{amssymb}

\usepackage{pifont}
\algsetup{linenodelimiter=.}

\newcommand{\ie}[0]{\textit{i.e.},~}
\newcommand{\eg}[0]{\textit{e.g.},~}
\newcommand{\etal}[0]{\textit{et al.}}

\newcommand{\vs}[0]{\textit{vs.}~}

\newcommand{\squishlist}{
 \begin{list}{$\bullet$}
  { \setlength{\itemsep}{0pt}
     \setlength{\parsep}{3pt}
     \setlength{\topsep}{3pt}
     \setlength{\partopsep}{0pt}
     \setlength{\leftmargin}{1.5em}
     \setlength{\labelwidth}{1em}
     \setlength{\labelsep}{0.5em} } }

\newcommand{\squishlisttwo}{
 \begin{list}{$\bullet$}
  { \setlength{\itemsep}{0pt}
     \setlength{\parsep}{0pt}
    \setlength{\topsep}{0pt}
    \setlength{\partopsep}{0pt}
    \setlength{\leftmargin}{2em}
    \setlength{\labelwidth}{1.5em}
    \setlength{\labelsep}{0.5em} } }

\newcommand{\squishend}{
  \end{list}  }

\begin{document}
\title{The Multiple Instances of Node Centrality and their Implications on the Vulnerability of ISP Networks}
%\vspace{0.12in}
%\Large Technical Report}
\author{\IEEEauthorblockN{George Nomikos\IEEEauthorrefmark{1} \hspace{20pt} Panagiotis Pantazopoulos\IEEEauthorrefmark{1} \hspace{10pt}  }
\\
\IEEEauthorblockA{\IEEEauthorrefmark{1}
 Department of Informatics and Telecommunications\\
 National \& Kapodistrian University of Athens\\
 Ilissia, 157 84 Athens, Greece\\
 Email: \{gnomikos, ppantaz, ioannis\}@di.uoa.gr}
\and
\IEEEauthorblockN{\hspace{10pt} Merkourios Karaliopoulos\IEEEauthorrefmark{2} \hspace{20pt} Ioannis Stavrakakis\IEEEauthorrefmark{1}}
\\
\IEEEauthorblockA{\IEEEauthorrefmark{2}
Centre for Research and Technology - Hellas \\
Information Technologies Institute\\
38334 Volos, Greece \\
 Email: mkaraliopoulos@iti.gr}
\thanks{\IEEEauthorrefmark{2} The major part of this work was carried out while Dr. Karaliopoulos was
affiliated with the Department of Informatics and Telecommunications, UoA.

This work has been partially supported by EINS, the Network of Excellence in Internet Science through
EC's Grant Agreement FP7-ICT-288021.}}

% %+++++++++++++++++++++++++++++++++++++++++++++++++++=========================
% %2222
% \author{\IEEEauthorblockN{George Nomikos~\hspace{12pt} Panagiotis Pantazopoulos~\hspace{12pt}Merkourios Karaliopoulos\IEEEauthorrefmark{2} \hspace{10pt} Ioannis Stavrakakis}
% \IEEEauthorblockA{Department of Informatics and Telecommunications,
% University of Athens, Ilissia, 157 84 Athens, Greece\\
% %National \& Kapodistrian
% Email: \{gnomikos, ppantaz, mkaralio, ioannis\}@di.uoa.gr}
%
% \thanks{
% \IEEEauthorrefmark{2} The major part of this work was carried out while Dr. Karaliopoulos was affiliated with the Department of Informatics and Telecommunications, UoA.
%
% This work has been supported by EINS, the Network of Excellence in Internet Science through
% EC's Grant Agreement FP7-ICT-288021.}
% }

\maketitle
%
%\thispagestyle{empty} 3
%2\pagestyle{empty}2

\begin{abstract}
The position of the nodes within a network topology largely determines the level of their involvement in various networking functions. Yet numerous \emph{node centrality indices}, proposed to quantify how central individual nodes are in this respect,
yield very different views of their relative significance.
%Clearly, the relative significance assigned to the nodes generally varies, depending on the specific index employed.
%Moreover, the appropriateness of an index to drive
Our first contribution in this paper is then an exhaustive survey
and categorization of centrality indices along several attributes including the type of information (local \vs global) and processing complexity required for their computation. % related to their computation.

We next study the seven most popular of those indices in the context of Internet vulnerability to address issues that remain under-explored in literature so far.
%: a) relating the network vulnerability %assessment
%with the traffic carrying capacity; b) systematically experimenting with router-level topologies.
First, we carry out a correlation study to assess the consistency of the node rankings those indices generate over ISP router-level topologies. For each pair of indices, we compute the full ranking correlation, which is the standard choice in literature, and the percentage overlap between the $k$ top nodes. 
%insisting more
%on the correlation of the six \textit{global} indices with the only \textit{local} one, the degree centrality. 
Then, we let these rankings guide the removal of highly central nodes and assess the impact on \textit{both} the connectivity properties and traffic-carrying capacity of the network. %is %also %evaluated.
Our results confirm that the top-$k$ overlap predicts the comparative impact of indices on the network vulnerability better than the full-ranking correlation. %is better predicted by their %values.
%Moreover, it can be used as a criterion for deciding when synthetic \emph{hybrid} rankings, including the top nodes according to a local and global index, can make the impact of node removals more dramatic.
Importantly, the locally computed degree centrality index approximates %employing the degree to approximate
closely the global indices with the most dramatic impact on the traffic-carrying capacity; whereas, its approximative power  %with the degree
in terms of connectivity is more topology-dependent. 

\end{abstract}

\section{Introduction}
\label{intro}
%Sociological studies: the origin of centrality
\textit{Social Network Analysis}~(SNA) constitutes a highly interdisciplinary theoretical framework
that seeks to process social information and analyze existing social structures~\cite{Faust}.
SNA draws heavily on graph models that map individual actors within the social
network to the graph vertices and their relationships to the graph (weighted) edges. It then leverages graph-theoretic concepts,
metrics and results to answer various questions about %the %social network such as
the relative importance of the actors for the network
%influential each is
or the way that information (or innovations) flow (resp. spread) across it.
%to assist the understanding of the behavior of

The centrality concept, to the best of our knowledge, dates back to the work of Bavelas~\cite{bavelas}. By that time significant sociological research was directed to the area of professional networks addressing
how the position and power of individual actors relate to their social interconnections and the way
they interact with the rest of the network. Such sociological studies motivated the
introduction of various sociological indices, which sought to quantify the \textit{importance} of
nodes and their relationships. 
%Research on the centrality concept, to the best of our knowledge,
%dates back to the work of Bavelas~\cite{bavelas}. 
Bavela's work appears to be the first to have given a formal definition
of node centrality in connected graphs as the sum of its own geodesics (shortest-path distances)
to all other nodes. %graph nodes. %Their ultimate goal was to understand how the communication level between pairs in human
%groups or even organizations differentiates through their internal connections.

This work triggered a large research thread and a huge number of publications in the area of centrality indices. Many of them proposed new indices~\cite{Katz} or adaptations of existing ones that expanded their applicability in a broader range of scenarios~\cite{Beauchamp, Anthonisse}. The vast majority of work was heuristic and only a few of them attempted to come up with axiomatic definitions of centrality indices and the properties they should satisfy~\cite{sabidussi}. 
%whereas fewer attemp largely devoted to the definition of new and adaptation of existing indices and, more rarely, to the  
The highly-cited work of Freeman in \cite{Freeman} appears to have served as a turning point for this first wave of work, by reviewing a number of centrality indices and promoting three of them, \ie the closeness, degree, and betweenness, as the most representative ones.  About the same time Bonacich had established the eigenvector centrality as a fourth, distinctly different but equally popular, index~\cite{power}. %ADDREF(Bonacich). 

The research interest in the centrality concept revives in late 90's and early 2000, primarily through the works of physicists such as D. Watts and M. Newman. They use centrality indices to explore the vulnerability and community structure, respectively, of general network instances. SNA techniques and centrality, in particular, find applicability to research work across a broader set of disciplines beyond sociology. In the case of computer scientists,
%and telecommunication network engineers
insights from centrality indices are primarily exploited in the design of more effective protocols for communication networks~\cite{Daly09,cacheICN}. The trend is only catalyzed by the
broader expectations about the evolution of a \textit{Network Science}~\cite{NetScience}, which could serve as
the theoretical foundation for a unified treatment of all network types.
%-Centrality, probably the most celebrated concept in the SNA toolbox, seeks to quantify a node's relevant social standing.
% -The introduction of centrality indices by sociologists dates back to the 50's. %It was the sociological that
% Through the last 30 years it has attracted the interest of physicists/biologists and
% lately has been been utilized in network protocol design.

% -A plethora of alternative formulations of the centrality concept has emerged;
% the degree centrality is clearly the easy-to-obtain index and therefore,
% the prominent one for networking applications.

% -Can we identify relations between all those indices? How do they compare in terms of correlation?
% On a more practical note, we need to know whether globally-determined metrics
% can be somehow related to the local one(s).
% Network topologies of particular interest for these questions are the Internet graphs where
% protocol instances that will rely on centrality indices, are designed to operate.
%
% -The relevance of centrality indices to the resilience properties of communication networks is open to question.
% How do centrality indices compare in terms of identifying critical Internet nodes to be attacked ?

\textit{Motivation and objective}: The relevance of centrality indices to the communication network (\ie Internet topologies)
robustness, in particular, is the motivation for this study. 
Our main objective is \textit{to quantify how much information is
embedded in centrality indices about the relative importance of Internet nodes for different network operations}.
Given that the different formulations of centrality proposed in literature are heuristic, the questions that naturally
arise are how do these formulations compare in their assessments/predictions about the relative importance of network
nodes and which one(s) may be the ``right one(s)'' to consider as reference for more reliable predictions of network robustness.

The paper seeks to \textit{systematically} address these questions by undertaking a three-step study with various instances of methodological innovation. The first step involves a thorough survey and novel classification of the variety of centrality indices proposed in literature over the last sixty years. This classification is then used to select the seven most popular and representative indices for carrying out the two experimental steps of the study. Hence, as a second step, we derive the node rankings these indices induce over more than 40 \textit{router-level} snapshots of network topologies and study their correlation. The correlation strength is assessed by the mainstream rank/linear correlation coefficients but also less widespread measures such as the percentage overlap in the lists of the $k$ most central nodes. 
Finally, we compare the seven indices with respect to their capacity to reveal the network vulnerability to node removals; we let the indices dictate the most central nodes to-be-removed and assess how the \textit{network connectivity} properties but also its \textit{traffic-carrying capacity} are affected. 

Our results identify certain index pairs with consistently high full rank correlation across all datasets we experiment with. However, they also warn against the interpretation of its high values showing that significant part of this correlation is due to nodes at the bottom of the rankings. As a result, the percentage overlap of the $k$ most central nodes for the same pairs assumes clearly smaller values.
Among the noteworthy results is that when the nodes removals are driven by the single index that can be computed through local-only information (\ie the Degree Centrality index), the impact on the network traffic serving capacity approximates closely the maximum over the seven indices. The hint for network vulnerability studies is that the added complexity of global indices may be circumvented when an estimate of what is the worst-case impact on the network is needed.

The remainder of the article is structured as follows: 
In Section~\ref{sec:survey}, we summarize a survey of a broad range of proposed centrality indices over the last sixty years and the classification scheme we adopted. Note that a detailed description of the various centrality formulations appears in~\cite{thesis}.
%We describe the inputs needed for our study in Section~\ref{sec:method} and subsequently 
We then turn to the question of how much information different centrality indices entail regarding the importance of network nodes and the resulting network vulnerability. First, we select a subset of seven popular indices and carry out a correlation study in Section~\ref{sec:correlations}. 
Then, in Section~\ref{sec:robustness} we let the seven indices drive targeted node attacks over the Internet graphs 
and experimentally assess their impact in both connectivity and traffic-carrying capacity terms. Related literature is summarized in Section~\ref{sec:related}. Finally, we conclude the paper with a summary of the main messages out of our study in~\ref{sec:conclusions}. %in Section~\ref{sec:conclusions}. %\textbf{(REVISIT-ED)}

\section{A novel Classification of centrality indices}\label{sec:survey}
%(\textit{\textbf{The size and amount of details appearing in this Section are subject to the targeted venue}})
% The work in  presents a systematic survey of the rich literature on centrality indices.
% Numerous indices that have appeared in different disciplines over the last sixty years
% are presented in detail and categorized along a number of well-defined attributes.
% %Those are, herein, briefly presented and some indicative centrality indices

\begin{figure}[ht]
\centering
\includegraphics[scale=0.35]{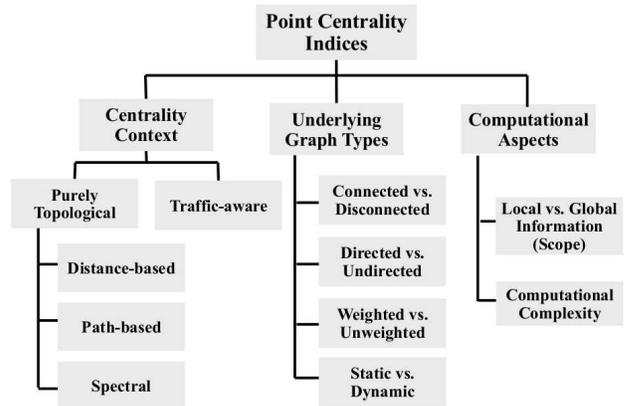}
  %\hspace{0.5cm}
\caption{Classification tree of point centrality indices.}
\label{fig:pointCentrality_tree}
\end{figure}

In this report, we attempt to summarize this classification, pointing the interested reader to \cite{thesis} for a much more detailed description of the indices and the context within which they were originally proposed.
%that come under
%of the proposed classification scheme. %, are discussed.
At a first-level the reviewed indices are split between \textit{node} (point) centrality and \textit{graph} centrality indices.
The former, which are addressed by the vast majority of the literature, characterize individual nodes, whereas the latter are derived for whole graphs as functions of the individual node centrality indices.
Then, node centrality indices are further characterized in line with the
attributes shown in Fig.~\ref{fig:pointCentrality_tree}. These include the network properties that are reflected in the index formulation (topological~\vs~flow-aware), the type of underlying graph over which an index is computed, as well as computational aspects such
as the amount of information (local~\vs~global) and complexity involved in the index computation.

\subsection{Node centrality indices}
%The proposed classification of node centrality indices is illustrated in Figure~\ref{fig:pointCentrality_tree}.
The first broad category groups point centrality indices that have been, at least originally, proposed for connected,
binary, non-directed graphs.
%Then we identify a set of indices that expand
%the centrality concept over disconnected, weighted directed and dynamic graphs.
%Finally, we classify these indices along their scope (local or global), and their computational complexity.

\subsubsection{Context: Purely topological \vs~flow-aware}
\label{}
%\noindent
%\textit{Pure topological indices}
\paragraph{Pure topological indices}
This set of centrality indices takes into account only the network topology, \ie the nodes and the links between them.
%\subparagraph{Distance-based centrality}
Topological indices may reflect two different aspects of a node's position in a network.
%\begin{list}{\labelitemi}{\leftmargin=1.4cm}
%\item Distance-based centrality: The corresponding indices measure how distant a node is from all other network nodes.
      %Indices that fall in this category are the closeness centrality and Eccentricity~\cite{thesis}.
%\item Path-based centrality: The indices of this type assess to what extent a node lies on paths
%connecting other nodes in the network. Degree and betweenness centrality are some of the relevant indices~\cite{thesis}.
%\end{list}

\emph{Distance-based centrality:} The corresponding indices measure how distant a node is from all other network nodes. Indices that fall in this category are the Closeness Centrality~\cite{Freeman} and Eccentricity~\cite{eccentrcity}.
%Indices that fall in this category are the Closeness Centrality and Eccentricity~\cite{thesis}.

\emph{Path-based centrality:} Indices of this type assess to what extent a node lies on \emph{paths}
connecting other nodes in the network. Degree and Betweenness Centrality are some of the relevant indices~\cite{Freeman}.%~\cite{thesis}.

Both types of indices can be further differentiated as to whether the distance (resp. path) definition accounts only for geodesics
(\ie shortest paths) between node pairs or a broader set of paths connecting them.
%The latter extension, especially when considered for the path-based indices,
Therefore, indices such as the Katz index~\cite{Katz} and random walk betweenness~\cite{RW_BC} essentially relax the underlying assumption that information flows only through shortest path routes.

A third category of topological indices, which has been shown to be closely related to path-based centrality, are spectral centrality indices. Common to these indices are their dependence on the eigenstructure of a matrix related to the network in question and their computation through linear-algebraic manipulations. The indices, sometimes also called ``prestige'' measures of centrality~\cite{Faust}, have the special feature that the centrality index of a node is a function of the centralities of the nodes it is (directly) connected to. The eigenvector centrality index~\cite{power} is the most popular one in this family.

\paragraph{Flow-aware indices}
The so far considered centrality indices rank the graph nodes taking into account the network topology only.
A separate thread of work has attempted to factor the traffic that a network is expected to serve in
the computation of the centrality indices. The traffic-aware betweenness~\cite{trafficBC} and the weighted conditional
betweenness centrality~\cite{ITC} are two relevant indices.

% \begin{list}{\labelitemi}{\leftmargin=1.4cm}
% \item traffic-aware indices
%   \item flow-aware indices
% \end{list}

%\subsubsection{Purely topological indices}

% \begin{itemize}
% \item Distance-based centrality
% \item Distance-based centrality beyond geodesics
% \item Path-based centrality
% \item Path-based centrality beyond geodesics
% \item Spectral graph-theoretic indices
% \end{itemize}

\renewcommand\arraystretch{1.7}
\begin{table*}[ht]
\centering \caption{Properties of seven popular centrality indices under a novel Classification scheme} %\vspace{-0.1 in}
\advance\leftskip-0.4cm
\label{tab:seven_indices} {\tiny
\begin{tabular}{ |r||  c |c |c | c||  c  |c | c| c|  c|  c| c || c| c| l|| l|}
%\hline
% \multicolumn{1}{|c||}{}&\multicolumn{4}{c||}{}&\multicolumn{7}{c||}{}&\multicolumn{3}{c|}{} \\
\multicolumn{1}{c}{}&\multicolumn{4}{c||}{\ssmall{\textbf{Context}}}&\multicolumn{7}{c||}{\ssmall{\textbf{Type of underlying graph}}}&\multicolumn{3}{|c}{\ssmall{\textbf{Computational aspects}}}&\multicolumn{1}{c}{} \\
% \multicolumn{1}{|c||}{}&\multicolumn{4}{c||}{}&\multicolumn{7}{c||}{}&\multicolumn{3}{c|}{} \\
\cline{2-15}
\multicolumn{1}{c||}{\ssmall{\textbf{Centrality Index}}}&\multicolumn{3}{c|}{\textbf{Topology }}&\multicolumn{1}{c||}{\textbf{Flow}}&\multicolumn{2}{c|}{\textbf{Binary/}}&\multicolumn{2}{c|}{\textbf{Directed/}}&\multicolumn{1}{c|}{\textbf{Dynamic}}&\multicolumn{2}{c||}{\textbf{Connected/}}&\multicolumn{2}{c|}{\textbf{Information }}&\multicolumn{1}{l||}{\textbf{Complexity}}&\multicolumn{1}{c}{\ssmall{\textbf{Definition}}}\\
\multicolumn{1}{|c||}{}&\multicolumn{3}{c|}{\textbf{aware}}&\multicolumn{1}{c||}{\textbf{aware}}&\multicolumn{2}{c|}{\textbf{weighted}}&\multicolumn{2}{c|}{\textbf{Undirected}}&\multicolumn{1}{c|}{\textbf{ }}&\multicolumn{2}{c||}{\textbf{Disconnected}}&\multicolumn{2}{c|}{\textbf{(local/global)}}&\multicolumn{1}{c||}{}&\multicolumn{1}{c|}{}\\
\hline
\multicolumn{1}{|c||}{}&\multicolumn{1}{c|}{path}&\multicolumn{1}{c|}{distance}&\multicolumn{1}{c|}{spectral}&\multicolumn{1}{c||}{}
&\multicolumn{1}{c|}{B}&\multicolumn{1}{c|}{W}&\multicolumn{1}{c|}{D}&\multicolumn{1}{c|}{U}&\multicolumn{1}{c|}{D}&\multicolumn{1}{c|}{C}
&\multicolumn{1}{c||}{D}&\multicolumn{1}{c|}{L}&\multicolumn{1}{c|}{G}&\multicolumn{1}{c||}{}&\multicolumn{1}{c|}{}\\%&\multicolumn{
\cline{2-15}
\textbf{Betweenness (BC)}   & \ding{51}    &            &              &   & \ding{51}   &  \ding{51}    &\ding{51}    &\ding{51}        & \ding{51} & \ding{51}         &\ding{51}   &              &\ding{51}   &  $O(V E)$         & $c_{i}^{BC}=\frac{2}{(N-1)(N-2)} \sum_{j \neq k \neq i} \frac{d_{j,k}(i)}{d_{j,k}}$             \\
\textbf{Closeness (CC)}     &              & \ding{51}  &              &   & \ding{51}   & \ding{51}     & \ding{51}   &\ding{51}        & \ding{51} & \ding{51}         &            &            &\ding{51}   &  $O(V(logV) E)$   & $c_{i}^{CC}=\frac{N-1}{\sum_{j \in G, j \neq i} d_{i,j}}$           \\
\textbf{Degree (DC)}        & \ding{51}    &            &              &   & \ding{51}   &  \ding{51}    & \ding{51}   &\ding{51}        & \ding{51} &  \ding{51}        &\ding{51}   & \ding{51}  & &  $O(V^{2})$                  & $c_{i}^{DC}=\frac{deg(i)}{N-1}$            \\
\textbf{Eccentricity (ECC)} &              & \ding{51}  &              &   & \ding{51}   &   \ding{51}   &\ding{51}    & \ding{51}       &           &  \ding{51}        &            &             &\ding{51}   & $O(V(logV) E)$    & $c_{i}^{ECC}=\frac{1}{max_{j \in V} d_{i,j} }$           \\
\textbf{Eigenvector (EC)}   &              &            &\ding{51}     &   & \ding{51}   &  \ding{51}    &             &\ding{51}        &           & \ding{51}         &            &            & \ding{51}          & $O(V^{3})$ &$c_{i}^{EC}=\frac{1}{\lambda} \sum_{j \in G} \alpha_{i,j} \cdot c_{j}^{EC}$                     \\
\textbf{Harmonic (HC)}      &              & \ding{51}  &              &   & \ding{51}   &  \ding{51}    &\ding{51}    &\ding{51}        &           & \ding{51}         &\ding{51}   &           & \ding{51}          &  $O(V(logV) E)$ & $c_{i}^{HC}=\frac{1}{N-1} \sum_{j \in G, j \neq i} \frac{1}{d_{i,j}}$        \\
\textbf{PageRank (PG)}      &              &            &\ding{51}     &   & \ding{51}   &               & \ding{51}   &                 &           & \ding{51}         & \ding{51}  &           & \ding{51}  & $\Omega(\frac{E^{2}}{ln(1/(1-d))})  $ & $c_{i}^{PG}=\frac{1-d}{N}+d\sum_{v \in B_{i}} \frac{c_v^{PG}}{L_{v}}$    \\
\hline
\multicolumn{16}{|c|}{\textbf{$N$: Total number of nodes,  ~~~~$d_{i,j}$ : Shortest path length from $i$ to $j$,~~~~$d_{j,k}(i)$ : Shortest path length via $k$, ~~~~$\alpha_{i,j}$: Adjacency matrix element  ~~~$d$ : Damping factor, ~~~$B_{i}$: Set of nodes linked to $i$, ~~~$L_{v}$: Out-degree of node $v$}}   \\
\hline
\end{tabular}}
\end{table*}
\renewcommand\arraystretch{1.0}% * * * back to normal * * *

%\subsubsection{Traffic-aware centrality indices}
% \begin{itemize}
% \item Traffic-aware betweenness centrality
% \item Weighted Conditional Betweenness Centrality
% \item Flow-aware betweenness centrality
% \end{itemize}

\subsubsection{Underlying graph types}
\label{subsec:graph_types}
Most, if not all, of the considered centrality indices are defined over \textit{connected, undirected, binary, static} graphs. In what follows, we relax in turn each one of these four graph attributes and
discuss how the centrality indices are adapted to the resulting types of graphs.

\paragraph{extensions of centrality indices for disconnected graphs}
Most of the centrality metrics have been formulated and proposed with connected networks in mind, \ie
there are finite paths between every pair of nodes in the network that together form a single giant connected component.
Much less attention has been paid to centrality metric formulations for disconnected graphs featuring more than one
connected component and/or isolated nodes. Notably, some metric definitions are such that they can directly generalize
for disconnected graphs without any additional care. For instance, this is the case with the degree
and betweenness centrality
while
%On the other hand,
the closeness centrality metric as defined by Beauchamp~\cite{Beauchamp} and Freeman~\cite{Freeman77}
%definition by Beauchamp [26] and Freeman [22]
does not trivially generalize into disconnected graphs.
%Typical example is the harmonic centrality which essentialy extends the closeness centrality index.

%since non-connected node pairs contribute infinite-value terms to the sum in the denominator
%of (2).
\paragraph{extensions of centrality indices for directed graphs}
%expand the original distance centrality definition to both directed and disconnected networks and
%were described in 2.2.1,
The main body of work that proposes centrality indices appropriate for
%much of the work on
directed graphs, %centrality indices
evolves around the spectral ones.
PageRank, one of the most discussed implementation on the Web, %developed by Brin and Page [41],
delineates the basic model to effectively manage graph-based structures composed by directed (either inbound or outbound) links~\cite{pagerank}.

\paragraph{extensions of centrality indices for weighted graphs}
Adaptations of centrality metrics for weighted graphs have been mainly proposed for the three most common centrality indices, the Degree, Betweenness and Closeness Centrality; in the last two instances, only geodesics are considered. The intuitive way to expand the node degree centrality definition is by replacing the sum of the node's neighboring links with the sum of their weights~\cite{weightedNets}. Likewise, the notion of link distance (or cost) underlying both the betweenness
and closeness centrality indices is captured by its (inverse) weight and geodesics are estimated accordingly~\cite{Brandes_variants}.

Regarding spectral indices, to derive the Eigenvector centrality variant for weighted graphs it suffices to substitute the binary elements of the adjacency matrix involved in the eigenvector computations with the edge weights~\cite{weightedNewman}.
The extension of PageRank is somewhat more involved; %that has attracted much of the interest.
the index originally designed to rank Web pages exploits the binary graph based nature of the Web. However, treating equally all inbound and outbound links remains restrictive when measuring the importance of each page. Therefore, proposed extensions of PageRank over weighted networks assign to each outlink page a value proportional to its
popularity~\cite{wPageRank_alg} or use factors to modulate how rank scores are distributed to neighbors~\cite{AuthorRank,journal_status}.
%By assigning weights to
With hyperlink weights, the surfer can now express preferences %ing behaviors
among pages instead of uniformly jumping to arbitrary ones.

\paragraph{extensions of centrality indices for dynamic graphs}
The extension of standard complex network indices, including centrality ones, to networks that vary over time is a more recent thread.
There are more than one graph representations for dynamic networks and many more terms that are used to denote
them such as temporal graphs~\cite{kostakos}, evolving graphs~\cite{evolving_graphs}, space-time graphs~\cite{spacetime_graphs} or time-varying graphs~\cite{time-varying}.
To the best of our knowledge, there have been two main studies that have proposed adaptations of centrality indices for temporal graphs. The first one draws on the notion of temporal path~\cite{temporal_paths} over a sequence of graph instances, whereas the second relies on a time-expanded graph representation to define temporal
Betweenness and Closeness indices~\cite{time_ordered}.
%[65, 70],
%time-ordered graphs [67], dynamic graphs [66].
% \begin{list}{\labelitemi}{\leftmargin=1.4cm}
% \begin{itemize}
% \item Extensions of centrality indices for disconnected graphs
% \item Extensions of centrality indices for directed graphs
% \item Extensions of centrality indices for weighted graphs
% \item Extensions of centrality indices for dynamic graphs
% \end{itemize}

\subsubsection{Computational Aspects}
\paragraph{Index scope (local \vs global)}
%It would be useful to examine the proposed centrality measures from a different point of view.
Centrality indices can be separated into local and global ones, depending on the extent of topological information that is required to compute them. For instance, since Degree Centrality is a function of the number
of direct (one-hop) neighbors, it is a local index. On the other hand, Betweenness
%indices of
%Freeman [15] as well the
and Closeness Centrality are global indices in that they rely on geodesic paths computed all over the network.
% Relevant implementations in terms of local knowledge of connectivity but completely
% differentiated from Freeman’s interpretation, represent the random walk definitions [28], [50].
% They consider probabilistic motions around the network, driven by direct neighbors each time as highlighted under
% the “Path-based centrality beyond geodesics” section.

One typical way to control the scope of (path-based) centrality computations
is through the sociological notion of the ego-network. The ego network of a node $v$ is the subgraph involving $v$, called the ``ego'' node, its 1-hop neighbors, and their inter-connections. The ego network (centered-graph~\cite{centered} in graph theoretic terms) is used sometimes to derive a local approximation of an otherwise global centrality index. Betweenness Centrality is a typical index
that lends to ego-centric approximations~\cite{ppantaz}.
Another way to control the scope of centrality indices, this time purely path-based indices, is by controlling
the length $k$ of paths that are taken into account in their computation. Indices such as $k$-path~\cite{k_path}
and %geodesic $k$-path %centrality
%edge-disjoint k-path
and vertex-disjoint k-path
centrality~\cite{k-BC} are examples of this category.

\paragraph{Computational cost of the index}
Of particular interest for embedding centrality indices in communication network protocols is their computational complexity.
%\textbf{Regarding
The centrality interpretation by Freeman~\cite{Freeman77} does not
seem to require special computing power to be applied on large network structures.
%proportionally to the number of participants
%exist in the network.
On the contrary, the solution proposed by Bonacich~\cite{Bonacich91} %[18]
to correlate the point centralities with graph centralities,
could be considered computationally heavy as the network size increases.
%THIS SENTENCE...??
Consequently, he suggested a way to control this limitation using a new kind of matrix
known as ``overlapping'' instead of adjacency matrix before the eigenvector calculation. Also, Moxleys'
solution~\cite{Moxley} for (un)connected graphs seems to encounter the same serious problem with their AIC (Adjusted Index of Centrality)
metric when analyzing rich datasets. As a result, they managed to
represent the connections of each element in a vector based structure to efficiently compute the adjacency matrix
and measuring the centrality for every reached or unreached point.

%
% The computation of the k-betweenness metric can be carried out by a generalization of the Brandes algorithm [61]
% to account for dependencies that involve paths of length d(s,t)+k, where d(s,t) is the length of the shortest path
% between s and t. Computing the k-Betweenness centrality may be tractable only for small values of k since the complexity
% for this generalization requires O() time.

\subsection{Graph centrality indices}
Centrality indices have also been proposed as single numbers for the whole graph (\ie graph centrality indices).
If point centrality indices essentially generate rankings of nodes within a given graph,
graph centrality indices seek to rank different graphs.
%O\textbf{ur classification of the graph centrality indices is illustrated in
%Figure~\ref{fig:graphCentrality_tree}.} WHAT DOES THIS ADD TO THE STORY AND WHERE IS THIS USED?
Note that the first studies on point centrality indices by Bavelas~\cite{bavelas} and Beauchamp~\cite{Beauchamp} address
graph centrality indices as well.
%More specifically, Bavelas defined the
Graph centrality was initially defined as the sum of point centralities over each network node.
Later on, more complex axiomatic definitions appeared in literature~(\eg~\cite{sabidussi}), therefore assigning different notions
to the graph centrality indices. A more insightful categorization attempt is given by H{\o}ivik~\cite{Hoivik} and recognizes three graph centrality concepts: %t hreads
\subsubsection{Integration} This is a measure of how centrally located are the nodes of a graph as a whole.
It is measured by the sum of individual node centrality indices. This is what Freeman calls \emph{compactness} in his definitions of graph centrality indices out of the Degree, Betweenness and Closeness Centrality indices \cite{Freeman}.
\subsubsection{Unipolarity} Reflects whether there is a very central node and is taken equal to the maximum node centrality.
\subsubsection{Centralization}  %reflects whether there is a very central node and is taken equal to the maximum node centrality.
Captures the dispersion of the node centrality values and is taken equal to the sum of differences of point centralities from the minimum point centrality value.
% connected vs disconnected

% \begin{figure}[ht]
% \centering
% \includegraphics[scale=0.39]{./figs/tree2_new}
%   %\hspace{0.5cm}
% \caption{Classification tree of graph centrality indices.}
% \label{fig:graphCentrality_tree}
% \end{figure}

%\subsection{Interpretation/Rationale }

%In this subsection we briefly discuss the various interpretations \ie the concept that some
%centrality expresses %by the  %that have appeared in literature
%and comment on the appropriateness of an index with respect to the information flow type in the network.

%A number of centrality measures have been proposed to analyze graph-theoretic concepts due to networking phenomena.
%Even though there is a plethora of point centrality variations, most of them seek to approximate degree centrality, closeness,
%betweenness and eigenvector centrality. %as provided by Freeman [15][22] and Bonacich [19].
As a final note, the interpretations of centrality indices are multiple. To refer to some of those,
Freeman noted that central parties may generally affect the communication, facilitating or even distorting whatever
flows in the network. Particularly, from the perspective of degree centrality, he argued that a point with relative high
degree serves to control the communication activity inside the topology having the advantage of binding together flow processes~\cite{thesis}. Bonacich agreed saying that a central firm has the possibility to acquire satisfying information rapidly
and therefore with high probability.
%Regarding betweenness point centrality, Freeman considered it as a potential index of participation in communication potency.
%Also, betweenness-like measures are capable to roughly assess how much
%stress an individual can suffer from.
%
% Borgatti [2], [32] briefly reviewing a bunch of centrality measures he confirmed that when a central node being
% isolated can more influence the speed of flow information throughout the network than if a peripheral node (a node with
% low point degree centrality) is going to be removed. He also supported that a degree based centrality is practically a key of
% how fast a message can be distributed all over the network.
% 	
% Regarding closeness point centralities, Freeman [15], [22] interpreted closeness as an index of the expected time to reach an intermediate node
% for any pair of source-destination. On the other hand, he assumed that the eigenvector centrality [19] is a measure to express the centrality
% of a node as a function of the connectivity of its incident or indirect neighbors. This exhibits a strong dynamic relationship between a point’s
% eigenvector centrality and a non-static networking topology.
%
%In social network analysis
Borgatti on the other hand, tried to shed more light on the question of which centrality index is appropriate
for which occasion~\cite{Borgatti_05}. He introduced general flow typologies over networks using two criteria \ie
%The first one refers to
the way the flow is realized in the network (\ie point-to-point transfer, serial and parallel duplication) 
%second one to
and the kind of graph-theoretic path (\ie walk, trail, and path) that is relevant.
%Particularly, he examined how fast the information flows for different kinds of traffic into the network and demonstrated that upon the
%context of each situation an appropriate measure has to be used as it is shown in the following table.
%Then, in Section we let the rankings of nodes introduced by each index, determine the attacks over network nodes and assess
%the impact on the Internet graphs connectivity and throughput.

\subsection{Selecting centrality indices for experimentation}\label{employed_indices}

In summary, the survey work in this Section has shown that there
is a plethora of centrality index formulations, many of them capturing different properties of network nodes. For the experimentation that follows, we select seven indices that include the most popular ones, \ie those that are repeatedly considered in the literature, and, at the same time, are highly representative of the attributes discussed in the survey:
%We then compute the node rankings they impose over a wide
%range of ISP network topology datasets and correlate them.
%We have selected to work with seven characteristic
%Those interpretations of point centrality are
the Degree (DC)~\cite{Freeman}, Betweenness (BC)~\cite{Freeman}, Closeness (CC)~\cite{Freeman},
Eigenvector (EC)~\cite{power}, Harmonic Centrality (HC)~\cite{harmonic}, Pagerank 
(PG, with the damping factor $d$ set to $0.85$ as typically used in literature)~\cite{pagerank} and
Eccentricity (ECC)~\cite{eccentrcity}.

In Table~\ref{tab:seven_indices} we characterize them according
to the aforementioned classification attributes. Also the formal normalized definitions are recalled for each index 
%The computational properties of these indices,~\ie 
along with the running time of the algorithms utilized for their computation, 
%are listed in the third column of the table 
as function of the node $V$ and edge $E$ sets of a graph $G=(V,E)$.
% as a function
%of the adjacency matrix of the graph $G=(V,E)$.
%As detailed in subsection~\ref{subsec:topologies}, the employed Internet topologies are connected, undirected and unweighted graphs $G(V,E)$ of $|V|$ nodes and $|E|$ edges.

%\input{methodology}

\section{Correlation study of centrality indices}\label{sec:correlations}
In almost all instances, where centrality indices inform communication network protocols, what matters is the \emph{ranking} of nodes that is induced by the indices rather than the absolute centrality values. These rankings are subsequently used in the decisions made by the respective protocols.
For example, in~\cite{Daly09,BubbleRap}, the rankings determine whether a Delay Tolerant Network (DTN) node will forward a message to another DTN node it encounters; in~\cite{cacheICN}, whether a content item will be cached at a Information-Centric Networking (ICN) node or not; and in~\cite{adamic} whether to search for a file in a given unstructured Peer-to-Peer(P2P) network node or not.
Likewise, in vulnerability analysis of the service migration protocol in \cite{ITC}, it is the \emph{set} of the $k,~k<|V|$ most highly-ranked nodes that matters, again irrespective of their actual centrality values. The question that plausibly arises in every case is how similar are the rankings generated by each centrality index.

In this section, we carry out a thorough correlation study of these rankings, as computed over a broad set of ISP router-level topologies.
The study proceeds in two steps. First, (\textbf{Step 1}) we calculate for each network topology and each node in it the seven centrality index (see subsection~\ref{employed_indices}) values, thus generating seven different node rankings per topology.
%Referring to point rankings, we consider sorted orders of node IDs with respect to their centrality values.
Then, (\textbf{Step 2}) we compute pairwise correlation measures over these rankings. Two different measures are considered, one accounting for the full node rankings and the other only for the highly-ranked nodes.
%rank (\ie Spearman, Kendall) and linear (\ie Pearson) correlation
%coefficients with a 95\% confidence, to assess the statistical dependence between metrics.
%Given the favorable computational properties of the degree centrality, we are especially concerned with the way it correlates with other harder-to-obtain indices.

\subsection{Index correlation measures and router-level topologies}

\subsubsection{Index correlation measures}
%or, that a protocol would not need to synchronize all, potentially highly heterogeneous, nodes with respect to the index they compute and use.
%As such we are mostly concerned with the correlation of nodes rankings (rather than the indices' absolute values).
%The correlation coefficients that become relevant are those that capture similarity of full node rankings
%Alternatively, we may focus only on the smaller subsets of $k$ nodes that exhibit the top-centrality
%values~\cite{ITC} and explore the overlap of these sets, as determined by different centrality metrics.
The first correlation measure is the nonparametric Spearman's rank-correlation coefficient, $\rho$, and is computed over the full node rankings. The coefficient assesses how well a monotonic function can describe the rankings induced by the two centrality indices on the network nodes. For a given network topology node set $V$, it is given by:
% We capture the rank correlation in the
% non-parametric Spearman measure of correlation, $\rho$, which
% assesses how monotonic is the relationship between the ranks of the two
% centrality variables.
%  and is computed as follows:

{\small
\begin{equation}\label{eqn:top_k}
%betweenness centrality
\rho_V(C_1,C_2) =1-\frac{6\sum\limits_{u \in V}(r_{C_{1}}(u)-r_{C_{2}}(u))^{2}}{|V|(|V|^{2}-1)} \nonumber %\sum_{s=1}^{|V|}\sum_{t=1}^{s-1}
\end{equation}}

\normalfont
\noindent
where $r_{C_1}(u)$ and $r_{C_2}(u)$ are the ranks of node $u$ in line with centrality indices
$C_1$ and $C_2$, respectively. The coefficient values lie in $[-1,1]$, with high positive (negative) values denoting strong positive (resp. negative) correlation\footnote{We have also computed the other popular rank-correlation coefficient, Kendall's $\tau$. In general, these two non-parametric coefficients, Spearman's and Kendall's, produce similar results, as will be reported later in Figure~\ref{fig:graphs}
and in more details in the Appendix.
%~\cite{thesis}.
%, the latter being computationally heavier (\textbf{???}).
Finally, for the sake of completeness we have also computed the well-known Pearson coefficient $r$ that assesses how linear is the relationship between
the actual values of the indices rather than the rankings they induce.
Typically, the highly rank-correlated centrality pairs have also been found to exhibit
considerable linear correlation, yet of lower strength~\cite{thesis}.}.

The second correlation measure is the percentage overlap between the sets of the $k$ most highly ranked (top-$k$) nodes that are generated by two centrality indices.

{\scriptsize
\begin{equation}\label{eqn:overlap}
%betweenness centrality
%\hspace{-0.3in}
ov_V(C_1,C_2;k)=\frac{|\{v\in V:r_{C_1}(v)\leq k \}\bigcap \{v\in V:r_{C_2}(v)\leq k\}|}{k}\cdot 100 \% \nonumber
\end{equation}}

\normalfont
\noindent
Contrary to the Spearman's coefficient, the percentage overlap is computed over a subset of the full node rankings and takes values in
$[0,100]$.

The relevance of the two measures depends on the usage context of centrality-based ranks. The decisions that relate to the DTN forwarding, CCN caching and P2P node search examples rely on full node rankings; whereas, vulnerability analysis is usually concerned with the subset of nodes that are important (``central'') for the network. High correlation between the rankings of two indices implies that a computationally complex or intractable index can be approximated by a simpler one without significant penalties for the intended protocol operation or the conclusions of the vulnerability analysis.
%Scaling this result across all pairs of indices, would mean that the node ordering is not sensitive to actual centrality index.

%Then, we let the rankings of nodes introduced by each index, determine the attacks
%over network nodes and assess the impact on the Internet graphs connectivity and throughput.

%In what follows we briefly present the centrality indices,
%the statistical tools and the Internet topologies we have used
%in our experimental study of Sections~\ref{sec:correlations}~and~\ref{sec:robustness}.

%\subsection{Network topologies}
%\label{subsec:topologies}
 %graphs for our experiments.
%Undirected and unweighted graphs
%We briefly the under the two broad categories
%The former serve as the basis for computing centrality indices, correlations and
%assessing the impact of node attacks to the Internet connectivity; %at the router-level (preseneted in subsection~\ref{ISPs})
%the latter are used for the relevant assessment in terms of network throughput. %the latter %a second smaller

\subsubsection{Router-level ISP topologies}\label{ISPs}
% Over the last decade or so, a number of projects managed to discover commercial
% (or not) ISP networks, unveiling information about their structure.
All experiments in this paper are carried out over datasets collected in the context of
%For the needs of the Section~\ref{sec:robustness} experiments, we exploited the
four projects. Four of them relate to measurement projects and are referred to as Rocketfuel~\cite{rocketFL}, CAIDA~\cite{caida}, and mrinfo (Tier-1 and Transit)~\cite{PAM10} datasets, respectively.
They report \emph{binary} router-level graphs\footnote{Many of the original
network topology files, as released in a raw trace-based format, miss some edges.
%, there may be more than one connected components in a given AS dataset.
We have therefore used a wellknown linear-time algorithm~\cite{tarzan} to retrieve the
giant connected component (GCC).} for different Internet ASes. On the contrary, the last dataset, called the Topology Zoo dataset, contains \emph{capacitated} topologies at the router- and Point-of-Presence (PoP) level~\cite{zoo}, as collected directly by network operators of primarily academic and research networks.
The basic properties of the all datasets are summarized in Tables~\ref{tab:nets} and \ref{tab:ZOOnets} in the Appendix.
%Overall, our main focus has been on the largest topologies available in each dataset.
%composed by undirected edges. The final snapshots contain for each AS, a symmetric adjacency matrix
%of the giant connected component

\paragraph{Rocketfuel dataset}
The Rocketfuel dataset~\cite{rocketFL} is the chronologically oldest dataset,
drawn with the help of the \verb|traceroute| active measurement tool.
The Rocketfuel engine collected raw \verb|traceroute| data from public BGP tables,
processed them and extracted router-level networks by mapping diverse ISP routers to ASes.
Ten diverse ISPs across the world were mapped utilizing approximately 800
\verb|traceroute| sources hosted by nearly 300 servers.

\paragraph{CAIDA dataset}
CAIDA topologies (IDTK 2011-10)~\cite{caida}, the most recent release out of
our datasets, were obtained by means of an active measurement infrastructure known
as Archipelago; it performed \verb|traceroute| probes to randomly-chosen destinations,
located in 29 countries worldwide within the interval of Oct 24 to Nov 3, 2011.
At first, publicly available BGP dumps were used to map IP addresses to ASes
relying on several tools for alias resolution. Then, heuristic rules~\cite{caida} properly
assigned each router to the AS it belongs.
%To obtain the adjacency matrix for each AS available in the dataset
%we have first collected the ID of the nodes that are operated by each AS
%(of interest) using the available ``.nodes.as'' file. The node-AS file assigns an AS to
%each node found in the nodes’ file as inferred by using the Election+Degree assignment heuristic.
%We next parsed the trace to identify whether there is a link or not between each pair of the set of
%nodes operated by the corresponding AS.

\paragraph{Mrinfo datasets}
The Mrinfo~\cite{PAM10}, dataset was collected during the period 2005-2008 and contains 264 Tier-1, 244 Transit,
and 342 Stub ISP network topology files. To cope with \verb|traceroute| inaccuracies, the dataset
collection was extracted by the new \verb|mrinfo| tool which silently crawls IPv4 addresses only,
based on the Internet Group Management Protocol (IGMP). The advantage of this tool is twofold;
it can efficiently discriminate interconnections between ASes without suffering from IP alias resolution
problems and, besides layer-3 devices, detect the presence of level-2 hardware (switches) between
interconnections of routers for each AS. In our study we considered only the largest available
snapshots corresponding to two datasets, Tier-1 and Transit, leaving aside the small-sized Stub topologies.

\paragraph{Topology Zoo dataset}
Whereas previous studies employ a number of route discovery tools to reveal the Internet connectivity, the Topology Zoo gathers the maps of more than 140 real-world topologies directly from the network operators, including layout views of the optical fibers used for both commercial (COM) and Research \& Education (REN) networks.
As the resulting maps (topologies and associated attributes) come from the owner and/or manager of the network, they are claimed to reflect an accurate network view circumventing any errors due to biases of measurement techniques.
Out of the 232 network graphs included in the Zoo, we have carefully singled-out
the largest capacitated router-level snapshots (see Table~\ref{tab:ZOOnets} in the Appendix) representing
the topologies\footnote{Especially for Uninett I and II networks, there are three snapshots of the same topology but with
different link capacities; the original Uninett networks have a couple of links which are tagged with capacities given
in min-max format and therefore we have reproduced the three snapshots by taking the minimum, maximum and mean value of
the given capacity interval. } of 11 different European, one Asian and one cross-European research networks as traced during
the period 2008-2011.
%The attractiveness of these datasets lies in the fact that their links are capacity-tagged.
We will use these topologies to evaluate the traffic-serving capacity of networks (see Section~\ref{subsec:flow}).
%entitled ``max'', ``min'' and ``mean'', respectively.)

\subsection{Results}

\subsubsection{Full-ranking correlation over binary graphs}
\label{subsec:full_correl}

\begin{figure*}%[ht]
% \vspace{-0.22 in}
\begin{center}
\begin{tabular}{cccc}
\includegraphics[width=0.25\textwidth]{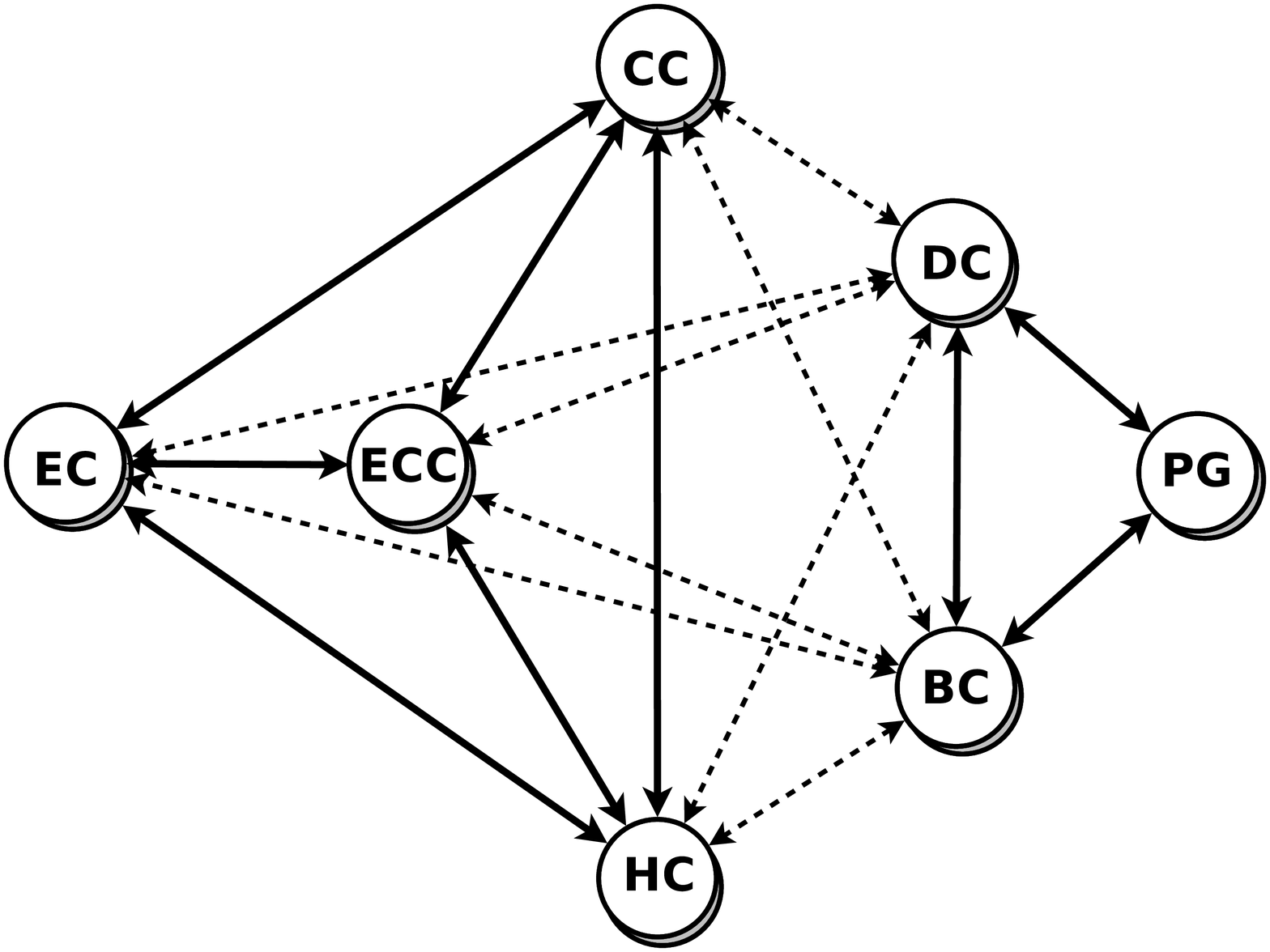} % arxika sto 0.17
&
\includegraphics[width=0.25\textwidth]{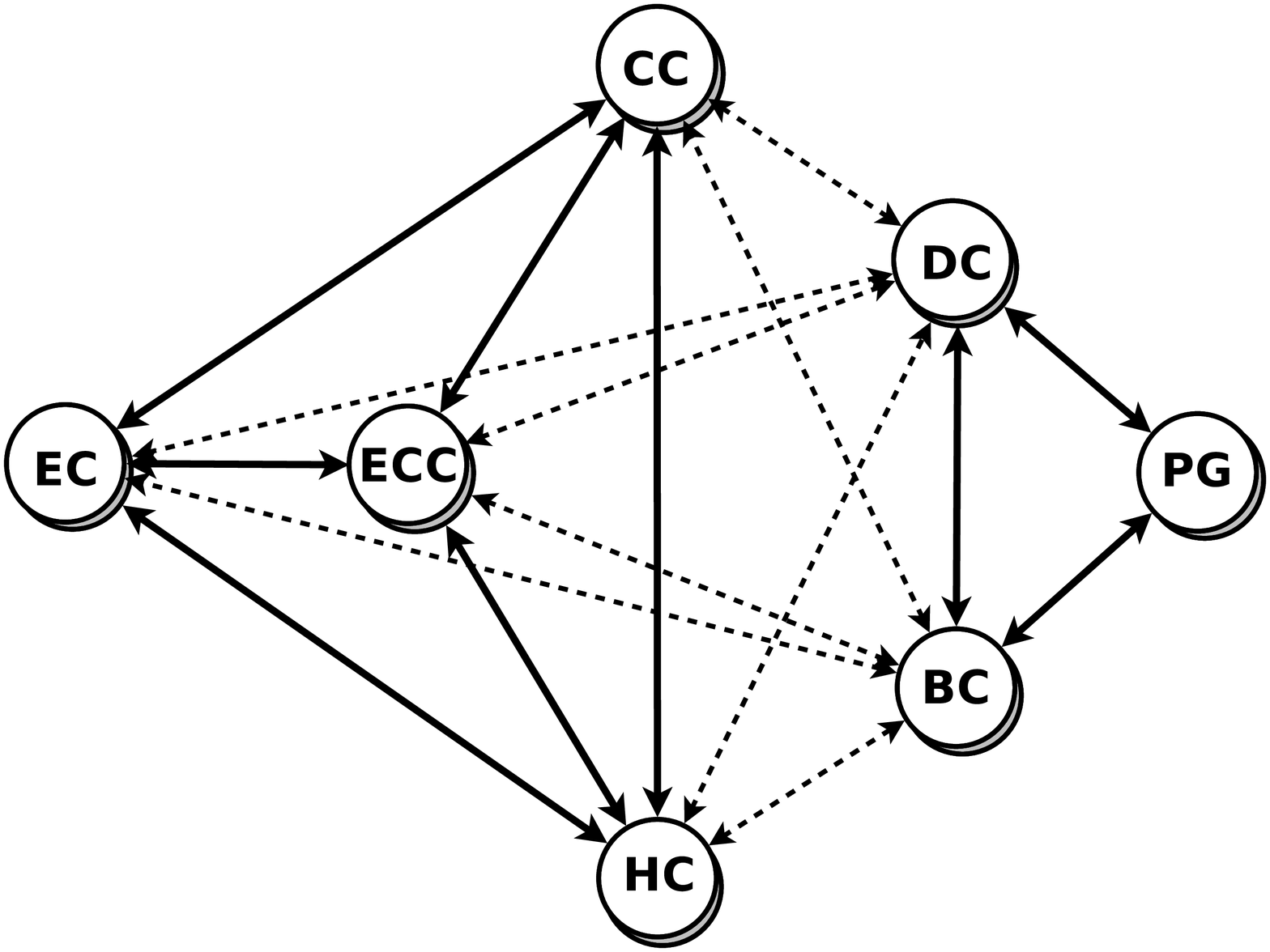}
&
\includegraphics[width=0.25\textwidth]{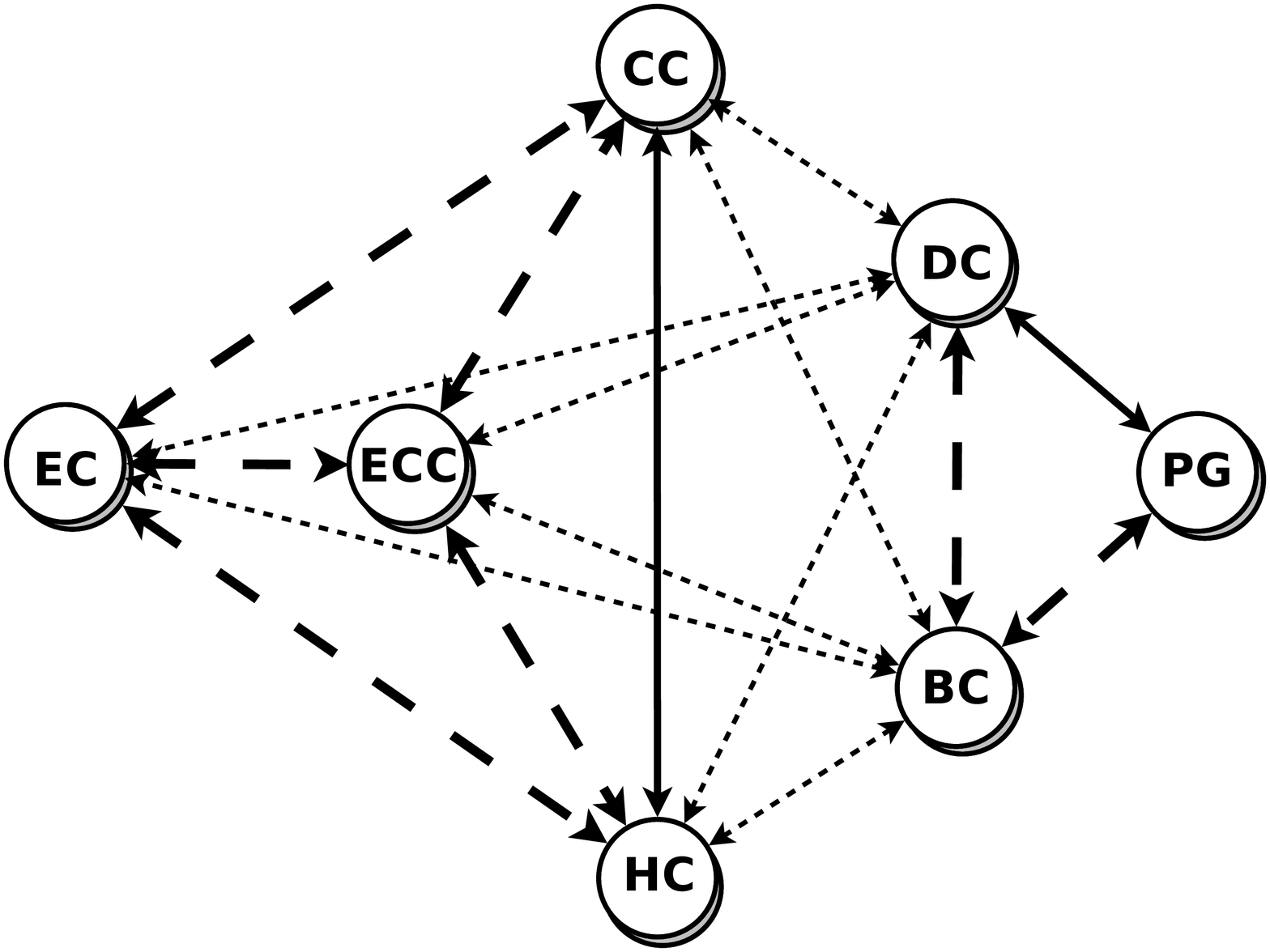}\\
\begin{scriptsize} a.  \end{scriptsize}  &
\begin{scriptsize} b.  \end{scriptsize} &
\begin{scriptsize} c.  \end{scriptsize}
\end{tabular}
\end{center}
\caption{Graph-based illustration for the average values of the Spearman (a), Kendall (b) coefficients and top-5\% overlap (c) among centrality indices. In (a) and (b) solid bold and dashed plain lines denote coefficients in the intervals [0.7-1],[0.3-0.7), respectively. In (c), solid bold, dashed bold and dashed plain lines denote overlap value higher than 70\%, between 40-70\%, and lower than 40\%, respectively. A special note involves the BC-CC and BC-HC pairs that exhibit increased values compared to this rule.}
%plain solid lines denote coefficients/(\%)overlap  . Dashed bold show moderate correlation }
\label{fig:graphs}
%\vspace{-0.15 in}
\end{figure*}

The results follow a similar trend over the four datasets
%==================================================
%\footnote{Tier-1 and Transit ISP network topologies are considered separately in this section-WHY?.}
% the are considered separately in the whole study
%==================================================
so that the rank correlation between the studied indices can be summarized in the graphs of Fig.~\ref{fig:graphs}. 
This graph-based illustration represents all AS topologies\footnote{The summarizing graphs have been derived 
considering the corresponding averages of every coefficient for each dataset. The relevant Kendall coefficient and top-$5\%$ overlap averages (per dataset) appear, respectively, in Table~\ref{kendal_correl_results} and~\ref{tbl:topk_av} of the Appendix.} since they exhibit similar coefficient values for every distinct centrality pair.  
%===================================================================
% further support to that claim may provide 4 tables (in the Appendix)
% with correlation values per dataset
%===================================================================
Notably, none of the possible centrality pairs have been found to be negatively correlated over any of the studied topologies. With this in mind, we empirically characterize the correlation strength as \textit{high} and \textit{low} when the corresponding coefficient exhibits a value in the interval [0.7,1] and [0.3,0.7), respectively; the two indices are actually uncorrelated when coefficients lie in [0-0.3).
%Note that in all relevant experiments we only observe positive correlations.
%we consider two intervals of coefficient
%values where , respectively;
%the first \ie higher than \textbf{XX} and lower than \textbf{XX}.
Accordingly, bold edge-lines (solid or dashed) denote high correlation values between two centralities, whereas plain edge-lines denote low values. We have omitted the connections for those index pairs %f indices
that do not exhibit any kind of correlation.
%However,

In what follows, index pairs of interest are discussed in more detail. Where appropriate we draw links to studies reporting similar or different results on different kinds of networks.
%Unless otherwise noticed, we will focus on the Spearman rank-correlation;
%the relevant values appear in
In Table~\ref{rank_correl_results} each row (\ie top to bottom) in every box reports averages measured over the CAIDA, Rocketfuel, MrInfo -Tier1 and -Transit datasets, respectively. The interested reader is referred to~\cite{thesis} for the full set of results while %is provided in.
in the Appendix (\ie Table~\ref{kendal_correl_results}) she can find %, we have chosen to present the average
%Spearman (Table~\ref{rank_correl_results}) and
the respective table when the Kendall $\tau$ is used as correlation measure.
%estimated over each of the three datasets.

%\vspace{-0.1in}
\begin{table}[ht]
\centering \caption{Averages of Spearman coefficients for all datasets}
\label{rank_correl_results} {\ssmall %\scriptsize %\
\begin{tabular}{|r| c c c c c c c l|}
\hline
\multicolumn{1}{|r|}{}&\multicolumn{1}{c}{CC}&\multicolumn{1}{c}{HC}&\multicolumn{1}{c}{EC}&\multicolumn{1}{c}{ECC}&\multicolumn{1}{c}{DC}&\multicolumn{1}{c}{BC}&\multicolumn{1}{c}{PG}&\multicolumn{1}{l|}{dataset}\\
% & & &
 %\cline{1-7}
%  & & ego-network \textbf{(r=1)} & ego-network \textbf{(r=2)}\\
\hline
CC   & 1   &    &  &  & &    &    &CAIDA\\
     & 1   &    &  &  & &    &    &RocketFuel\\
     & 1   &    &  &  & &    &    &MrInfo-Tier1\\
     & 1   &    &  &  & &    &    &MrInfo-Transit\\
\hline
HC & 0.99 &     1 &    &  &  &  & &-//-\\
   & 0.98 &     1 &    &  &  &  & &\\
   & 0.95 &     1 &    &  &  &  & &\\
   & 0.99 &     1 &    &  &  &  & &\\
\hline
EC & 0.93 &   0.95 & 1 & &  &  &     &-//-\\
   & 0.80 &   0.83 & 1 & &  &  &     &\\
   & 0.66 &   0.69 & 1 & &  &  &    &\\
   & 0.86 &   0.88 & 1 & &  &  &    &\\
\hline
ECC & 0.84& 0.84 & 0.84 & 1 &  &  & &-//-\\
    & 0.73& 0.67 & 0.56 & 1 &  &  & &\\
    & 0.80& 0.69 & 0.52 & 1 &  &  & &\\
    & 0.89& 0.88 & 0.75 & 1 &   &  & &\\
\hline
DC &0.28& 0.28 & 0.29 & 0.25&1&     & &-//-\\
   &0.48& 0.53 & 0.45 & 0.38&1&     & &\\
   &0.43& 0.59 & 0.47 & 0.30&1&     & &\\
   &0.50& 0.55 & 0.49 & 0.45&1&     &  &\\
\hline
BC &0.29 & 0.29 & 0.28 & 0.27 & 0.90&1& &-//-\\
   &0.45 & 0.50 & 0.40 & 0.37 & 0.94&1& &\\
   &0.50 & 0.61 & 0.30 & 0.38 & 0.69&1& &\\
   &0.54 & 0.58 & 0.48 & 0.47 & 0.88&1& &\\
\hline
PG &0.04& 0.05 & 0.05 & 0.04 & 0.86 & 0.80 & 1&-//-\\
   &0.25& 0.30 & 0.14 & 0.20 & 0.83 & 0.80 & 1&\\
   &0.34& 0.49 & 0.30 & 0.24 & 0.90 & 0.74 &1&\\
   &0.40& 0.44 & 0.36 & 0.35 & 0.92 & 0.88 &1&\\
\hline
\end{tabular}}
%\vspace{-0.12 in}
\end{table}
\normalfont

 % (either solid or dashed)

\textbf{Betweenness \vs~Degree centrality}
% The Degree centrality provides, at least phenomenally, a completely different notion of centrality than the Betweenness centrality does. Specifically, DC takes into account only the node's local neighbors, whereas BC considers the position of the node within the whole network topology. Yet, the two are found consistently highly \textbf{rank-correlated}. A number of earlier studies~\cite{ppantaz,correl_complex,AS_properties} report positive \textbf{Pearson} correlation between the DC and BC indices over a wide range of network structures such as random graphs and real-world complex networks. This correlation is intuitive to some extent since high-degree nodes have better chances to be parts of the shortest paths linking node pairs. % of vertices.
% %Despite the fact that degree is much
% %easier to calculate compared to betweenness,
% However, in some cases the DC index can evaluate nodes' position very differently than BC; it may overestimate the importance of nodes belonging to isolated subgraphs (high DC-low BC) or underestimate the role of nodes
% %vertices
% acting as bridges between groups of nodes (low DC-high BC).
% %being connected with only a small number of other nodes.
Degree centrality (DC) captures, at least phenomenally, a completely different notion of centrality than Betweenness (BC). DC takes into account only the node's local neighbors, whereas BC considers the position of the node within the whole network. Therefore, in some cases DC can evaluate nodes' position very differently than BC; it may overestimate the importance of nodes belonging to isolated subgraphs (high DC-low BC) or underestimate the role of nodes acting as bridges between groups of nodes (low DC-high BC). On the other hand, high-degree nodes have better chances to be parts of the shortest paths linking node pairs.
%being connected with only a small number of other nodes.
In our datasets, the two indices are found consistently highly correlated, in agreement with earlier studies~\cite{ppantaz,correl_complex,AS_properties} that report positive \emph{Pearson} correlation between DC and BC over a wide range of networks such as random graphs and real-world complex networks. % of vertices.
%Despite the fact that degree is much
%easier to calculate compared to betweenness,
%\newline

\textbf{Pagerank \vs~Degree centrality.} Another interesting result, that is immediately apparent from Figure~\ref{fig:graphs}, is the strong correlation between Pagerank and Degree centrality. Pagerank is principally defined for digraphs discriminating between incoming and outgoing connections at each node.
For undirected general graphs, Grolmusz~\cite{PG_undirect} shows that Pagerank is
statistically close to the degree distribution but not identical. %(FOR WHAT GRAPHS IS THIS SHOWN?).
Furthermore, taking into account the aforementioned strong BC-DC
correlation, a triangle-like schema emerges and may be of practical importance as it relates the only local index (\ie Degree) with globally-determined ones (\ie Pagerank and Betweenness centrality). Interestingly, significant positive correlation between these three indices (PG-DC-BC), with $\rho$ values in [0.66, 0.95] for all three index pairs, is also reported by Yan and Ding~\cite{coauthorship} over coauthorship real-world data (directed graphs).
%for their experiments in the field of library and information science.
%They report that the same group of centralities (PG-DC-BC) exhibits significant
%positive rank correlation.
%at the 0.01 level. Specifically, they estimate

% \begin{figure}[ht]
% \centering
% \includegraphics[scale=0.23]{./figs/damping_factor}
%   %\hspace{0.5cm}
% \caption{Rank-correlation scaling as the Pagerank increasingly depends on DC and BC for AS1299.}
% \label{fig:PG_DC}
% \end{figure}

\begin{figure}[ht]
\begin{center}
%\advance\leftskip-1.4cm
\begin{tabular}{ccc}
 \includegraphics[width=0.22\textwidth]{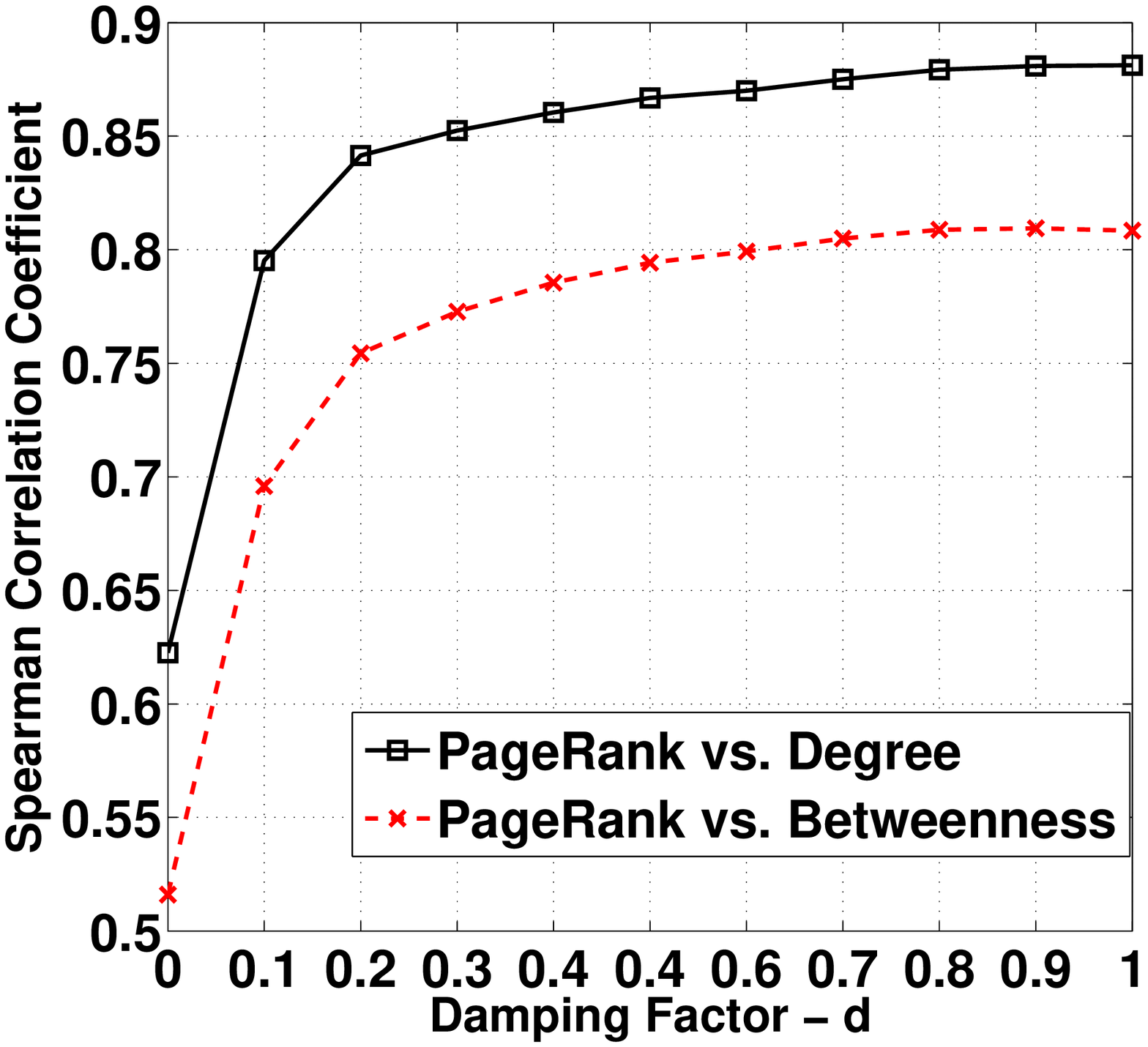} % arxika sto 0.17
 &
 \includegraphics[width=0.23\textwidth]{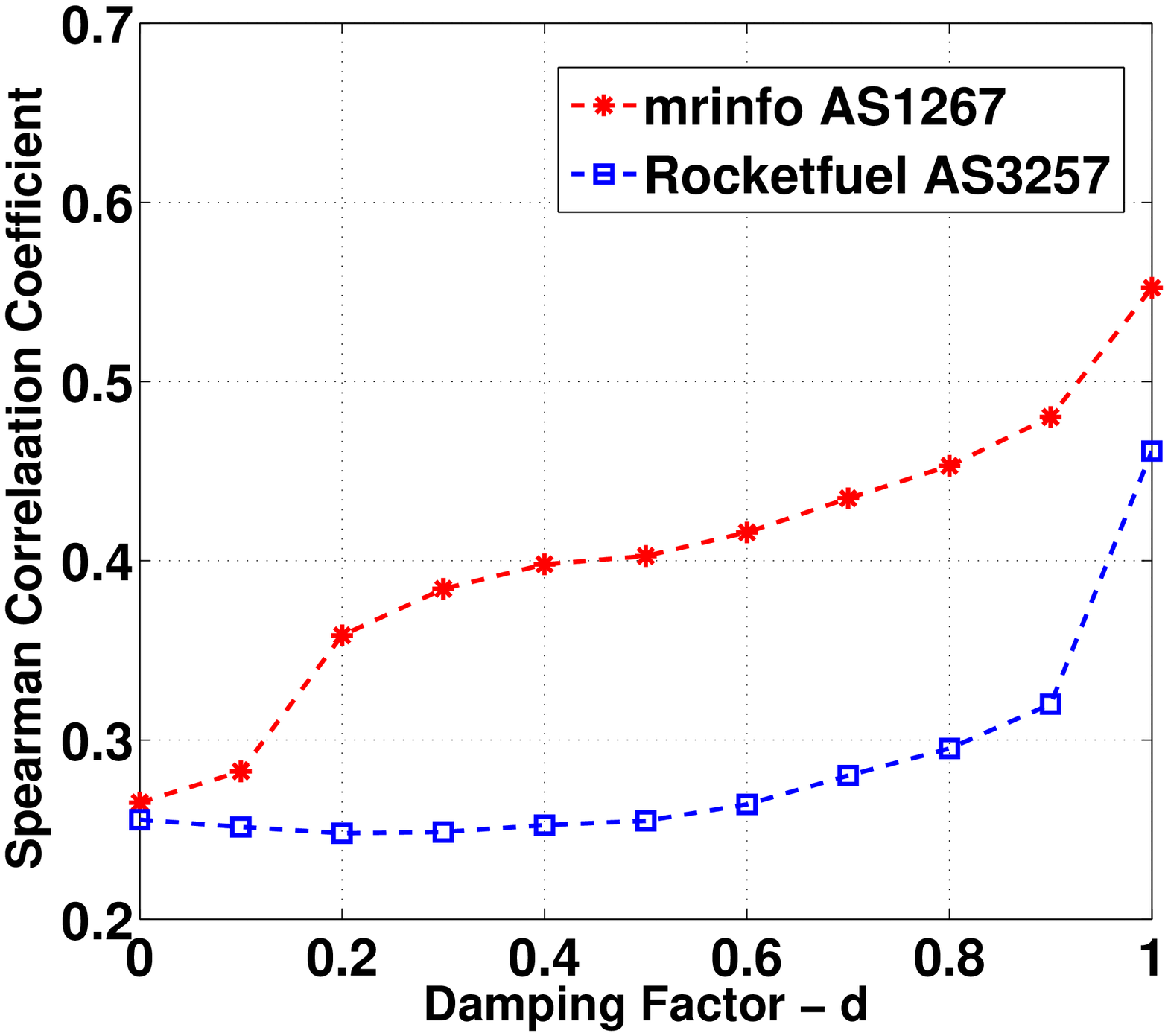}\\
%  &
%  \includegraphics[width=0.28\textwidth]{./figs/PDF_IFconnectivity}\\ %untitled
 \begin{scriptsize} a.   \end{scriptsize}  &
\begin{scriptsize}  b.   \end{scriptsize} \\%&
%\begin{scriptsize}  c. All datasets \end{scriptsize} \\
\end{tabular}
\end{center}
\caption{a) Rank-correlation scaling as the Pagerank increasingly depends on DC and BC for AS1299.
b) Rank-correlation between EC and PG as a function of the damping factor $d$ for indicative ASes.}
\label{fig:PG_DC}
%\vspace{-0.15 in}
\end{figure}

%From a different point of view,
Figure~\ref{fig:PG_DC}.a describes the monotonic increase of the PG-DC correlation
with the damping factor $d$ of PG.
%For all values of $d$ starting from zero, the correlation level increases and remains constantly positive until $d$
%reaches unity.
Pagerank~\cite{pagerank} is often approached as the steady-state distribution of the frequency of visits to the network nodes by a random walker who each time either jumps towards another arbitrary network node with probability (1-d)/N, where N the cardinality of the network node set, or randomly
follows an outbound link towards a neighboring node. As $d$ increases %Figure~\ref{fig:PG_DC} only demonstrates that as $d$ increases, the PG operation resembles that of
%option tend to be similar with the degree centrality notion as the damping factor increases.
the walker's long jumps get rarer and only 1-hop steps are feasible.
%By setting $d$ equal to 1, the random surfer cannot move utilizing long jumps at random destinations in the network. Instead, he disciplines over the direct neighbors of its current position.
Figure~\ref{fig:PG_DC}.a (dashed line) also shows similar association between
the damping factor and the PG-BC $\rho$ values.

\textbf{Pagerank \vs~Eigenvector centrality.} 
PG, and EC centrality are the two spectral indices we experiment with. Both express the stationary probability of a random surfer to reside on some page while moving on the Web graph. Hence, one would expect some positive correlation between these indices. However, our results indicate the absence of such a relationship. A possible cause is that their actual interpretation differs as, contrary to EC, the PG Centrality utilizes the damping factor $d$ to determine the ``jump'' probability. However, as a couple of indicative experiments suggest (Fig.~\ref{fig:PG_DC}.b), the rank correlation between the two metrics increases yet does not reach very high values as $d$ moves to unity \ie the surfer moves only to neighboring pages. 
It seems then that $d$ can only partially justify the poor PG-EC correlation strength; 
as the PG formula suggests (Table~\ref{tab:seven_indices}) a node's (\ie Web page) PG rank value 
is evenly divided ($L_u$ term) over its neighbors, which for the case of undirected graphs corresponds to its DC value.  
The fact that DC index is found to be weekly correlated with EC (Table~\ref{rank_correl_results}) can further distort any anticipated PG-EC correlation.

\textbf{Eccentricity \vs~Closeness centrality.} Another strong correlation that we observe in our correlation study involves the Eccentricity and Closeness centrality indices.
%To acquire a better view, we attempt to further explicate this relation by showing that when a node $n_1$ exhibits a greater eccentricity value than a node $n_2$,
%we expect $n_1$ to have also a greater closeness centrality value than node $n_2$.
Recalling the definitions of the two indices (ref. Table \ref{tab:seven_indices}), there is absolute positive ECC-CC correlation if it holds:
\begin{equation}
\label{eq:ecc}
ECC(n_1) > ECC(n_2)
\end{equation}
whenever
\begin{equation}
\label{eq:cc}
CC(n_1) > CC(n_2)
\end{equation}
We can rewrite eq.~\ref{eq:ecc} as
\begin{equation}
\label{eq:max}
max_{j \in V} d_{n_2,j}> max_{j \in V} d_{n_1,j}
\end{equation}
and eq.~\ref{eq:cc} as
\begin{equation}
\label{eq:avr}
\sum_{j \in V} d_{n_2,j}> \sum_{j \in V} d_{n_1,j}
\end{equation}
Hence, the question becomes when the order in maximum index values (eq.~\ref{eq:max}) is also preserved for their averages (eq.~\ref{eq:avr}) over a certain graph. This holds in several trivial graphs (\eg line graph, rectangular grid) but not in all graphs. One simple counterexample is the 4-node star network with a 2-node line graph attached to one of its leaf nodes (compare the two indices for the hub node and the leaf node, where the line graph is attached).

\textbf{Additional remarks.} There exist further centrality pairs yielding positive correlations, which are less straightforward to reason about. For instance, in our results, high rank correlation has been observed for pairs such as Eigenvector-Harmonic and Eigenvector-Closeness centrality.
%while the corresponding linear correlation is found weaker, with
%Pearson $r$ lying around 0.5.
%with previous works
These findings seem consistent with previous results.
%Precisely,
Iyer~\etal~\cite{attack_robustness} have noticed that synthetic scale-free networks
(whose degree distribution follows a power law, at least asymptotically) present moderate positive Pearson CC-EC correlation. Higher values ($r$=0.61) are reported for the case of networks with exponential degree distribution.

\begin{figure}[ht]
\advance\leftskip-0.2cm
\begin{minipage}[b]{.5\linewidth}
\centering
\includegraphics[width=4.7cm]{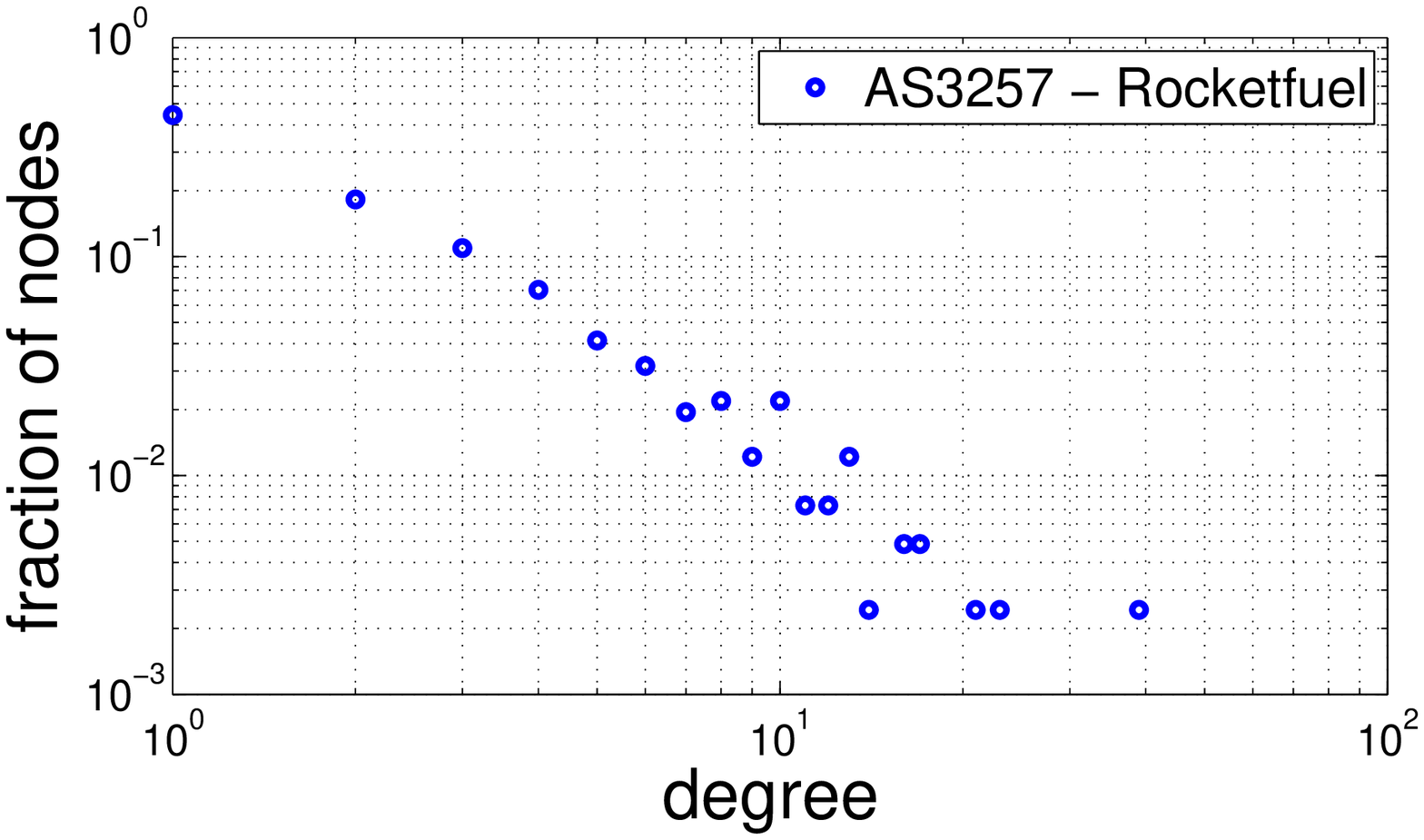}
\end{minipage}
\hspace{0.1cm}
\begin{minipage}[b]{0.3\linewidth}
\centering
\includegraphics[width=4.7cm]{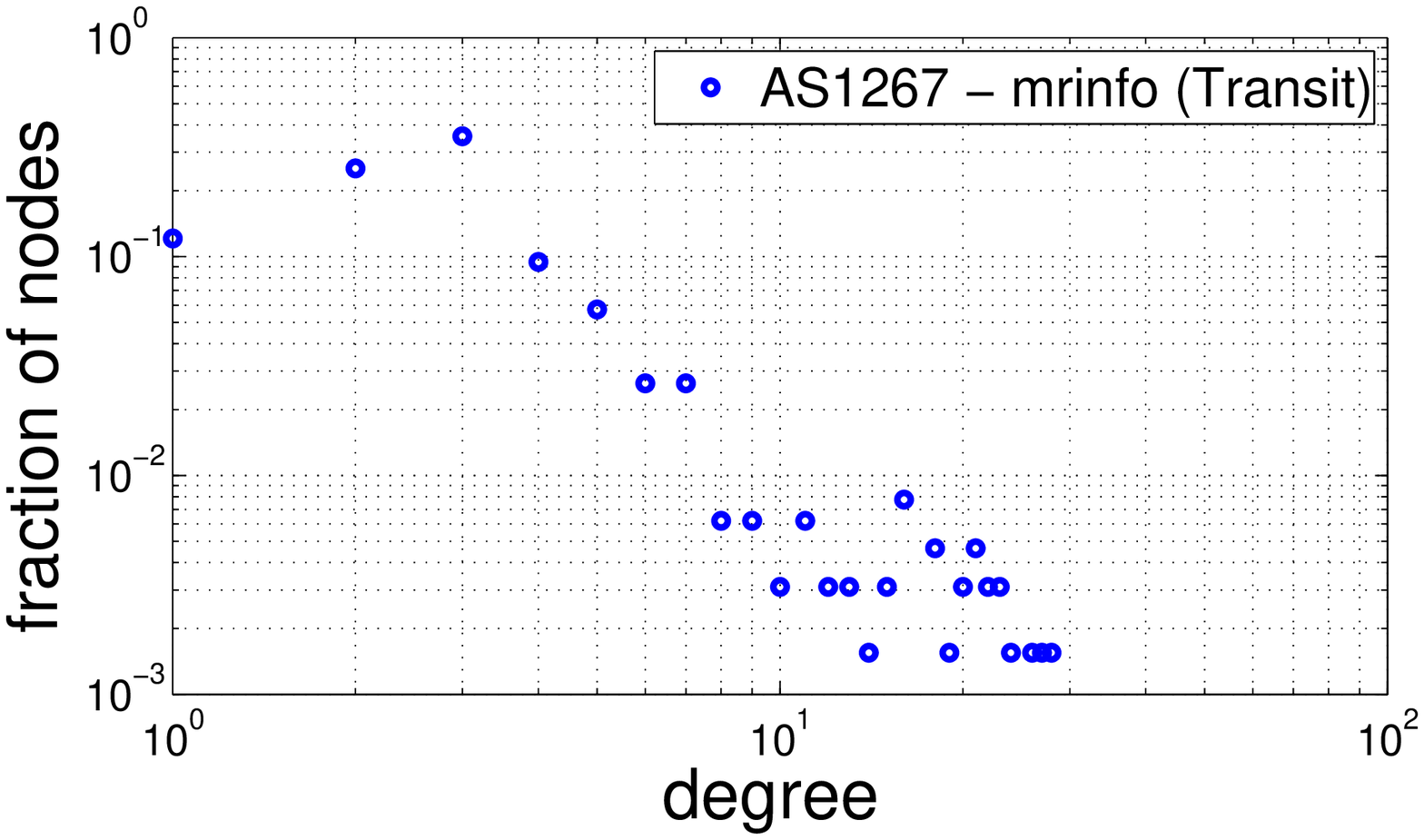}
\end{minipage}
\caption{Degree distribution for two indicative snapshots.}
\label{fig:DDs}
%\vspace{-0.14cm}
\end{figure}

Elaborating more on this thread, we have tried to identify how the degree distribution relates to the EC-CC correlation. In Figure~\ref{fig:DDs} left, we plot in log-log scale the degree distribution of a 411-node large AS out of the RocketFuel datasets, as a representative sample, with positive Pearson correlation between EC and CC ($r$=0.65). The straight-line points to power-law degree distribution suggesting that %essentially confirming
%suggesting that
this may be beneficial for the positive correlation, as in~\cite{attack_robustness}.
%the remark of Iyer~\etal ~may be relevant also for ISP network topologies.
%
On the other hand, the scale-free property is not a necessary condition for high EC-CC correlation. %For instance,
In Figure~\ref{fig:DDs} right, the degree distribution of a 645-node large \verb|mrinfo| Transit AS clearly deviates from a power-law pattern, yet it features a considerably higher Pearson coefficient ($r$=0.78). Similar remarks hold for the EC-CC rank correlation over these snapshots (where we measure the corresponding Spearman $\rho_V$=$0.88$ for the former and $\rho_V$=$0.96$ for the latter one).
%\textbf{(xx and yy, respectively for the two topologies (ADD VALUES)}.
%Similar comments are due for the correlation between EC and HC.
%taking into account the foretold interconnection
%for the pair CC-HC.
% One more noteworthy observation is the lack of correlation between Eigenvector and Degree centrality.
% We would expect correlation since EC might be considered as a refined recursive version of DC~\cite{correl_complex}.
% However, Wang~\cite{correl_complex}, using simulated random graphs, experimentally showed that
% only in directed networks where few nodes both send and receive ties, EC and DC are highly correlated.

\subsubsection{Top-\textit{k} percentage overlap over binary graphs}
So far, our correlation analysis has taken into account the full rankings produced by the seven centrality indices. 
We now focus our attention on the top-5\% most central nodes identified by each index and investigate how large are the overlaps between different rankings. Our motivation for this set of experiments is the existence of network protocol instances that typically seek to exploit a small set of the top-central nodes~\cite{ITC}. Likewise, vulnerability 
studies of Internet graphs, as the one we carry out in Section \ref{sec:robustness},
are concerned with the subset of most central nodes.
%inline with the semantics each index assigns to centrality.

\begin{figure}[ht]
%\advance\leftskip-1.4cm
\centering
%\advance\leftskip-0.4 cm
\includegraphics[scale=0.24]{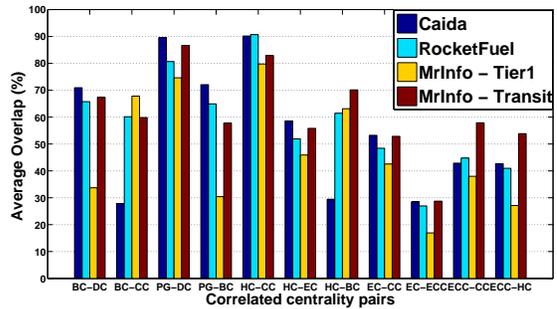}
  %\hspace{0.5cm}
\caption{Mean overlap (\%) of the top-5\% important nodes between centrality rankings over all ASes.}
\label{fig:top5_overlp}
\end{figure}

In Figure~\ref{fig:graphs}.c we show a summarizing graph-based illustration of the overlap scores among the seven centrality indices. The bold solid lines (\eg between CC-HC) denote what we consider as high top-5\% overlap between two centralities \ie beyond 70\%. The dashed solid lines (\eg between EC-HC) reflect overlap values between 40-70\%, whereas the dashed plain lines represent looser relations for the corresponding pairs. 
%\textbf{Moreover, different line styles roughly illustrate the strength of the relation among index pairs once reflected by the partial overlap of centrality rankings and the other by the correlation values}
%. Specifically, the double solid (\eg between HC-BC and CC-BC) represent an increase of the relation strength when passing from the rank correlation to the overlap measure. 
Additionally, figure~\ref{fig:top5_overlp} presents the average overlap of nodes over all ASes of each dataset for the most significant centrality pairs. On the one hand the overlap of some indices (such as BC-CC or HC-BC) appear to be highly sensitive to the considered topology, with differences that reach 40\% across different datasets. On the other, all pairs that are strongly correlated in terms of full rankings in~\ref{subsec:full_correl} appear to be more weakly associated in terms of overlap values\footnote{The characterization retains a loose empirical meaning; essentially, the comparison between a correlation coefficient and the \% overlap value is not straightforward.}. 
Exceptions to that rule are the HC-BC and CC-BC pairs that represent a slight increase of the relation 
strength when passing from the rank correlation to the overlap measure. 
Overall, only two of the centrality index pairs combine high overlap values with strong full rank-correlation (see Figure~\ref{fig:graphs}.a): PG-DC and HC-CC, both exhibiting larger than 80\% overlap in the top-5\% node rankings they induce across all datasets, whereas all the other pairs hardly exceed the 60\% value. This result should come as no surprise since rank correlation is determined over all network nodes rather than a subset of cardinality $k$.
%\textbf{Why is that?}

%??} \emph{THEY SHOULD BE READ IN BLACK and WHITE.}
\begin{figure}[ht]
\centering
\includegraphics[scale=0.37]{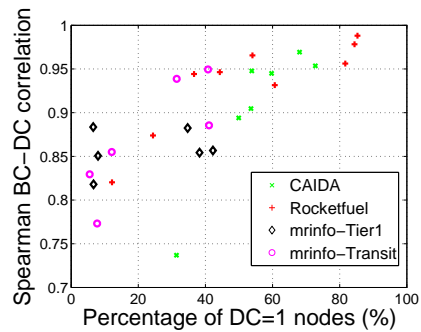}
  %\hspace{0.5cm}
\caption{Plot of the relation between the BC-DC rank correlation and the percentage of nodes with degree equal to one.
(A pair of outlier values for the mrinfo (Tier1) dataset have been ommitted.)}
\label{fig:scatter}
\end{figure}

\begin{table}[htp]
\centering \caption{Rank correlation strength vs. overlap (\%) between BC and DC}
 \label{tbl:corrVStopk} {\scriptsize
\begin{tabular}{c c c c }
\hline\hline
\multicolumn{1}{c}{Dataset-ID}&\multicolumn{1}{c}{BC-DC }&\multicolumn{1}{c}{Top-5\% }&\multicolumn{1}{c}{Fraction of nodes  }\\
\multicolumn{1}{c}{}&\multicolumn{1}{c}{Spearman Coefficient}&\multicolumn{1}{c}{ Overlap }&\multicolumn{1}{c}{having DC=1 }\\
%\cline{1-7}
\hline \hline
CAIDA-1557           &0.95  &53\%  &54\%   \\
RocketFuel-1239      &0.96  &85\%  &82\%   \\
MrInfo, Tier1-1239   &0.86  &54\%  &43\%   \\
MrInfo, Transit-3292 &0.94  &40\%  &32\%   \\
\hline\hline
\end{tabular}}
%\vspace{-0.12 in}
\end{table}

Let us look closer into the BC-DC pair. 
For these two indices, there is an apparent association between the nodes that are ranked 
in the last positions by the two indices; namely, nodes with the lowest DC value (\ie DC=1) exhibit as well the lowest BC value (\ie BC=0). Figure~\ref{fig:scatter} illustrates how the number of nodes with DC=1 %correspond to the
affects the rank correlation coefficients. It seems that (especially for the datasets of Caida and RocketFuel)
the Spearman values between the considered indices increase with the number of DC=1 nodes. These nodes are expected to positively contribute to the DC-BC correlation as they also exhibit the lowest-ranked betweenness value (\ie BC=0). At the same time, the ones with the top BC and DC values may not necessarily coincide as indicated in Table~\ref{tbl:corrVStopk}. The above results suggest that the high DC-BC correlation is mainly due to nodes of lowest ranks. This observation warns against the actual value of high Spearman rank correlation coefficients between two indices. On the other hand, the overlap measure does not suffer from similar biases. The repercussions of this will become clearer in the results of the Section~\ref{sec:robustness} experiments.

\subsubsection{Correlation/overlap results over capacitated graphs}
We have carried out a brief correlation study to identify
how the node rankings generated by the indices, relate over the topology Zoo dataset.
For determining the node rankings we had to carry out the centrality indices computations
over weighted graphs. This was mainly a question of computing shortest paths over weighted graphs.
Regarding the spectral indices, in the Topology Zoo experiments we only employ the EC index
that lends to a straightforward extension over the weighted graphs (see paragraph~\ref{subsec:graph_types}).
As such, we compute the Spearman correlation coefficients for the centrality pairs across all 18 snapshots.
Table~\ref{tbl:Zoo_correl} summarizes our results demonstrating the average
and variance values for the measured coefficients, respectively.
Those index pairs that were measured earlier to be strongly correlated
over the binary graphs (\ie Figure~\ref{fig:graphs}.a),
generally maintain similar relations over the capacitated Zoo networks.
A relevant comment involves the BC-DC correlation which is again found high yet not as close
to unity as before; ECC and EC appear in most cases highly correlated except for a few
topologies that contribute to %do not favor %and therefore
a high variance value for the coefficient average. %(\textbf{any insight?})
Clearly, these results are shaped by both the topology and the link capacity values
that are now taken into account for the corresponding index computations.

%\vspace{-10.1in}
\begin{table}[ht]
\centering
%\advance\leftskip-0.4 cm
\caption{Spearman Averages and variance for the topology zoo dataset}
\label{tbl:Zoo_correl} {\ssmall
\begin{tabular}{|c|| l l l l l l  |}
\hline
\multicolumn{1}{|l||}{}&\multicolumn{1}{l}{BC}&\multicolumn{1}{l}{CC}&\multicolumn{1}{l}{DC}&\multicolumn{1}{l}{EC}&\multicolumn{1}{l}{HC}&\multicolumn{1}{l|}{ECC}\\
% & & &
 %\cline{1-7}
%  & & ego-network \textbf{(r=1)} & ego-network \textbf{(r=2)}\\
\hline
\hline
BC & 1             &               &              &             &             & \\%     & \\
CC & 0.68$\pm$0.01 &  1            &              &             &             &\\%      & \\
DC & 0.75$\pm$0.01 & 0.85$\pm$0.02 &  1            &             &             &\\%      & \\
EC & 0.59$\pm$0.03 & 0.94$\pm$0.01 &  0.79$\pm$0.04  & 1           &             & \\%     & \\
HC & 0.68$\pm$0.01 & 0.98$\pm$0.01 &  0.87$\pm$0.01  & 0.95$\pm$0.01  & 1           &\\%      & \\
ECC& 0.64$\pm$0.01 & 0.87$\pm$0.04 &  0.80$\pm$0.02  & 0.77$\pm$0.16  & 0.86$\pm$0.04 &1   \\
%    & 0.92 &     1 &  0.74  & 0.75 & 0.42 & 0.44 &0.30\\
% \hline
\hline
\end{tabular}}
%\vspace{-0.12 in}
\end{table}

\begin{table}[ht]
\centering
%\advance\leftskip-0.4 cm
\caption{Average top-$15\%$ overlap (\%) for the topology zoo dataset}
\label{tbl:Zoo_topk} {\ssmall
\begin{tabular}{|c|| l l l l l l  |}
\hline
\multicolumn{1}{|l||}{}&\multicolumn{1}{l}{BC}&\multicolumn{1}{l}{CC}&\multicolumn{1}{l}{DC}&\multicolumn{1}{l}{EC}&\multicolumn{1}{l}{HC}&\multicolumn{1}{l|}{ECC}\\
\hline
\hline
BC & 1     &       &         &        &             & \\%     & \\
CC & 75.55 &  1    &         &        &             &\\%      & \\
DC & 78.38 & 84.36 &  1      &        &             &\\%      & \\
EC & 67.40 & 84.07 &  78.06  & 1      &             & \\%     & \\
HC & 75.55 & 89.63 &  85.10  & 87.77  & 1           &\\%      & \\
ECC& 69.99 & 77.91 &  73.62  & 71.24  & 73.09 &1   \\
%    & 0.92 &     1 &  0.74  & 0.75 & 0.42 & 0.44 &0.30\\
% \hline
\hline
\end{tabular}}
%\vspace{-0.12 in}
\end{table}

In Table~\ref{tbl:Zoo_topk} we present our results for the overlap between the $k\%$ top central
nodes averaged over the whole topology Zoo dataset. As their size is relatively small, we have chosen to set $k=15\%$
in order to each time avail vectors of at least 5 nodes' size. Compared to the top$k$ overlap
measured over the binary graphs, we have found the same index pairs to exhibit high values;
one exception is the HC-DC pair which now appears of considerably high overlap. Still, we will see
in the experimentation section how theses overlap values reflect on the (similar) effects of
the corresponding node removals.

% %\vspace{-23.in}
% \begin{table}[ht]
% \centering \caption{Variance values for the Spearman averages  }
% \label{tbl:Zoo_correl_var} {\scriptsize
% \begin{tabular}{|c|| c c c c c c  |}
% \hline
% \multicolumn{1}{|c||}{}&\multicolumn{1}{c}{BC}&\multicolumn{1}{c}{CC}&\multicolumn{1}{c}{DC}&\multicolumn{1}{c}{EC}&\multicolumn{1}{c}{HC}&\multicolumn{1}{c|}{ECC}\\
% % & & &
%  %\cline{1-7}
% %  & & ego-network \textbf{(r=1)} & ego-network \textbf{(r=2)}\\
% \hline
% \hline
% BC     & 1      &           &        &       &        & \\%     & \\
% CC     & 0.01   &   1       &        &       &        &\\%      & \\
% DC     & 0.01   &  0.02     & 1      &       &        &\\%      & \\
% EC     & 0.03   &  0.01     & 0.04   & 1     &        &\\%      & \\
% HC     & 0.01   &  0.01     & 0.01   & 0.01  & 1      &\\%      & \\
% %PG     & 0.01   &  0.03     & 0.01   & 0.06  & 0.02  &  1   &\\
% ECC    & 0.01   &  0.04     & 0.02   & 0.16  & 0.04   &  1   \\
% %    & 0.92 &     1 &  0.74  & 0.75 & 0.42 & 0.44 &0.30\\
% % \hline
% \hline
% \end{tabular}}
% %\vspace{-0.12 in}
% \end{table}

\section{Centrality and network vulnerability}\label{sec:robustness}

The correlation study yields a first indirect indication of how different centrality indices compare and whether they could be interchanged in the context of a network protocol or analysis that draws on node rankings. The ultimate reply to this question is, however, protocol/analysis-dependent. In this section, we seek to come up with a reply in the context of the network vulnerability to node failures. More specifically, we ask how much different are the conclusions about the network vulnerability when its most central nodes are removed in line with the rankings induced by the different centrality indices\footnote{In this report, nodes are removed simultaneously after being ranked in order of decreasing centrality values.
An alternative, lying at the core of what is often called \emph{sequential} targeted attack strategy,
%Under the sequential targeted attack, the network
is to recalculate the rankings of the residual nodes after each node removal.
%and seeking to identify the next most central one (\ie($k$+1)-th).
As intuitively expected and shown in ~\cite{attack,attack_robustness, vulnerWeighted} the impact of such sequential node removals upon network connectivity properties is more dramatic.
%In line with intuition, previous studies have shown that the
%sequential attack that heuristically (\ie \textit{w.r.t} centrality indices) selects nodes to attack
%can be more harmful than the simultaneous one~\cite{attack,attack_robustness, vulnerWeighted}, as its
%decisions are more informed regarding the topology changes.
Expanding our study to the sequential node removal case is straightforward.}.
%However, this attack effectiveness due to recalculations requires questionable information
%availability to be obtained, and clearly comes with significantly higher computational cost.
%As such, it is deemed as a hard-to-implement option.}.

The network vulnerability analysis is of interest to various parties.
A potential attacker would like to know which centrality index can reveal the node set, whose removal has the the most significant impact on the network performance, so as to orchestrate the most effective attack.
%i.e., identify the nodes, whose removal causes the most significant impact on the
%network performance.
From the network operator's side, the dual aim is to identify and secure %learn which nodes how to build or even guard
the most critical network nodes that, when shut-down by an attack, result in maximum network performance degradation.
In this paper, we relate the term ``performance'' to fundamental connectivity and traffic capacity properties of the network rather than the scores achieved by specific protocols/applications. This way we get away with their engineering details that shape the end impact and place the emphasis on the network topologies as such.

\subsection{Centrality-driven node removals and connectivity}
\label{subsec:connectivity}

%Three connectivity metrics are computed on the residual ISP network graphs (after each node removal) :
Our study evaluates the impact of node removals on three different connectivity measures of the Internet graphs:
(i) the size of the giant connected component; (ii) the total number of connected components in the graph; and (iii) the average shortest path length\footnote{Shortest paths are computed only for node pairs residing
in the same connected component.} between all nodes. Experiments are carried
out over the binary datasets described in the subsection~\ref{ISPs}.

\begin{figure*}[ht]
\begin{center}
\advance\leftskip-1.4cm
\begin{tabular}{ccc}
 \includegraphics[width=0.36\textwidth]{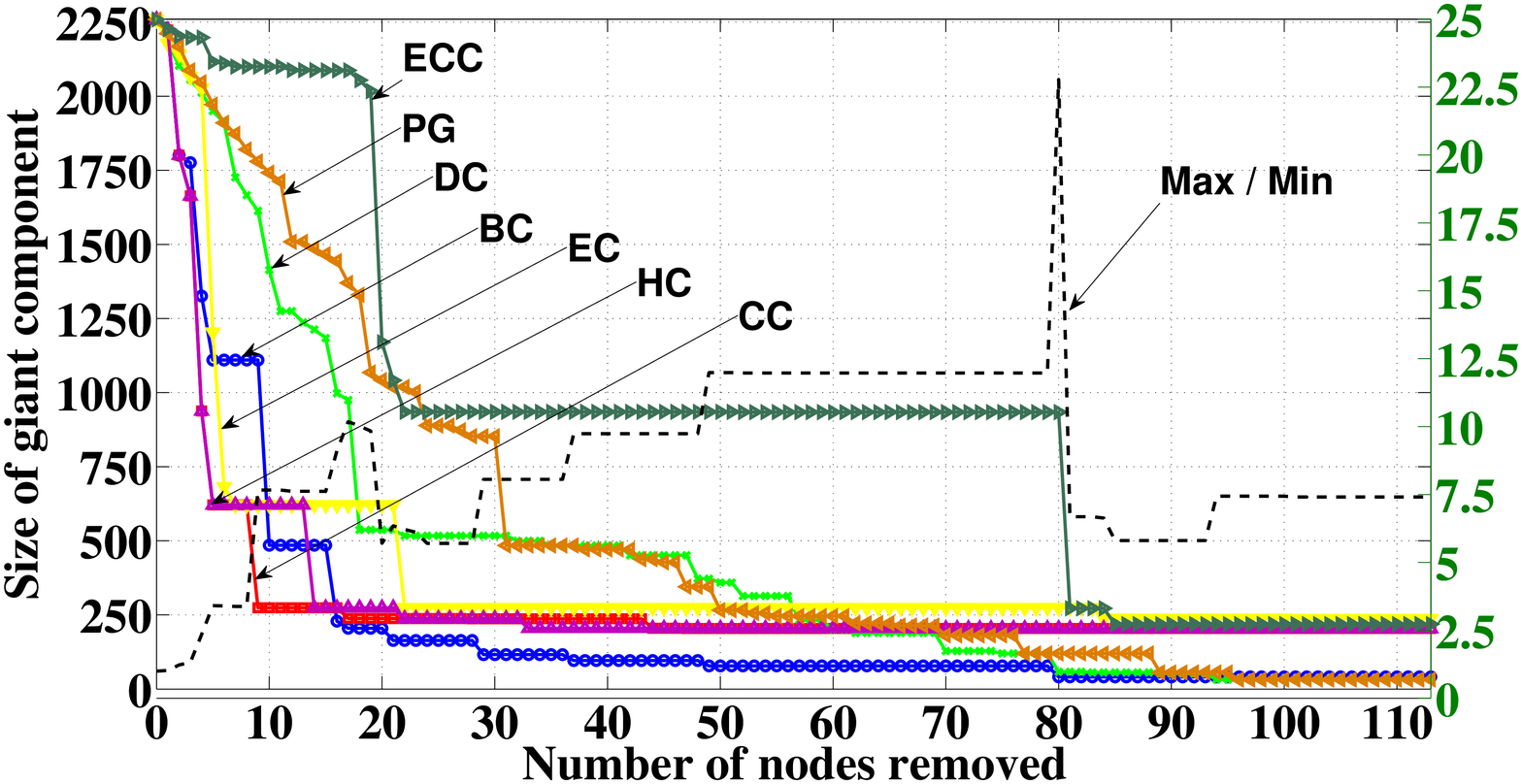} % arxika sto 0.17
 &
 \includegraphics[width=0.36\textwidth]{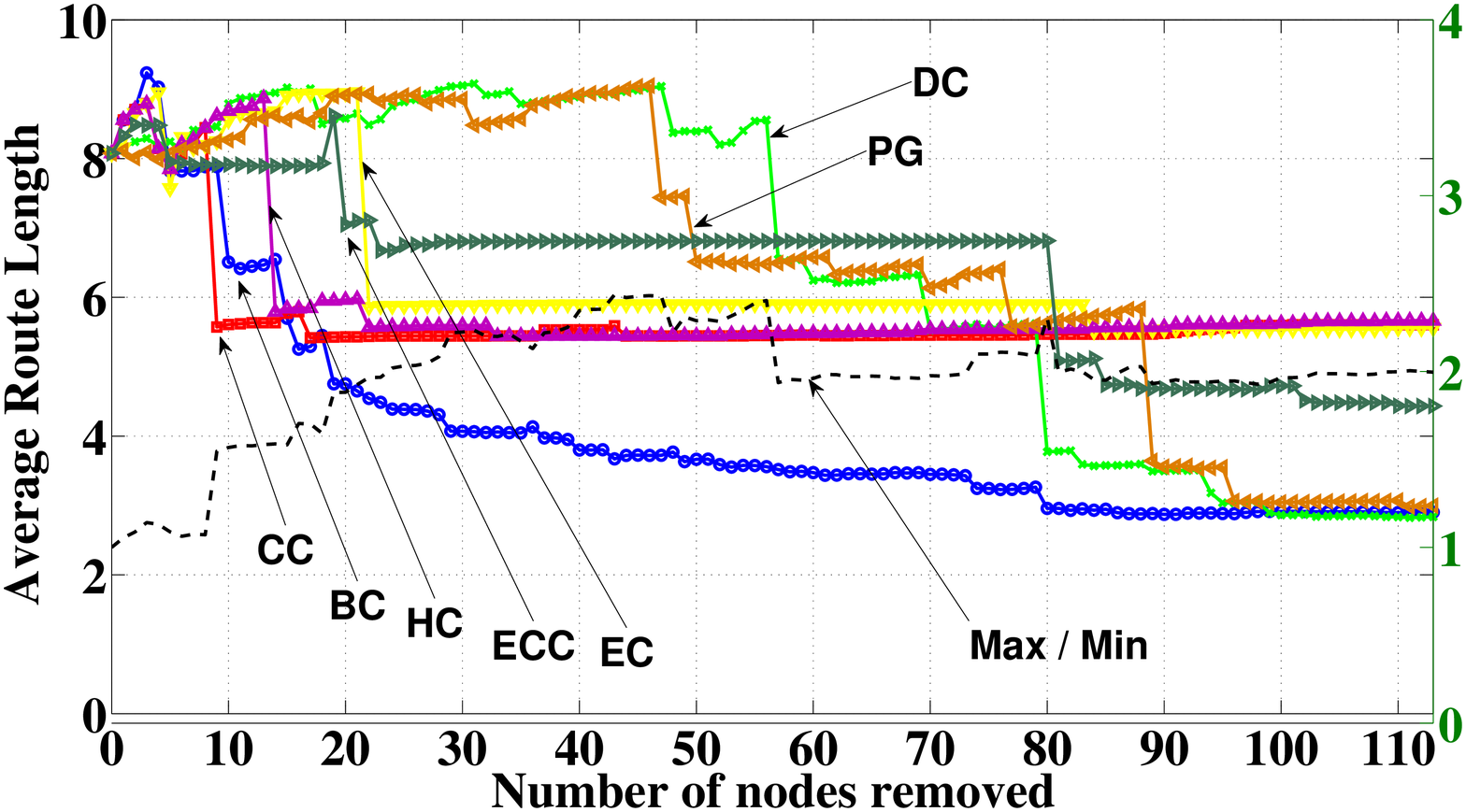}
 &
 \includegraphics[width=0.36\textwidth]{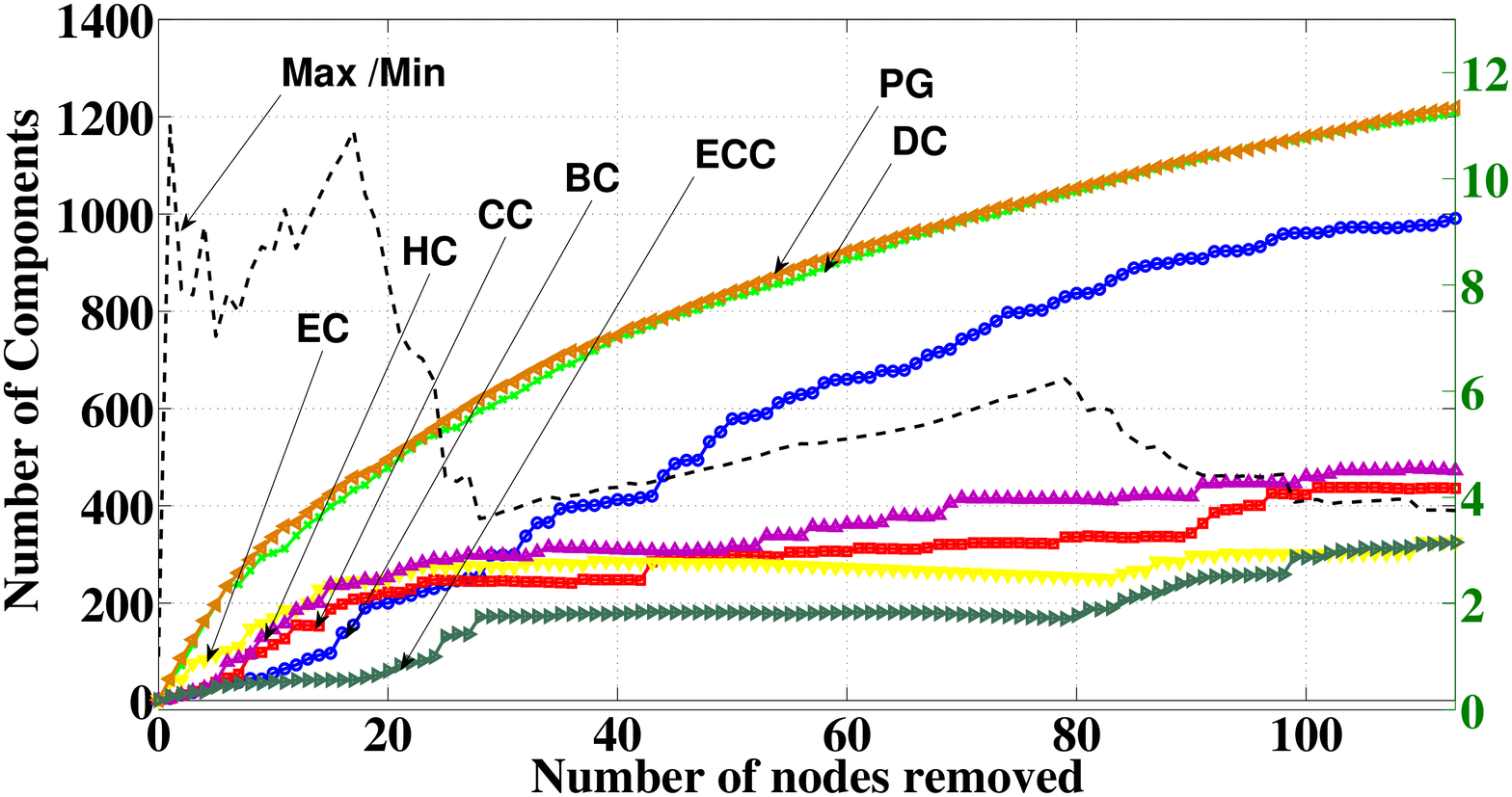}\\
 \begin{scriptsize} a. Caida-AS786  \end{scriptsize}  &
\begin{scriptsize}  b. Caida-AS786  \end{scriptsize} &
\begin{scriptsize}  c. Caida-AS786 \end{scriptsize} \\
% &
 \includegraphics[width=0.36\textwidth]{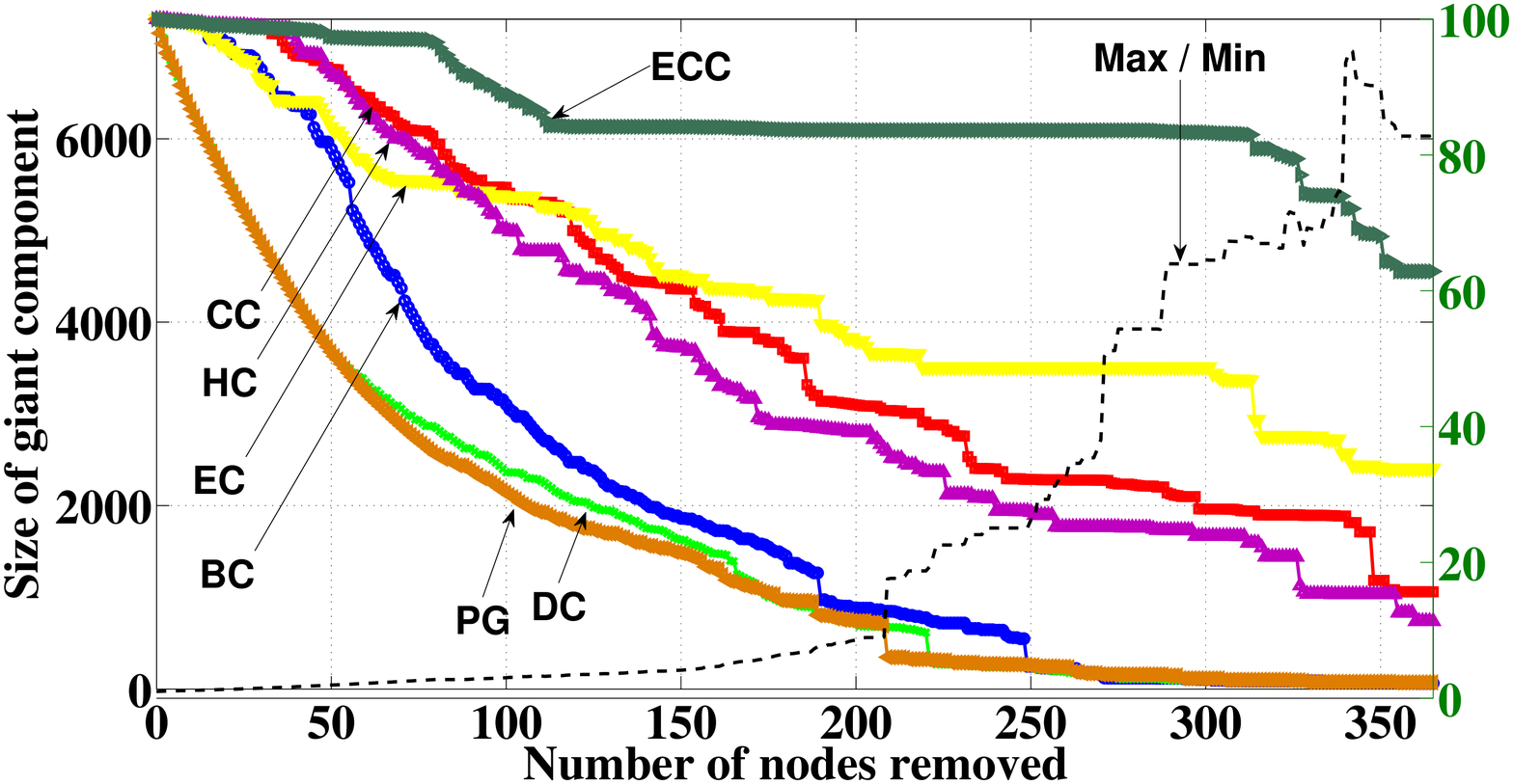}
 &
 \includegraphics[width=0.36\textwidth]{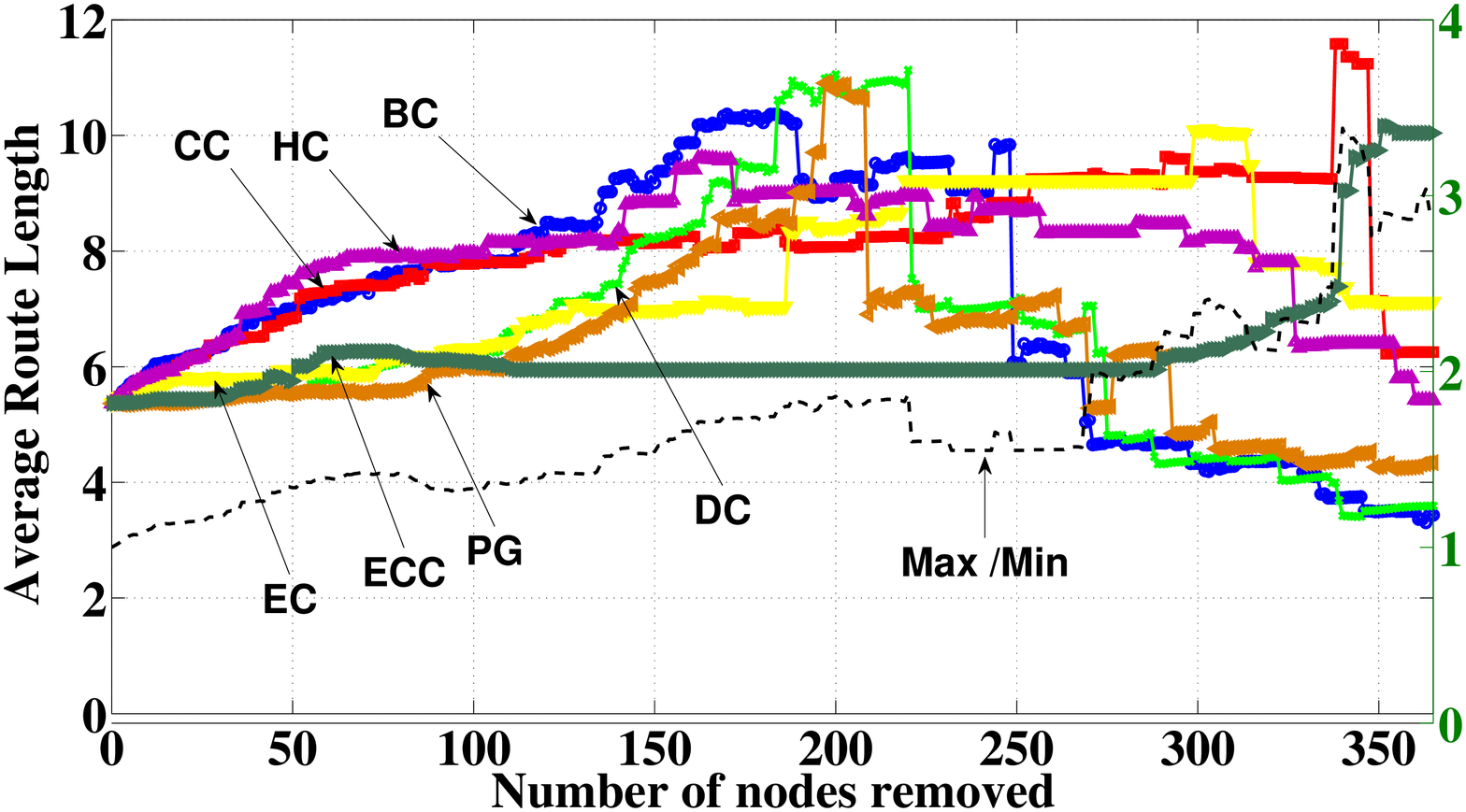}
 &
\includegraphics[width=0.36\textwidth]{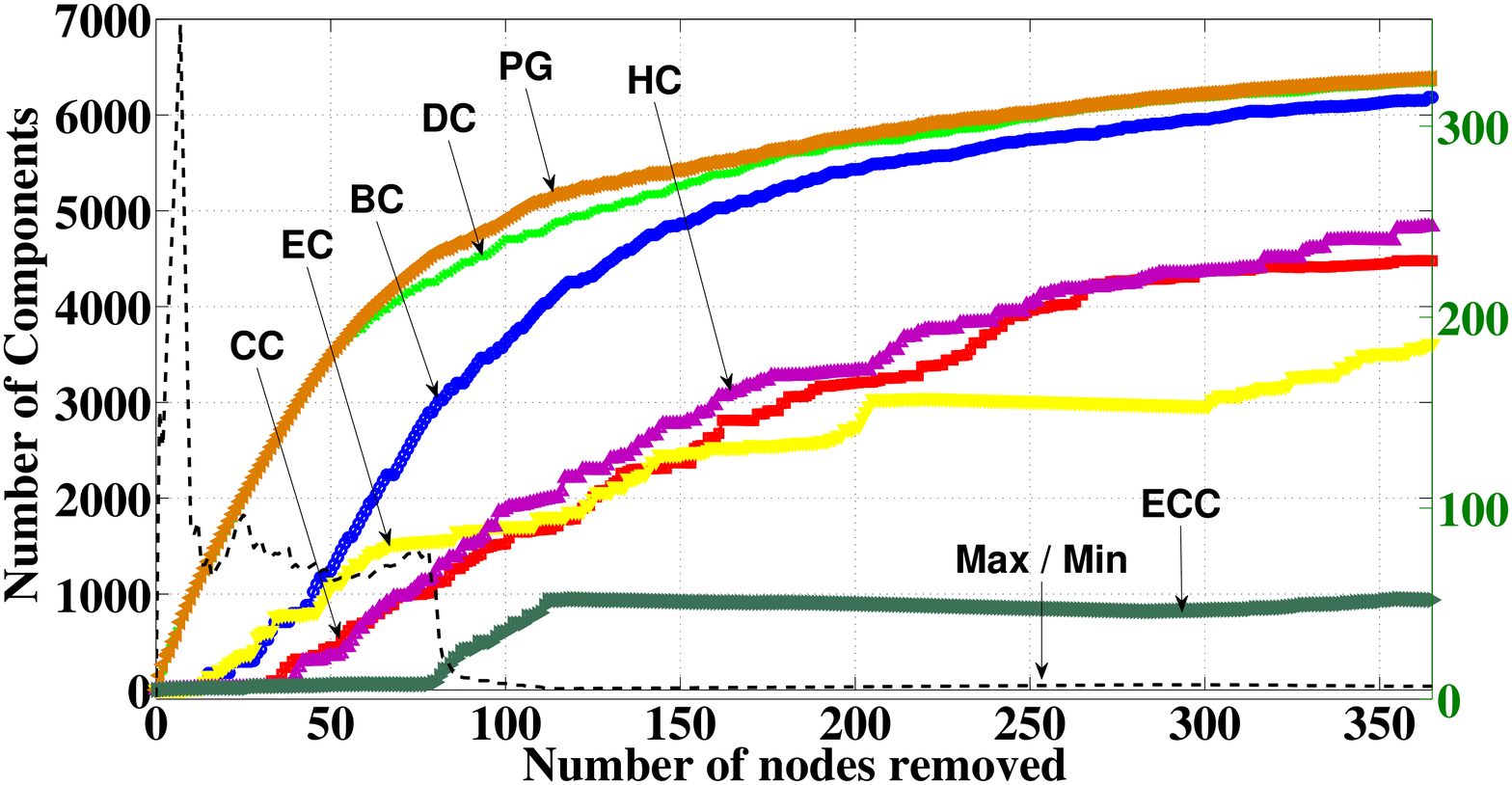}\\
\begin{scriptsize} d. Rocketfuel-AS1239  \end{scriptsize} &
\begin{scriptsize} e. Rocketfuel-AS1239 \end{scriptsize} &
\begin{scriptsize} f. Rocketfuel-AS1239 \end{scriptsize} \\
\end{tabular}
\end{center}
\caption{Effects of node removals on three network connectivity measures \ie the size of the giant-connected component (a,d),
the average path-length (b,e) and the number of components (c,f) for two indicative ASes.}
\label{fig:robustness_results}
%\vspace{-0.15 in}
\end{figure*}

Figure~\ref{fig:robustness_results} presents a representative set of results %(more plots can be found in~\cite{thesis})
showing the impact of centrality-driven node removals on the network connectivity.
%An AS snapshot has been selected for each of the four datasets,
%comparing the seven preferred centralities (BC, DC, CC, EC, HC, PG, ECC).
%It is important to note that we present
Experimental points correspond to removals up to 5\% of the network size
since the connectivity properties tend to stabilize thereafter.
%====================================================================================
%\textbf{(ARE YOU SURE?)}. Nomikos has detected that
%====================================================================================
Apart from the three connectivity measures %that assess the effectiveness of each one of the seven centrality
%indices,
we also compute %for every connectivity measure,
the Max/Min ratio (plotted in dashed line and measured on right Y axis); % of Figure~\ref{fig:robustness_results});
%this is the fraction between the maximum and minimum value of the current networking property among the seven
%centrality indices for every node removal
this is the fraction between the maximum and minimum value of the connectivity measure as obtained over all centrality indices. The Max/Min ratio essentially seeks to quantify the effectiveness loss in terms of network connectivity between the optimal
and worst choice out of the considered indices.

In view of the reported correlation results, one may expect that any two highly rank-correlated indices should have similar impact over the network connectivity,
when used to drive node removals. Interestingly, our results suggest that this is rarely the case for all three connectivity measures. In what follows we comment on the
experimentation outcomes seeking to relate them to the earlier observed correlations;
conclusions are mainly drawn with respect to four highly-correlated
index pairs (\ie BC-DC, PG-DC, HC-CC and PG-BC) as well as a couple of weakly associated ones (\eg HC-BC or BC-CC).

%====================== conclusion for older version =======
% The only two pairs that still affect the connectivity in similar ways are the
% Harmonic-Closeness and PageRank-Degree centrality
% %and PG-DC centrality
% (see Figure~\ref{fig:robustness_results}) which were earlier found
% to exhibit high correlation strength and top-5\% overlap.
% %This is an evidence for as we noticed in the previous Chapter in Figures 4.1 and 4.6.
%======================================

\begin{figure*}[ht]
\begin{center}
%\advance\leftskip-1.4cm
\begin{tabular}{ccc}
 \includegraphics[width=0.26\textwidth]{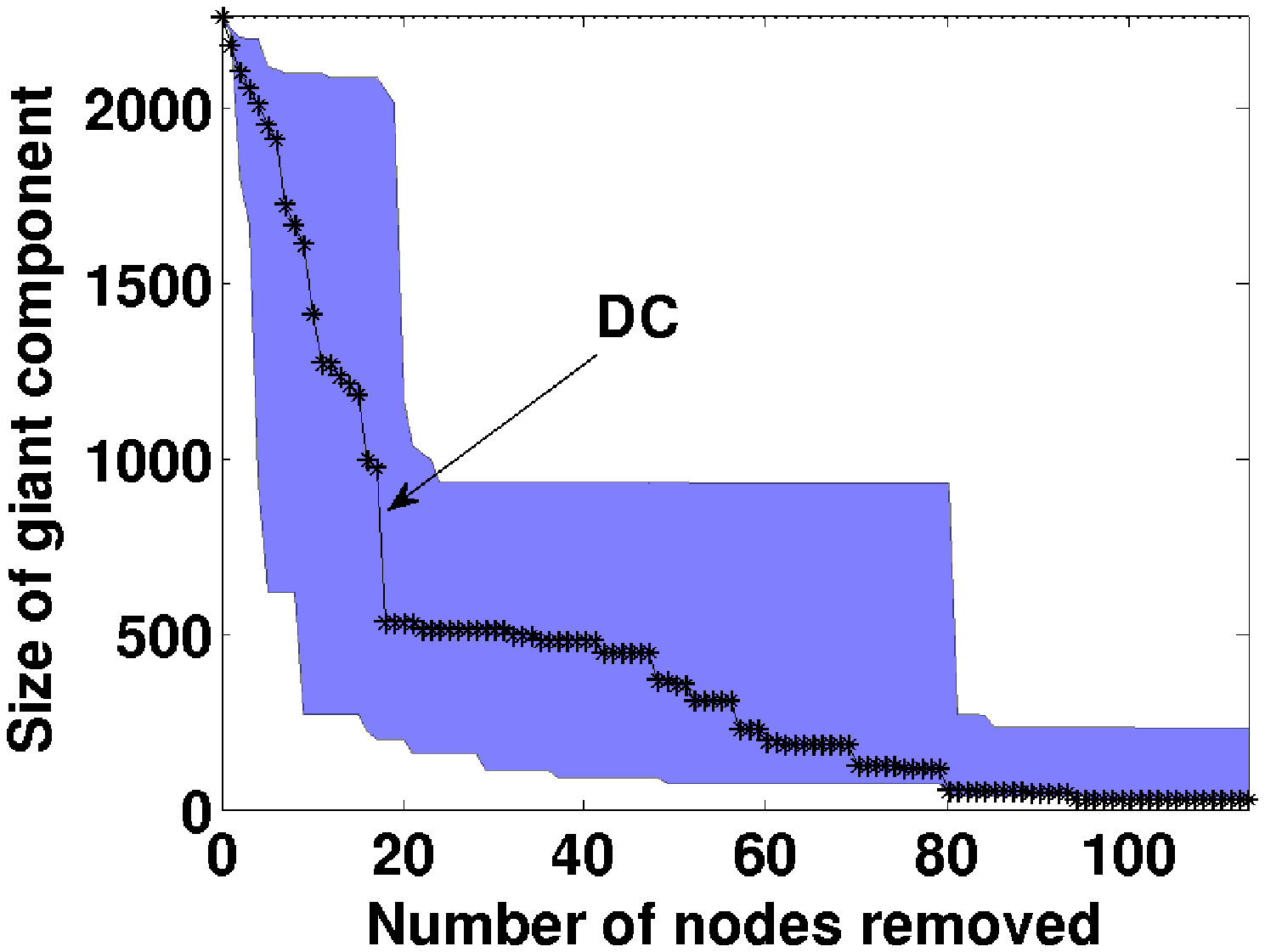} % arxika sto 0.17
 &
 \includegraphics[width=0.26\textwidth]{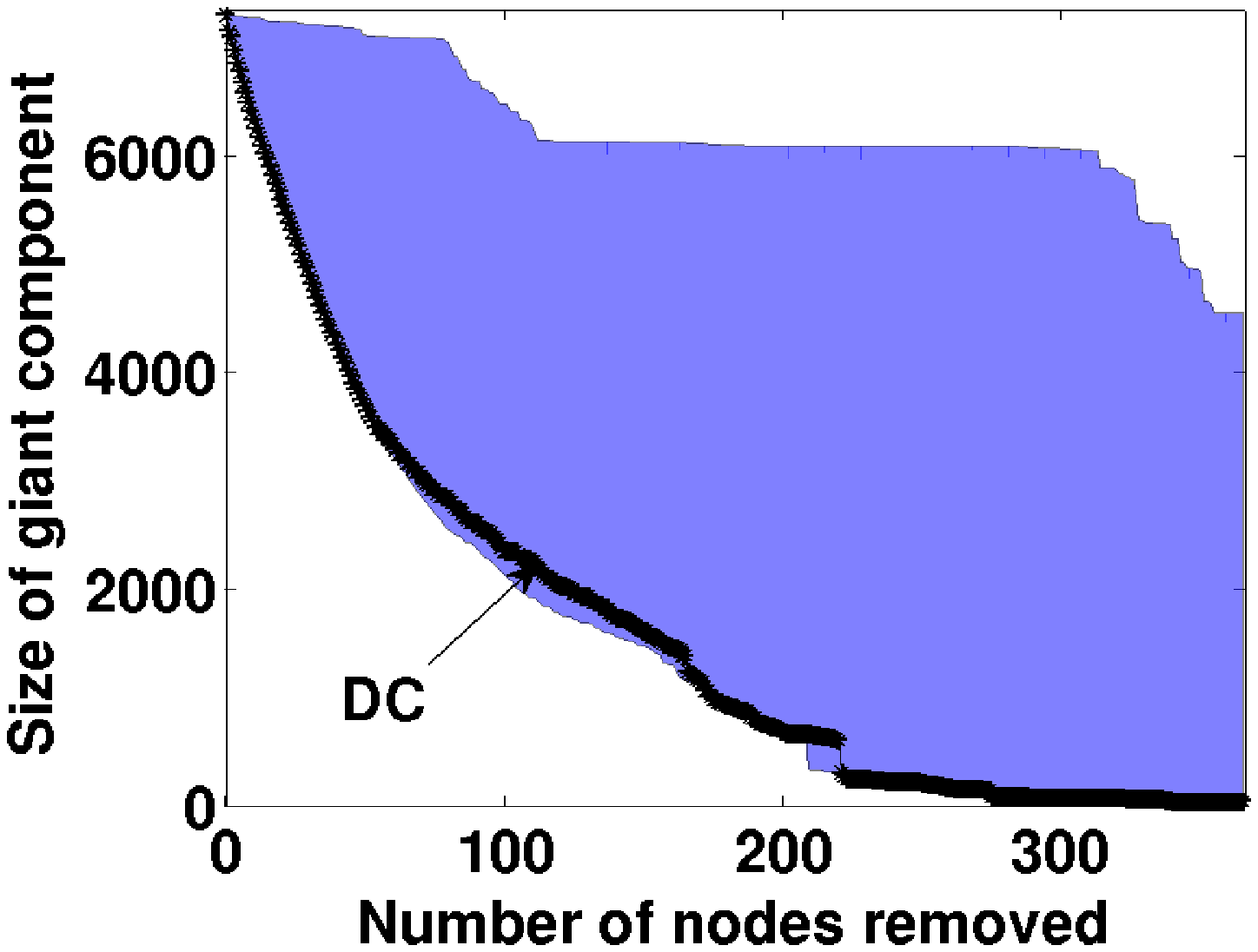}
 &
 \includegraphics[width=0.28\textwidth]{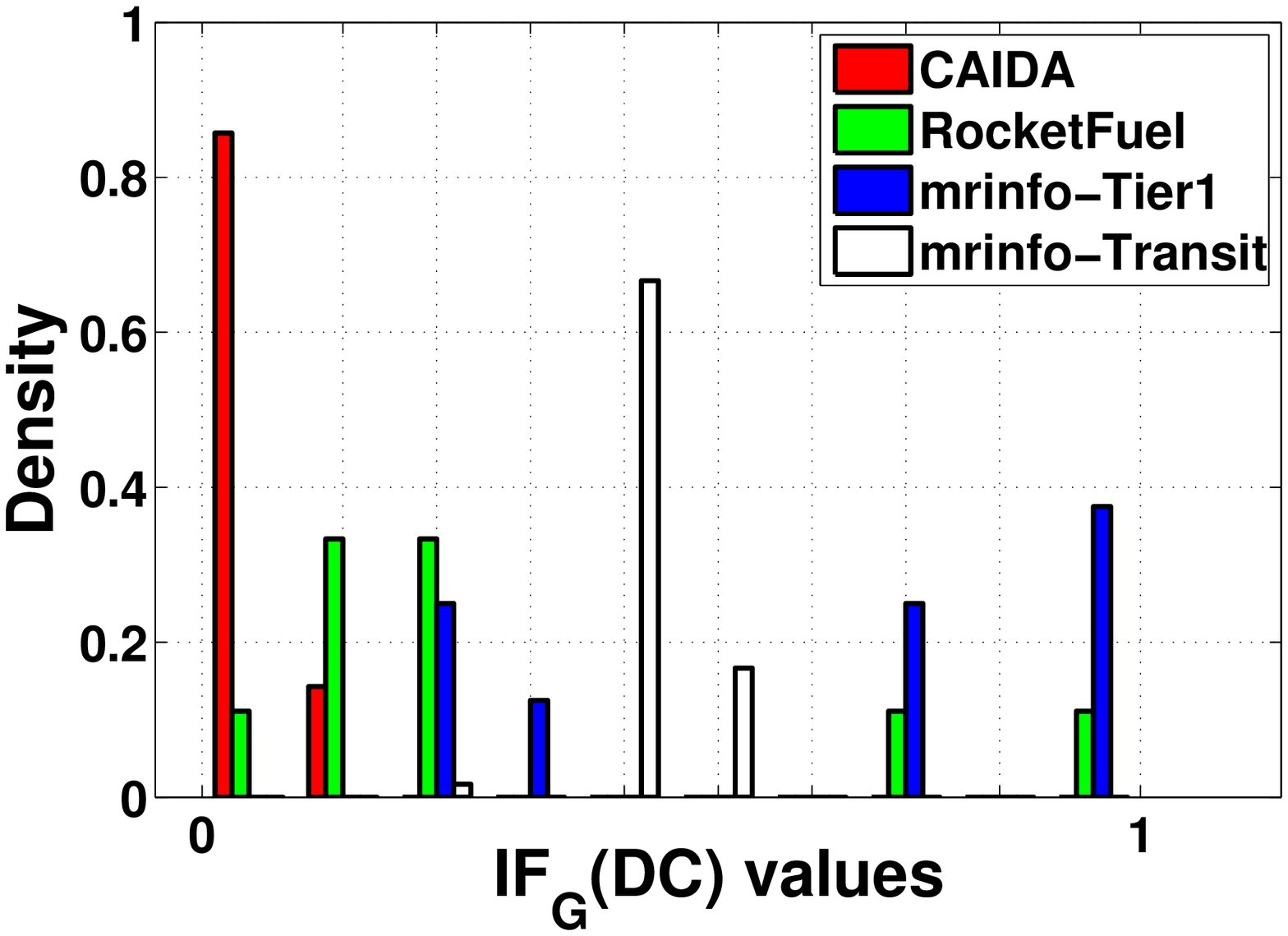}\\ %untitled
 \begin{scriptsize} a. Caida-AS786  \end{scriptsize}  &
\begin{scriptsize}  b. Rocketfuel-AS1239  \end{scriptsize} &
\begin{scriptsize}  c. All datasets \end{scriptsize} \\
\end{tabular}
\end{center}
\caption{a,b)Envelope plots of the DC-based node removal effects on the size of the giant-connected component for two
indicative ASes. c) Empirical probability mass function of the $IF_G(DC)$ measured with respect to the size of the giant component.}
\label{fig:envelopes_PDF}
%\vspace{-0.15 in}
\end{figure*}

\subsubsection{Size of giant connected component (GCC)}

The size of Giant Connected Component (GCC) reflects the number of nodes being able to communicate with each other.
%yields \textbf{an upper bound on the number of nodes being able to communicate with each other}(WHY UPPER BOUND?) .
%With respect to the giant connected component,
The only consistent result here involves the least effective index. As Figs.~\ref{fig:robustness_results}.a and .d
suggest removing vertices as determined by the Eccentricity measure,
has the minimum impact on GCC. All other indices expose more quickly the vulnerability of the network. Yet, we cannot identify any dominance relationship among them across all datasets. However, the behavior of certain %the behavior of
pairs such as the HC-CC and PG-DC is in good agreement with the earlier
observed strong associations, both in rank-correlation and top-$k$ overlap
(\ie $\rho_v$ and $ov_V$ higher than 0.85 and 85\%, respectively); %the corresponding curves in
indeed, the corresponding curves in Figs~\ref{fig:robustness_results}.a and .d %(a.1, b.1, c.1, d.1)
appear to (partially) coincide or exhibiting small GCC size differences as nodes are removed. A closer look reveals that it is the top-$k$ overlap between two indices, rather than their rank-correlation, that essentially determines how similar is the impact of the corresponding removals. A relevant example is the index of BC and DC over AS1239 which in Fig~\ref{fig:robustness_results}.d demonstrate highly dissimilar impact; their measured top-$5\%$ overlap does not exceed 68\% while
the Spearman $\rho_v$ reaches 0.94.

Comparing the impact of the locally-determined, DC-driven removal against the globally-determined BC-driven one, Holme \etal~\cite{attack} showed that the two types
are equally harmful over synthetic (scale-free) topologies, while a distinct
real-world co-authorship network appears more vulnerable to BC-driven attack.
In our broad dataset of real-world topologies we do not witness expressions of the latter effect; on the contrary, the local DC index occasionally turns out to have more dramatic impact than the global BC.
%Finally, those indices that were measured with weak associations (\eg HC-BC in Fig.~\ref{fig:robustness_results}.d), appear in line with intuition to cause different impact on the GCC size.

% in Figure~\ref{fig:robustness_results}.a the size of giant component appears slightly more vulnerable
% to BC-driven attacks while in the case of Rocketfuel (Figure~\ref{fig:robustness_results}.d),
% appears more robust against BC-driven attacks.
Overall, the concluding note here would be that \textit{any two indices measured with high top-$k$ overlap values would appear to cause exactly the same impact
on the connected component for a certain sequence of node removals, and vice versa. The full-rank correlation values are not always in line with the experienced impact}.

\subsubsection{ Average shortest path length}
Regarding the average shortest path length, there is no index that constantly exhibits the best or worst performance while all present a twofold behavior over every topology. First, the average path length increases and then, suddenly follows a fast decay
(Figure~\ref{fig:robustness_results}.b and .e). This fluctuation seems to mainly depend on how fast the centrality-driven node removals lead to the total network fragmentation. Potentially, there exists an upper bound of node
removals that permits the giant component to maintain a relatively large size before its connectivity has been significantly diminished. Consequently, as long as the largest connected component maintains a significant size, the node removals result in increasingly longer paths between its node pairs; when the network has been broken down to several small clusters, further removals tend to create single isolated nodes and therefore, decrease the average shortest path.
%This inconsistency demonstrates that the shortest path length does not hold a constant pattern
%with respect to the required number of nodes that should be removed. In this case either
% Due to this non-monotonic decrease of the average shortest path length (with node removals), %this
% an attacker or a network administrator might need to avoid assessing the
% network vulnerability based on this metric.
Looking at how the correlation results predict the impact of removals,
we notice that the highly associated indices in both rank-correlation and top-$k$ overlap (\ie HC-CC, PG-DC) again have similar impact, whereas those of slightly weaker association in at least one association metric (\ie BC-DC, PG-BC), may affect the average shortest path length in considerably different ways.

\subsubsection{ Number of connected components}
As far as the the number of connected components is concerned, ECC is again found the dominant index in terms of the least effective removal (Figure~\ref{fig:robustness_results}.c and .f).
According to the ECC notion~\cite{eccentrcity}, a node is central when its maximum distance to any other node is close to the radius of the graph. Hence, %based on its definition,
a node can exhibit a significantly low ECC value when only a few other nodes lie far away (from it) in the topology. This sensitivity makes ECC assigning less significance to nodes considered highly-central with respect to other indices. So when removing the top-5\% ranked nodes
we may not actually refer to those holding prominent network locations %e most central nodes
and this prevents the fast fragmentation of the topology.
In sharp contrast, DC and PG are together the dominant in terms of effectively partitioning the topology, as
their earlier observed high association values suggest.
Interestingly, DC, a purely local index succeeds in removing nodes %topology hubs (i.e, high degree nodes)
that play critical role in connectivity as opposed to the other global and more complex ones (except for PG).
On the other hand, BC and DC which were also found strongly rank-correlated yet of weaker top-$k$ overlap,
cause different impact over the connected components.
Removing vertices according to DC, %a constantly higher increase of
the number of components increases constantly compared to the impact of the BC-driven node removals.
This implies that the network connectivity mainly relies on strategic hub-nodes rather than bridge
nodes that are typically of high BC.

\subsubsection{Local \vs global centrality indices}
%We have
Figure~\ref{fig:robustness_results} clearly shows that the removal of the most central nodes may have a
significantly varying impact depending on which centrality index is used to determine them. Conceptually,
for each number $k$ of removed nodes, one can identify best- and worst-case values, $m_{bc}(k)$ and
$m_{wc}(k)$ respectively, for all three performance metrics plotted in Fig.~\ref{fig:robustness_results}.
These values may be obtained by different centrality indices as the considered metric $m$ changes and
outline an envelope.
%==============================================================
%(\textbf{alternative term?}). % den einai standard term auto??
%==============================================================
Such envelopes define the shaded area in Figs~\ref{fig:envelopes_PDF}.a,b.
What we ask next is where in this envelope the metric values corresponding to the degree centrality, lie.
Essentially, we would like to quantify how close to the best-/worst-case is the impact of removals when
directed by the single locally computable centrality index.

%We introduce
To this end, for each centrality index $c$, topology $G$, number of removed nodes $k$ and performance
metric $m(k;c)$ we define a normalized distance measure, hereafter called impact factor $IF_G(k;c)$ as:

{\small
\begin{equation}\label{eqn:IF}
%betweenness centrality
IF_G(k;c) =\frac{| m(k;c)-m_{wc}(k)|}{|m_{bc}(k)- m_{wc}(k)|} \nonumber
%\sum_{s=1}^{|V|}\sum_{t=1}^{s-1}
%IF_{c}(m) =\frac{ m_{c}(k)-\max\limits_{1\leq i \leq N}\{f_{C_i}(m)\}}{\min\limits_{1\leq i \leq N}\{f_{C_i}(m)\}-\max\limits_{1\leq i \leq N}\{f_{C_i}(m)\}} \nonumber %\sum_{s=1}^{|V|}\sum_{t=1}^{s-1}
\end{equation}}
%\end{equation}
\normalfont

\noindent
Note that depending on the metric, the worst-case value may coincide with the minimum or
maximum value the metric gets over all indices. %if for some $m$ value holds $min_{1\leq i \leq N}\{f_{C_i}(m)\}=max_{1\leq i \leq N}\{f_{C_i}(m)\}$ \ie all $f_{C_i}$ curves coincide,
%then $DSF_{DC}(m)$ equals zero, by default.
It is then straightforward to derive a topology-average measure of the impact factor as:

{\small
\begin{equation}\label{eqn:IF_G}
%betweenness centrality
IF_G(c) =\frac{1}{|\mathcal{K}|}\sum_{k \in \mathcal{K}}\frac{| m(k;c)-m_{wc}(k)|}{|m_{bc}(k)- m_{wc}(k)|} \nonumber
%\sum_{s=1}^{|V|}\sum_{t=1}^{s-1}
%IF_{c}(m) =\frac{ m_{c}(k)-\max\limits_{1\leq i \leq N}\{f_{C_i}(m)\}}{\min\limits_{1\leq i \leq N}\{f_{C_i}(m)\}-\max\limits_{1\leq i \leq N}\{f_{C_i}(m)\}} \nonumber %\sum_{s=1}^{|V|}\sum_{t=1}^{s-1}
\end{equation}}
%\end{equation}
\normalfont

\noindent
where $\mathcal{K}$ is the set of $k$ values considered in the evaluation.
Clearly, both $IF_G(k;c), k \in \mathcal{K}$ and $IF_G(c)$ take values in $[0,1]$.
We are particularly interested in $IF_G(DC)$
%when it attains close-to-zero values the locally determined
%BC-based attack approximates closely the most damaging (globally-determined) one. %zero revealing that DSF value approximates zero
and Figure~\ref{fig:envelopes_PDF}.c plots the empirical probability mass function
of the $IF_G(DC)$ values over all topologies of a given dataset, %over
%for the $DSF_{DC}$
when the metric $m$ is taken to be the size of the giant connected component.
%\textbf{+comment on PDFs}
Despite its local nature, DC is proved in most cases to cause significant impact on the
giant connected component. To which extent this impact approximates %the impact of
the most effective removal appears to depend on the underlying network topology.
Over the CAIDA networks %Figure~\ref{fig:envelopes_PDF}.c shows that
DC can closely approximate the most effective index while it seems to
offer a less effective approximation over the Rocketfuel topologies.
%in the \verb|mrinfo| (Transit) dataset it offers a somewhat moderate solution.
Finally, in the \verb|mrinfo| (Tier-1) and (Transit) topologies, considerable mass appears
for medium and high $IF_G(DC)$ values, respectively. This renders DC as an option of very
low effectiveness for both datasets.

% Our results showed to which extent a centrality index can drive the attack and cause extensive
% fragmentation phenomena. Interestingly, only two pairs of indices (\ie HC-CC, PG-DC) that have
% been earlier measured of high rank correlation and top-$k$ overlap,
% appear to cause similar impact to the network graph for all three meausures.
%when serving as drivers for the attack. Other indices earlier measured to be highly rank-correlated
%do not bring about the same impact to the graphs when acting as drivers for the attacks.

\subsection{Centrality-driven node removals and traffic capacity}\label{subsec:flow}
We now turn our attention to a much less investigated topic, the comparative impact of centrality-driven node removals on the network traffic serving capacity.
%This s the impact of the centrality-driven node attacks over capacitated Internet graphs %employing a different criterion;
%We approximate calculate
%by measuring the reduction of the total traffic a topology can accommodate.
%Interestingly, only a few works appear in literature that carry out network vulnerability studies
%under this perspective.
%as we remove nodes realizing the
%attack strategy.

%analyze the throughput of these networks
%when the nodes or the edges are attacked using some of the above mentioned
%strategies

\begin{figure*}[ht]
% \vspace{-0.22 in}
\begin{center}
\advance\leftskip-1.5cm
\begin{tabular}{cccc}
\includegraphics[width=0.285\textwidth]{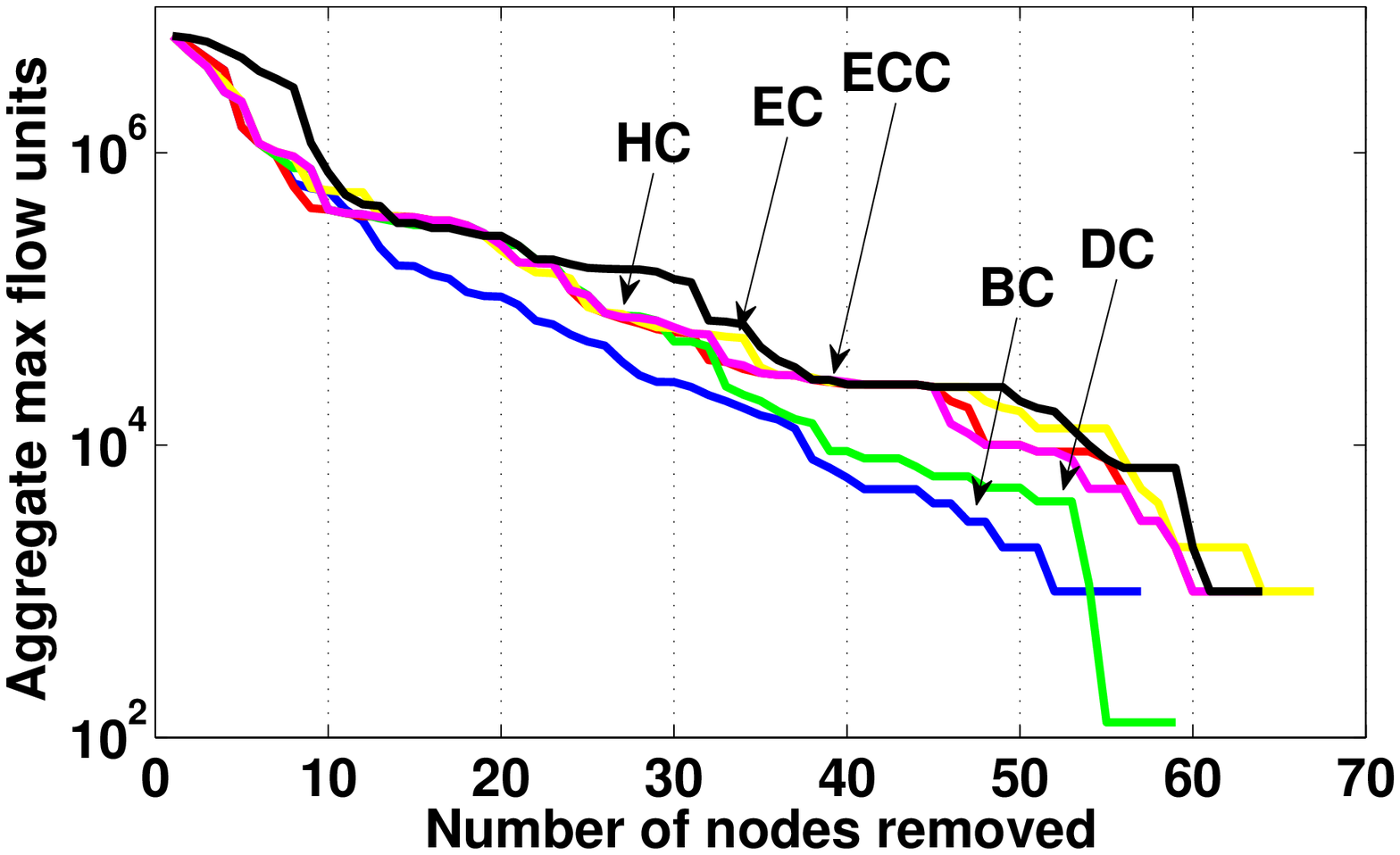}% UninettIImean_new
& \hspace{-0.23in}
\includegraphics[width=0.285\textwidth]{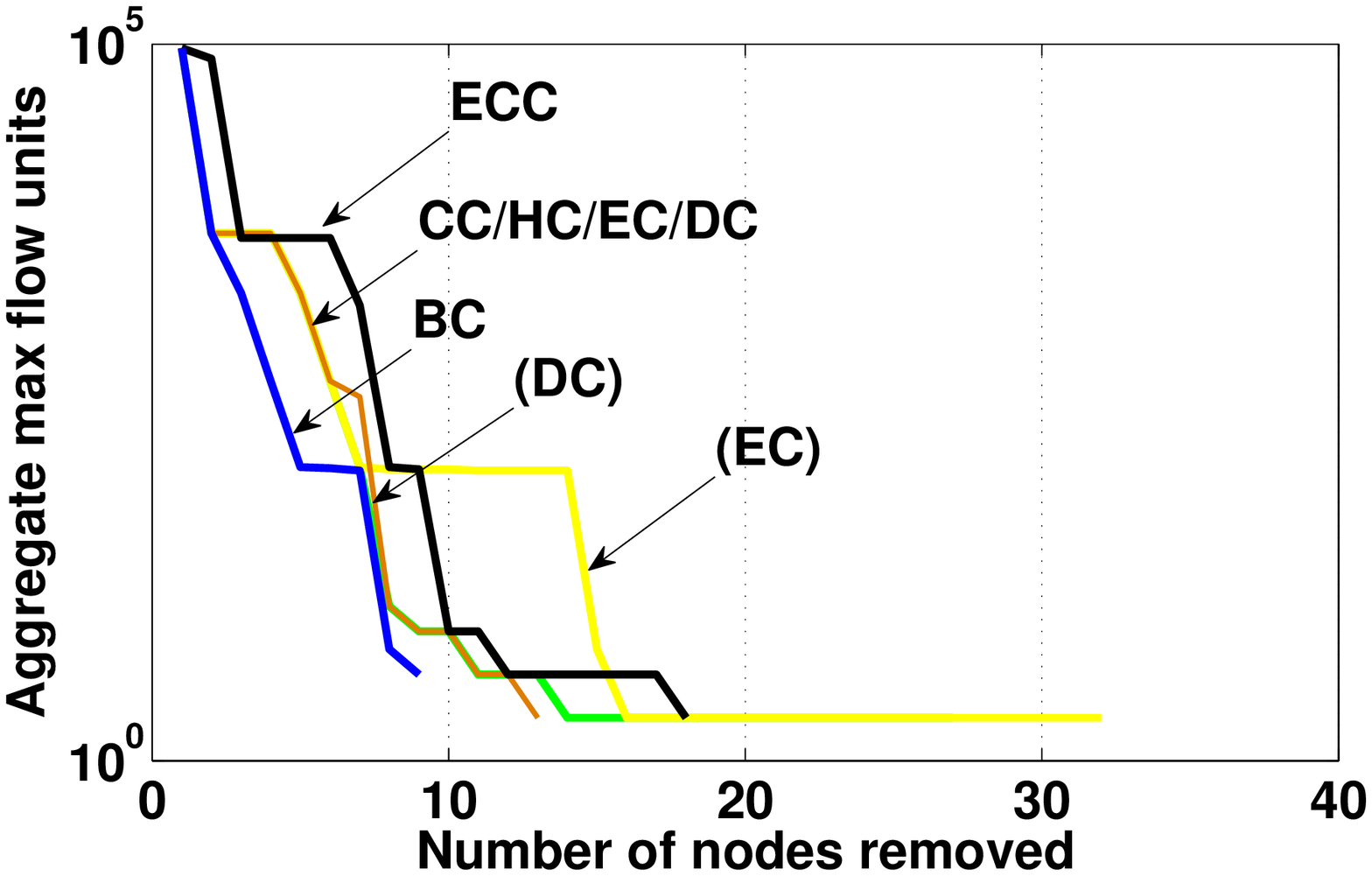} %Carnet  % 0.24
& \hspace{-0.23in}
\includegraphics[width=0.285\textwidth]{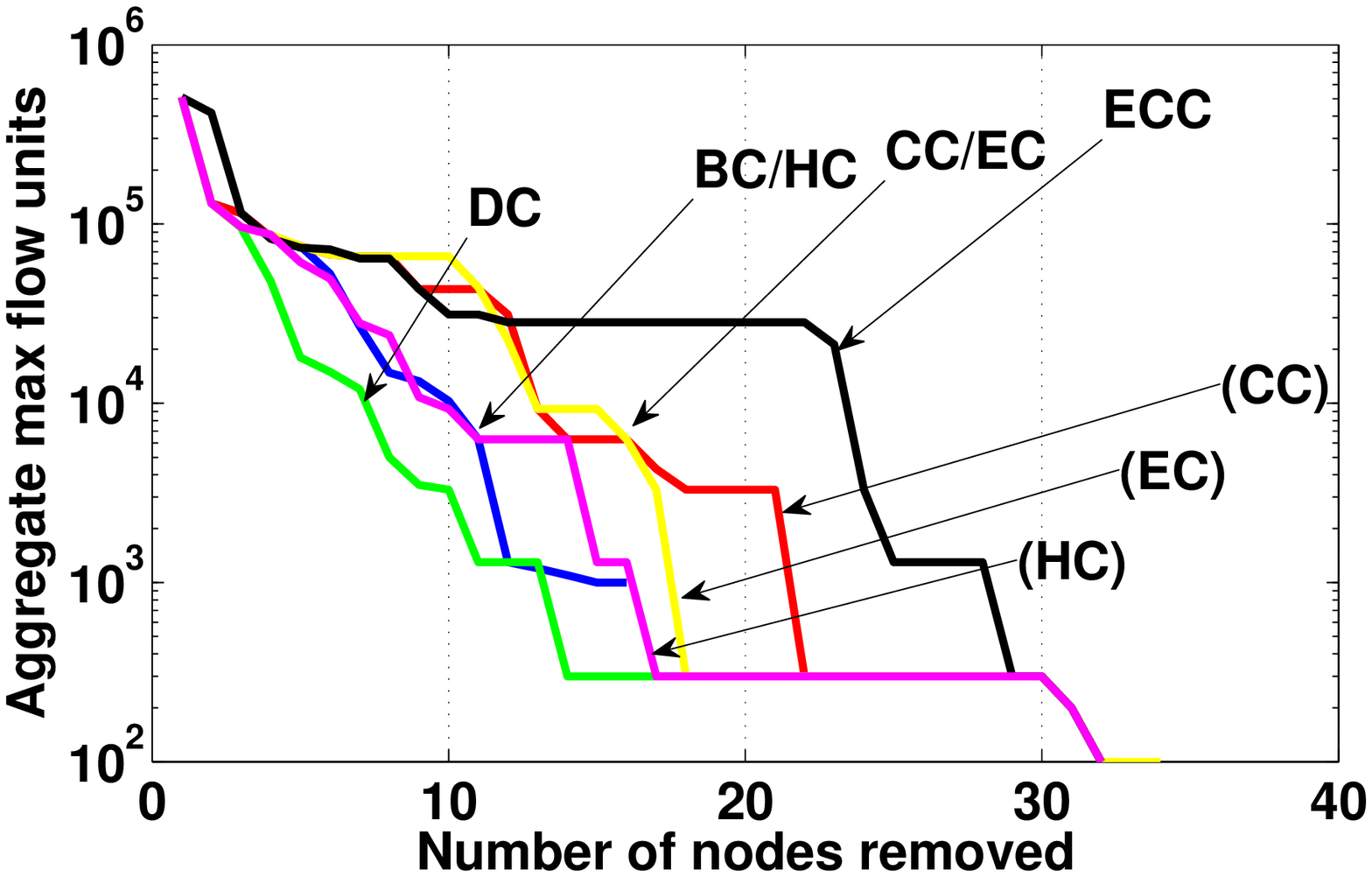} %
& \hspace{-0.23in}
\includegraphics[width=0.285\textwidth]{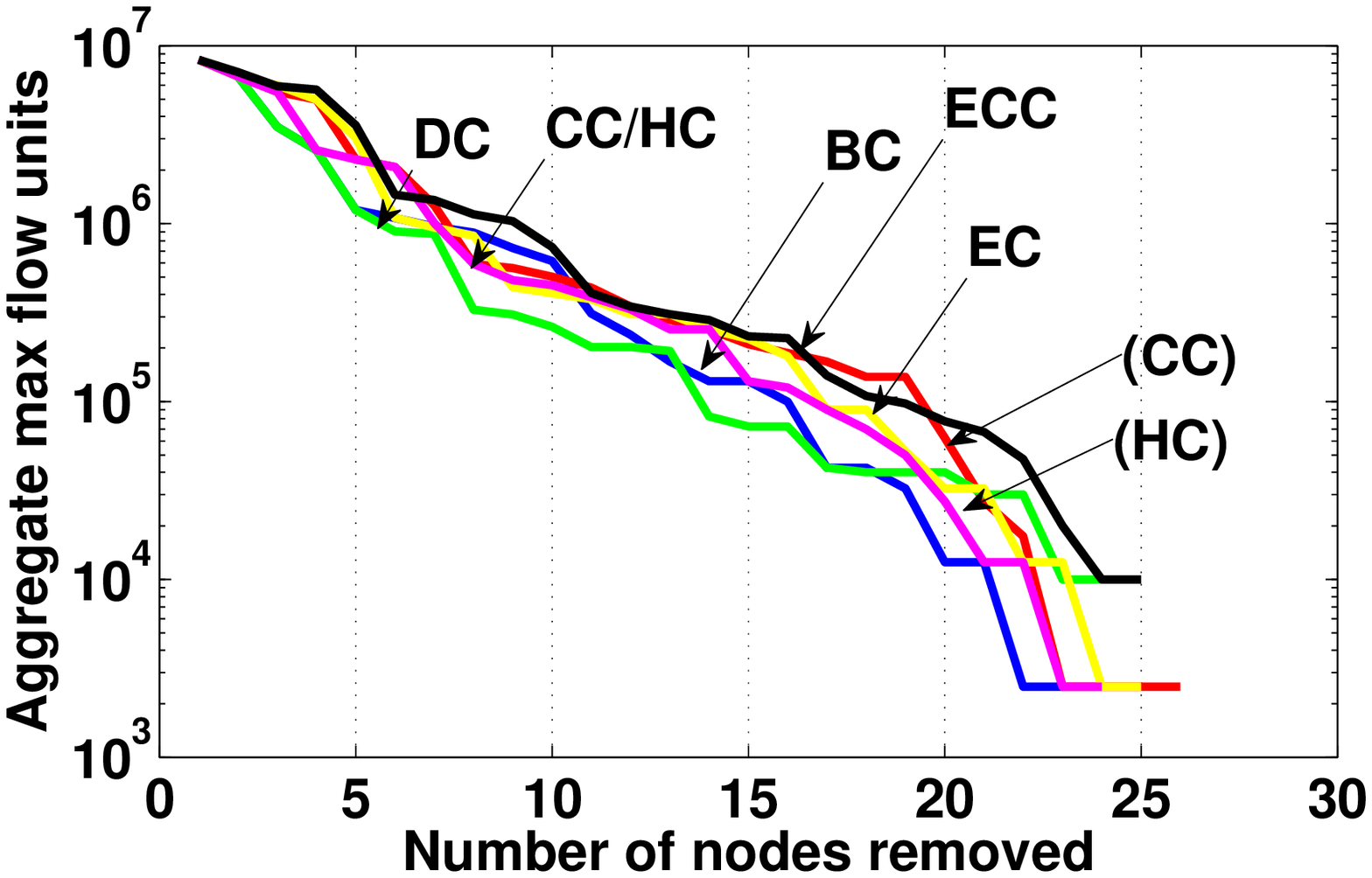}\\  %
\begin{scriptsize} a.  \end{scriptsize}  &
\begin{scriptsize} b.  \end{scriptsize} &
\begin{scriptsize} c.   \end{scriptsize} &
\begin{scriptsize} d.   \end{scriptsize}
\end{tabular}
\end{center}
\caption{Impact of node removal (in centrality-decreasing order) on the maximum flow the networks accommodate:
(a) the Uninett I\_mean, (b) Carnet, (c) Bren and (d) Geant. Wherever curves coincide, a single arrow identifier is used for
multiple indices and when later on they become separated, each one is pointed with the corresponding index in parenthesis.}%~\textit{(\textbf{Inset}}: Initial decrease of the maximum flow as the top-central nodes are removed)
\label{fig:flow_results}
%\vspace{-0.15 in}
\end{figure*}
%\subsubsection{ Traffic-serving capacity approximation}
Such a task is
%Assessing the traffic-serving capacity of the network is
not straightforward.
One approach would be to consider a given traffic matrix, determining either the node pairs that exchange traffic %(source-destination nodes)
only or the node pairs plus the average traffic loads that are (expected to be)
served for each one of them. Then, the traffic-serving capacity of the network could be given by the solution of some version of the multicommodity flow (MCF) problem~\cite{thesis}.
%, \eg the maximum multicommodity (MMF)~\cite{maxMcommodity} or the maximum concurrent
%multicommodity flow (MCF) problem~\cite{maxConcMcommodity}.
%The difference between those two versions is that in the MMF problem we need to find a multicommodity flow
%such that the sum of the flows of all the given commodities (one commodity is associated with
%a source-sink pair) is maximized. Whereas in case of MCF, there are also demands associated with each commodity and the
%objective is to satisfy the maximum possible proportion of all demands.
Yet doing so bears two significant challenges: First, the traffic demand
matrix is rarely known a priori and often varies broadly over different time scales.
%so that it becomes questionable whether a single or few instances of it should be used for characterizing the network vulnerability.
Secondly, and most importantly, the MCF problem is an NP-complete
problem~\cite{complexityMCF}, with the computational complexity raising fast with the number of commodities. %(pairwise traffic loads).

To overcome those limitations, we have taken a simpler approach and estimate the traffic serving capacity of the network as the sum of maximum flows over all network node pairs. Namely, we iterate over all node pairs and for each pair
we solve an instance of the maximum flow problem, \ie compute the maximum traffic load that can be served by the network when only the particular pair transfers traffic
across the network. Clearly, this sum is a (very) loose upper bound of the traffic load that can simultaneously be served by
the network. % --indeed it is not aimed as such a bound.
However, it provides a traffic load-neutral measure of what can the network carry and how is this affected when a variable number of nodes is removed.
%As an instance of
%broad interpretation and for the sake of brevity we call this measure ``network throughput''.
%The maximum flow problem has received a great deal of attention, and a number of efficient algorithms have been
%proposed. In our experiments
For the solution of the the maximum flow problem we have employed the Edmonds-Karp algorithm~\cite{EdmondsKarp} %which is a refined version
%of Ford-Fulkerson
with a $O(V\cdot E^2)$ polynomial-time complexity (where $V$ and $E$ is the total number of nodes and edges, respectively).

\subsubsection{Experimentation methodology and results}
Our experimental study was carried out over the Zoo Internet topologies with capacitated links, described in Section \ref{ISPs}. We remove nodes in decreasing order of centrality and measure the aggregate maximum flow over all node pairs.
%We implemented the same \ie simultaneous attack strategy and measured the network
%throughput degradation for every node removal.
%
% ==== moved the correlation results over the topology Zoo
% For determining the node rankings we had to carry out the centrality indices computations over weighted graphs.
% This was mainly a question of computing shortest paths over weighted graphs.
% Regarding the spectral indices, in the subsequent experiments we only employ the EC index that lends to
% a straightforward extension over the weighted graphs (see paragraph~\ref{subsec:graph_types}).

\begin{figure}[ht]
\begin{center}
\advance\leftskip-0.4cm
\begin{tabular}{ccc}
 \includegraphics[width=0.28\textwidth]{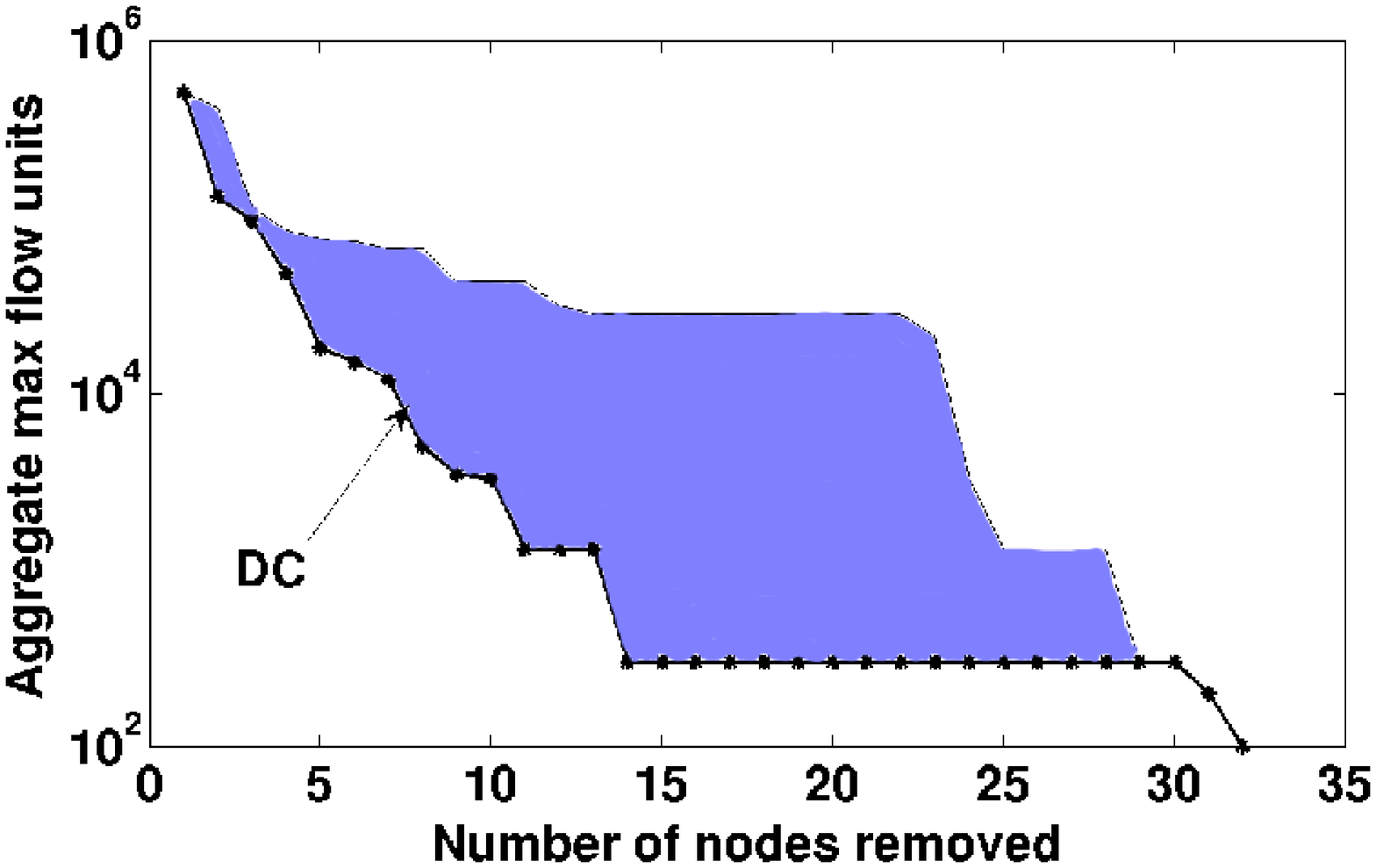} % arxika sto 0.17
 \hspace{-0.26in}
 &\hspace{-0.26in}
 \includegraphics[width=0.27\textwidth]{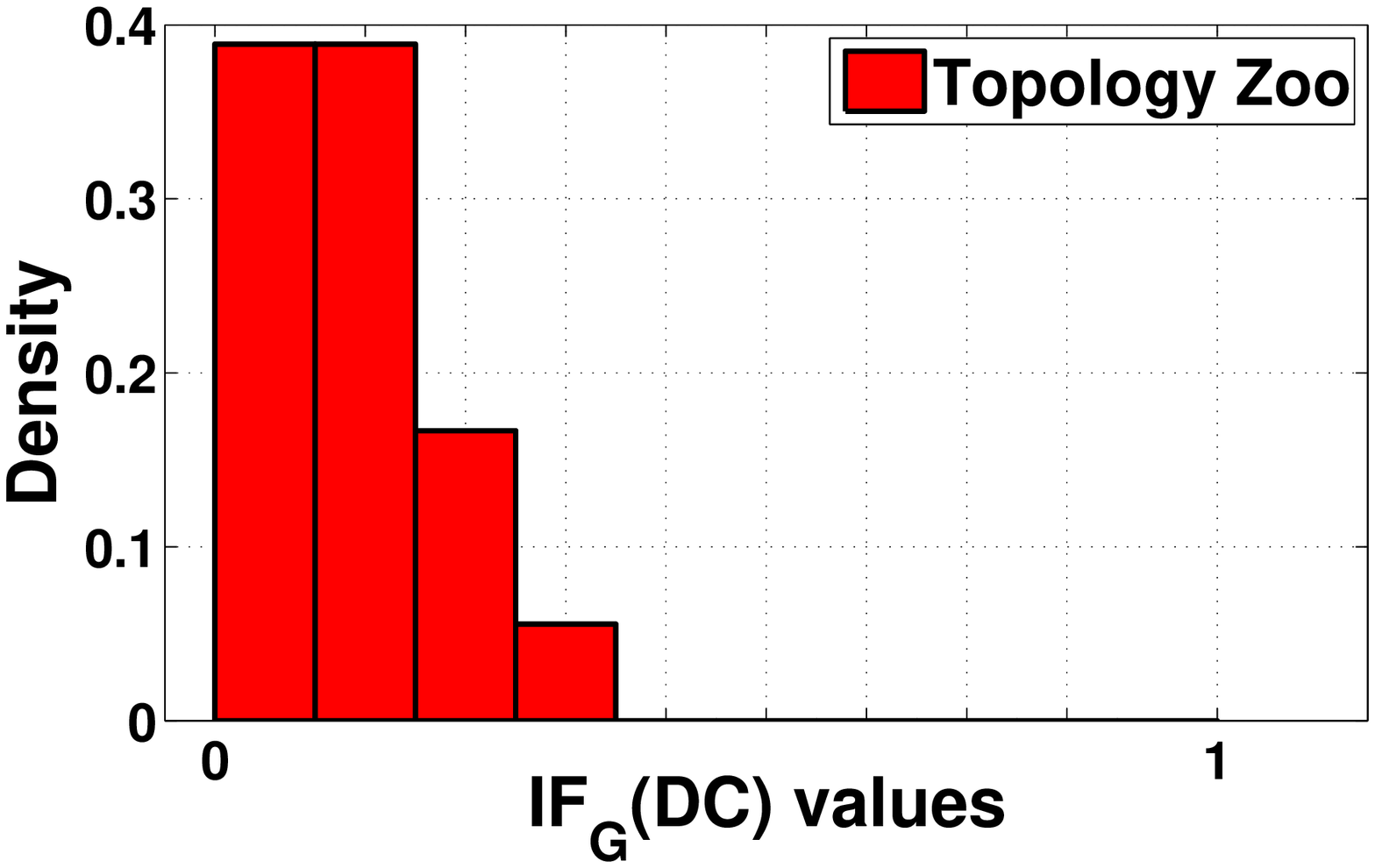}\\
%  &
%  \includegraphics[width=0.28\textwidth]{./figs/PDF_IFconnectivity}\\ %untitled
 \begin{scriptsize} a.   \end{scriptsize}  &
\begin{scriptsize}  b.   \end{scriptsize} \\%&
%\begin{scriptsize}  c. All datasets \end{scriptsize} \\
\end{tabular}
\end{center}
\caption{a) Envelope plot of the DC-based node removal effects on the max flow in the Bren topology.
b) Empirical probability mass function of the $IF_G(DC)$ measured with respect to the max flow the Topology Zoo dataset accommodates.}
\label{fig:PDF_flow}
%\vspace{-0.15 in}
\end{figure}

The computed aggregate maximum flow over an indicative set of networks is plotted in Fig.~\ref{fig:flow_results}. We have obtained similar results for the rest of the topology Zoo dataset (18 snapshots in total).
%Figure~\ref{fig:flow_results}(a) includes an additional subplot for a more detailed view of the network
%reaction as the most central nodes are initially removed.
The rate of aggregate max flow reduction with the fraction of removed nodes varies wildely, as shown in Fig.~\ref{fig:flow_results}.
%or more progressively (Figs.~\ref{fig:flow_results}.a,d) .
%In the beginning, a very fast breakdown of the aggregate max flow takes place, which eventually
%slows down after a small fraction of nodes have been removed.
This results in high best- to worst-case flow values and wide envelopes, as shown in Fig.~\ref{fig:PDF_flow}.a). Highly correlated index pairs, especially those with high top-$k$ overlaps, impact the accommodated flow in similar ways
(\ie intersection of corresponding curves).
In particular, certain index pairs that have been earlier measured with high rank-correlation, and most notably overlap of the top central nodes, yield similar curves over a sequence of removals; for instance, the highly associated pairs of EC-CC, EC-HC and HC-CC (see Tables~\ref{tbl:Zoo_correl}~and~\ref{tbl:Zoo_topk}) are seen in Figs.~\ref{fig:flow_results}.b and .c.
Similar impact of BC- and DC-driver node removals %of the attacks
has been reported over synthetic graphs (\ie Erd\H{o}s- R{\'e}nyi and small-world networks) in~\cite{vulner_flow}, yet in our case the impact of these two indices is typically different over the considered Internet snapshots. Finally, weakly correlated pairs such as EC-BC and ECC-BC, inline with intuition,
yield well-separated aggregate flow curves. %(\eg Figs.~\ref{fig:flow_results})

On a positive note, when node removals are driven by the DC index, the resulting aggregate maximum flow in most cases of Fig.~\ref{fig:flow_results} is very close to the worst achieved over all indices. This is more clearly shown in the empirical probability mass function of the $IF_G(DC)$ measure in Fig.~\ref{fig:PDF_flow}.b, %and in particular in .b plot
whose mass is highly concentrated in (very) low values close to zero. 
%means that DC systematically
%constitutes(approximates) the most effective centrality index in terms of max-flow degradation.
On the contrary, the considered networks exhibit their highest resilience against the ECC-driven node removals. This behavior can be explained along the same arguments employed earlier, when discussing the impact of node removals on the connected components. Having a single node \ie the furthest one determine the ECC value %of a node 
may result in some of the most central nodes not being included in the top positions of the ECC ranking.
% It is interesting that all the centrality rankings except for the ECC affect the throughput roughly
% in the same way. In other words, the globally-determined metrics do not appear as more attractive
% candidates to drive the attacks compared to the local one (\ie DC).
% ECC seems as before to deviate from the neutrality of the other indices regarding their impact on the
% network throughput.
% %any neither the
% any strong or the weak correlation ties that may be identified between the corresponding indices over the Zoo topologies.
%exhibited throughout the statistical analysis in the previous Chapter. In strongly correlated
%pairs, we would expect targeting vertices according to both centralities to be similar in effect in contrast to the
%weak related couples.
%Neutrality : the studied networks were found to be similarly harmed (in terms of flow degradation)
%by almost all the centrality-driven attacks. ECC is the only exception being clearly the less harmful one. 

\section{Related work}\label{sec:related}

We group relevant work in literature along two threads:

\textbf{Surveys and categorization of centrality indices: }
Our detailed survey of more than 30 different centrality indices appears in \cite{thesis}.
Similar studies are rare in literature despite %the existence of numerous indices and their
the extended use of centrality indices. % in social studies, physics and computer science.
One of the first relevant attempts dates back to 1979. Freeman
reviewed several centrality indices that had been introduced by that time and reduced them down to three fundamental
centrality notions, expressed by the degree, closeness and betweenness centrality~\cite{Freeman}.
More recently, Borgatti~\cite{Borgatti_05} introduced  %dynamic model-based
a typology of the different flows that may occur through a network
and accordingly associated the various centrality measures
with the flows that they are most appropriate for.
Whereas the graph-theoretic review in~\cite{k-BC} classifies centrality measures
on the basis of the requirements posed by their calculation.
Compared to these works, our approach reviews a great body of centrality indices, with very different origin and motivation, and seeks to classify them along multiple dimensions. At the same time we retain special interest for ``engineering'' properties 
such as the complexity related to the computation of the indices in distributed Internet environments.
%: % of the indices: % that seem to have been :
%a) the complexity related to the computation of the indices in distributed Internet environments;
%and b) how the centrality concept an index expresses relates to the communications facilitated by the underlying network.  

\textbf{Correlation of indices and use for network resilience studies: }Likewise limited are the correlation studies of centrality indices. Two studies we are aware of compute linear correlation values between the two most well-known indices (\eg degree and betweenness centrality); 
%\textbf{(is this correct?)} 
the first employs a random network and a couple of real-world topologies with a single router-level snapshot~\cite{correl_complex}, while  
the second presents experiments over three AS-level snapshots representing the same network along three 
consecutive years~\cite{AS_properties}. Neither of the two works assesses how the network is affected when 
different centrality indices are used to direct intelligent attacks. 
%Those works present correlation studies without investigating how the different metrics compare in terms of their effectiveness to
%drive network operations. 
%Yet, a systematic investigation of the statistcal relations between the indices
On the other hand, there is significantly richer literature with respect to attacks that are directed towards the most central network nodes. Most of them concern synthetic graphs and the attack impact is measured through purely topological measures. The scale free topologies have been found vulnerable to high-degree nodes~\cite{error}. In ~\cite{attack}, attacks target the high-degree and -betweenness nodes in a real-world AS-level topology. The two attacks are found equally harmful in terms of 
the inverse geodesic length and the number of connected components in the residual network. More recently, the work in~\cite{envelopes} considers both random node failure and centrality-driven node attacks in the context of a more general network topological robustness framework. Experimenting with families of random graphs, power grids, railway and co-authorship networks
the authors show that many centrality indices drive removals of similar impact and that DC and EC relate to the most harmful ones.
%produce similar targeted attacks while the degree and eigenvector centrality may be enough to indicate the worst-case behavior of networks under attack.
%We show that and that a combination

Similar experimental results on the vulnerability of the Internet router-level graphs
to centrality-driven node attacks seem to be missing, especially when the studied network property is the accommodated traffic.
In an earlier work~\cite{ppantaz}, we have related the correlation values between socio- and egocentric BC computations  with the effectiveness of the local centrality-driven content search over ISP networks. Here, we generalize the approach by considering an extended set of centrality indices/topologies and adding traffic-carrying capacity measures to the network resilience context.
% 
% In an earlier work~\cite{ppantaz}, we have sought to relate the correlation values between socio- and egocentric computations of betweenness centrality with the effectiveness of the local centrality-driven content search over ISP router-level topologies.
% In this paper, we essentially generalize the approach by considering an extended set of centrality indices and topologies and adding network capacity measures to the network resilience context.
%Moreover, the is concerned %, seems to be underexplored.
%In the subsection~\ref{subsec:flow} we experimentally investigate this thread.

%222222
%\input{flow}
%\vspace{1in}

\section{Conclusions}\label{sec:conclusions}

Our paper has iterated on the broad variety of indices that embody and quantify \emph{point centrality}, a popular concept borrowed from the Social Network Analysis and increasingly used in information network analysis and protocol design. Our starting point is a novel classification scheme that attempts to systematically characterize more than thirty indices proposed over the last sixty years %in various disciplines.
by sociologists, physicists, and, to a lesser extent, biologists and computer scientists. 

We have then chosen the seven most popular and representative of these indices and derived how they rank the nodes of more than 40 router-level topologies. We have found high rank correlations of certain index pairs such as DC-BC and DC-PG that persist across all four topology datasets we experiment with. Yet a significant part of the high full rank correlation is due to the nodes that are ranked last (\eg DC=1, BC=0) so that the association weakens when we consider the overlap between the sets of the top-5\% most central nodes only. In several cases, these findings stand in agreement with what has been 
reported in literature for other types of real and synthetic networks.

Finally, we have experimentally assessed %how those rankings affect the ISP networks
the impact of removing the most central nodes of the network, as determined by each index, on its \textit{connectivity} properties and \textit{traffic-serving capacity}.
As expected, it is the top-5\% overlap rather than the full rank correlation that can predicts more accurately when the node removals that are determined by two different indices have similar impact on the network. This is a warning against the widespread use of full rank correlation as a proxy for the ``equivalence'' of two indices.
In general, the use of different indices for the choice of to-be-removed nodes varies significantly the impact of the network. Whereas ECC is consistently the index with the least impact, the indices that induce the more dramatic changes on the network performance change with the topology and performance measure. However, and less intuitively, the single index that can be computed through local-only information (\ie DC) appears to approximate closely the worst-case impact on the network traffic capacity.

One hint for vulnerability analysis out of these results is that the added complexity of global indices may be escaped when we want an estimate of what is the worst-case impact on the network. A second hint towards attackers (network operators) is that it might be worth considering attacking (resp. better defending) a set of nodes that results from mixing the rankings of two indices, \eg one local and one global. In this case, the top-$k$ overlap measure between the two ranking could serve as criterion for the efficiency of this mixing: if it is high, then there is little more to gain by mixing; if it is low, then mixing might generate a node set, whose removal affects the network even more dramatically. 
%This is what we now explore.
This is a direction that we are currently investigating.

%\vspace{2in}
%\section*{Acknowledgemen2222ts}2222222222%
\vspace{.5in}
\bibliographystyle{IEEEtran}
\bibliography{CR_ref}

% Generated by IEEEtran.bst, version: 1.12 (2007/01/11)
\begin{thebibliography}{10}
\providecommand{\url}[1]{#1}
\csname url@samestyle\endcsname
\providecommand{\newblock}{\relax}
\providecommand{\bibinfo}[2]{#2}
\providecommand{\BIBentrySTDinterwordspacing}{\spaceskip=0pt\relax}
\providecommand{\BIBentryALTinterwordstretchfactor}{4}
\providecommand{\BIBentryALTinterwordspacing}{\spaceskip=\fontdimen2\font plus
\BIBentryALTinterwordstretchfactor\fontdimen3\font minus
  \fontdimen4\font\relax}
\providecommand{\BIBforeignlanguage}[2]{{%
\expandafter\ifx\csname l@#1\endcsname\relax
\typeout{** WARNING: IEEEtran.bst: No hyphenation pattern has been}%
\typeout{** loaded for the language `#1'. Using the pattern for}%
\typeout{** the default language instead.}%
\else
\language=\csname l@#1\endcsname
\fi
#2}}
\providecommand{\BIBdecl}{\relax}
\BIBdecl

\bibitem{Faust}
S.~Wasserman and K.~Faust, \emph{Social network analysis: Methods and
  applications}.\hskip 1em plus 0.5em minus 0.4em\relax Cambridge Univ Pr,
  1994.

\bibitem{bavelas}
A.~Bavelas, ``{A mathematical model of Group Structure},'' \emph{Human
  Organizations}, vol.~7, pp. 16--30, 1948.

\bibitem{Katz}
L.~Katz, ``A new status index derived from sociometric data analysis,''
  \emph{Psychometrika}, vol.~18, pp. 34--43, 1953.

\bibitem{Beauchamp}
M.~Beauchamp, ``An improved index of centrality,'' \emph{Behavioral Science},
  vol.~10, pp. 161--163, 1965.

\bibitem{Anthonisse}
J.~Anthonisse, ``The rush in a directed graph.'' \emph{Technical Report BN9/71,
  Stichting Mahtematisch Centrum}, Amsterdam, 1971.

\bibitem{sabidussi}
G.~Sabidussi, ``The centrality index of a graph,'' \emph{Psychomatrika},
  vol.~31, pp. 581--603, 1966.

\bibitem{Freeman}
L.~C. Freeman, ``Centrality in social networks: Conceptual clarification,''
  \emph{Social Networks}, vol.~1, no.~3, pp. 215--239.

\bibitem{power}
P.~Bonacich, ``Power and centrality: A family of measures,'' \emph{American
  Journal of Sociology}, vol.~92, no.~5, pp. 1170--1182, 1987.

\bibitem{Daly09}
E.~M. Daly and M.~Haahr, ``Social network analysis for information flow in
  disconnected delay-tolerant manets,'' \emph{IEEE Trans. Mob. Comput.},
  vol.~8, no.~5, pp. 606--621, 2009.

\bibitem{cacheICN}
W.~K. Chai, D.~He, I.~Psaras, and G.~Pavlou, ``Cache "less for more" in
  information-centric networks,'' in \emph{Proc. of the 11th IFIP Networking},
  Prague, Czech Republic, May 2012.

\bibitem{NetScience}
A.-L. Barabasi, ``Linked: The new science of networks,'' Perseus Pub., 2002.

\bibitem{thesis}
\BIBentryALTinterwordspacing
G.~Nomikos, ``Point centrality indices and {ISP} network vulnerability,''
  September 2013, {M}Sc Thesis, Dept. of Informatics \& Telecom., UoA.
  [Online]. Available: \url{http://anr.di.uoa.gr/index.php/theses}
\BIBentrySTDinterwordspacing

\bibitem{eccentrcity}
P.~Hagea and F.~Harary, ``Eccentricity and centrality in networks,''
  \emph{Social Networks}, vol.~17, no.~1, pp. 57--63, Jan. 1995.

\bibitem{RW_BC}
M.~J. Newman, ``A measure of betweenness centrality based on random walks,''
  \emph{Social Networks}, vol.~27, no.~1, pp. 39 -- 54, 2005.

\bibitem{trafficBC}
A.~Tizghadam and A.~Leon-Garcia, ``On traffic-aware betweenness and network
  criticality,'' in \emph{INFOCOM IEEE Conference on Computer Communications
  Workshops , 2010}, 2010, pp. 1--6.

\bibitem{ITC}
P.~Pantazopoulos, M.~Karaliopoulos, and I.~Stavrakakis, ``Centrality-driven
  scalable service migration,'' in \emph{23rd International Teletraffic
  Congress (ITC'11)}, San Francisco, USA, Sept. 2011.

\bibitem{Freeman77}
L.~C. Freeman, ``A set of measures of centrality based on betweenness,''
  \emph{Sociometry}, vol.~40, no.~1, pp. 35--41, 1977.

\bibitem{pagerank}
L.~Page, S.~Brin, R.~Motwani, and T.~Winograd, ``The pagerank citation ranking:
  Bringing order to the web,'' Stanford University, Technical Report;, 1998.

\bibitem{weightedNets}
A.~Barrat, M.~Barth\'{e}lemy, R.~Pastor-Satorras, and A.~Vespignani, ``{The
  architecture of complex weighted networks},'' \emph{Proceedings of the
  National Academy of Sciences of the United States of America}, vol. 101,
  no.~11, pp. 3747--3752, Mar. 2004.

\bibitem{Brandes_variants}
U.~Brandes, ``On variants of shortest-path betweenness centrality and their
  generic computation,'' \emph{Social Networks}, vol.~30, no.~2, 2008.

\bibitem{weightedNewman}
M.~E.~J. Newman, ``Analysis of weighted networks,'' \emph{Phys. Rev. E},
  vol.~70, no.~5, p. 056131, Nov. 2004.

\bibitem{wPageRank_alg}
W.~Xing and A.~Ghorbani, ``Weighted pagerank algorithm,'' in \emph{In Proc.
  Second Annual Conference on Communication Networks and Services Research,},
  May 2004, pp. 305 -- 314.

\bibitem{AuthorRank}
X.~Liu \emph{et~al.}, ``Co-authorship networks in the digital library research
  community,'' \emph{Inf. Process. Manage.}, vol.~41, no.~6, pp. 1462--1480,
  2005.

\bibitem{journal_status}
J.~Bollen, M.~A. Rodriquez, and H.~Van~de Sompel, ``Journal status,''
  \emph{Scientometrics}, vol.~69, no.~3, pp. 669--687, 2006.

\bibitem{kostakos}
V.~Kostakos, ``Temporal graphs,'' \emph{Physica A: Statistical Mechanics and
  its Applications}, vol. 388, no.~6, pp. 1007--1023, 2009.

\bibitem{evolving_graphs}
B.~B. Xuan, A.~Ferreira, and A.~Jarry, ``{Computing shortest, fastest, and
  foremost journeys in dynamic networks},'' \emph{International Journal of
  Foundations of Computer Science}, vol.~14, no.~2, pp. 267--285, 2003.

\bibitem{spacetime_graphs}
\BIBentryALTinterwordspacing
S.~Merugu, M.~Ammar, and E.~Zegura, ``{Routing in Space and Time in Networks
  with Predictable Mobility},'' \emph{Technical Report, Georgia Institute of
  Technology}, 2004. [Online]. Available:
  \url{https://smartech.gatech.edu/handle/1853/6492}
\BIBentrySTDinterwordspacing

\bibitem{time-varying}
A.~Orda and R.~Rom, ``Shortest-path and minimum-delay algorithms in networks
  with time-dependent edge-length,'' \emph{Journal of the ACM}, vol.~37, pp.
  607--625, 1990.

\bibitem{temporal_paths}
J.~Tang, M.~Musolesi, C.~Mascolo, V.~Latora, and V.~Nicosia, ``Analysing
  information flows and key mediators through temporal centrality metrics,'' in
  \emph{Proceedings of the 3rd Workshop on Social Network Systems}, ser. SNS
  '10.\hskip 1em plus 0.5em minus 0.4em\relax New York, NY, USA: ACM, 2010, pp.
  3:1--3:6.

\bibitem{time_ordered}
H.~Kim and R.~Anderson, ``{Temporal node centrality in complex networks},''
  \emph{Physical Review E}, vol.~85, 026107 (2012).

\bibitem{centered}
L.~C. Freeman, ``Centered graphs and the structure of ego networks,''
  \emph{Mathematical Social Sciences}, no.~3, pp. 291--234, 1982.

\bibitem{ppantaz}
P.~Pantazopoulos, M.~Karaliopoulos, and I.~Stavrakakis, ``On the local
  approximations of node centrality in internet router-level topologies,'' in
  \emph{the 7th IFIP IWSOS, Palma de Mallorca, Spain.}, May 2013.

\bibitem{k_path}
D.~S. Sade, ``{Sociometrics of Macaca Mulatta III: n-path centrality in
  grooming networks},'' \emph{Social Networks}, vol.~11, no.~3, pp. 273--292,
  1989.

\bibitem{k-BC}
S.~Borgatti and M.~Everett, ``{A Graph-theoretic perspective on centrality},''
  \emph{Social Networks}, vol.~28, no.~4, pp. 466--484, Oct. 2006.

\bibitem{Bonacich91}
P.~Bonacich, ``Simultaneous group and individual centralities,'' \emph{Social
  Networks}, vol.~13, no.~2, pp. 155 -- 168, 1991.

\bibitem{Moxley}
R.~Moxley and N.~Moxley, ``Determining point-centrality in uncontrived social
  networks.'' \emph{Sociometry}, vol.~37, no.~1, pp. 122--130, Mar. 1974.

\bibitem{Hoivik}
T.~H{\o}ivik and N.~P. Gleditsch., ``Structural parameters of graphs: A
  theoretical investigation,'' \emph{Quantitative Sociology.}, pp. 203--223,
  Academic Press, New York, 1975.

\bibitem{Borgatti_05}
S.~P. Borgatti, ``Centrality and network flow,'' \emph{Social Networks},
  no.~27, pp. 55--71, 2005.

\bibitem{harmonic}
Y.~Rochat, ``Closeness centrality extended to unconnected graphs: The harmonic
  centrality index,'' \emph{In proc. of Applications of Social Network
  Analysis}, Zurich, Switzerland (2009).

\bibitem{BubbleRap}
P.~Hui, J.~Crowcroft, and E.~Yoneki, ``Bubble rap: Social-based forwarding in
  delay-tolerant networks,'' \emph{IEEE Trans. Mob. Comput.}, vol.~10, no.~11,
  pp. 1576 --1589, nov. 2011.

\bibitem{adamic}
L.~A. Adamic \emph{et~al.}, ``{Search in power-law networks},'' \emph{Physical
  Review E}, vol.~64, no.~4, Sep. 2001.

\bibitem{rocketFL}
N.~T. Spring \emph{et~al.}, ``Measuring {ISP} topologies with rocketfuel.''
  \emph{IEEE/ACM Trans. Netw.}, vol.~12, no.~1, pp. 2--16, 2004.

\bibitem{caida}
\BIBentryALTinterwordspacing
The CAIDA UCSD Macroscopic Internet Topology Data Kit (ITDK) - [ITDK 2011-10].
  [Online]. Available:
  \url{http://www.caida.org/data/active/internet-topology-data-kit/}
\BIBentrySTDinterwordspacing

\bibitem{PAM10}
J.-J. Pansiot \emph{et~al.}, ``Extracting intra-domain topology from mrinfo
  probing,'' in \emph{Proc. PAM}, Zurich, Switzerland, April 2010.

\bibitem{tarzan}
R.~M. Karp and R.~E. Tarjan, ``Linear expected-time algorithms for connectivity
  problems (extended abstract),'' in \emph{ACM STOC '80}, Los Angeles,
  California, 1980, pp. 368--377.

\bibitem{zoo}
S.~Knight, H.~X. Nguyen, N.~Falkner, R.~A. Bowden, and M.~Roughan, ``The
  internet topology zoo.'' \emph{IEEE Journal on Selected Areas in
  Communications}, vol.~29, no.~9, pp. 1765--1775, 2011.

\bibitem{correl_complex}
\BIBentryALTinterwordspacing
C.-Y. Lee, ``Correlations among centrality measures in complex networks.''
  [Online]. Available: \url{http://arxiv.org/abs/physics/0605220}
\BIBentrySTDinterwordspacing

\bibitem{AS_properties}
A.~V\'azquez \emph{et~al.}, ``Large-scale topological and dynamical properties
  of the internet,'' \emph{Phys. Rev. E}, vol.~65, no.~6, p. 066130, Jun 2002.

\bibitem{PG_undirect}
\BIBentryALTinterwordspacing
V.~Grolmusz, ``A note on the pagerank of undirected graphs,'' May 2012.
  [Online]. Available: \url{http://arxiv.org/abs/1205.1960}
\BIBentrySTDinterwordspacing

\bibitem{coauthorship}
E.~Yan and Y.~Ding, ``Applying centrality measures to impact analysis: A
  coauthorship network analysis,'' \emph{J. Am. Soc. Inf. Sci. Technol.},
  vol.~60, no.~10, pp. 2107--2118, Oct. 2009.

\bibitem{attack_robustness}
\BIBentryALTinterwordspacing
I.~Swami, K.~Timothy, S.~Bala, and W.~Zhen, ``Attack robustness and centrality
  of complex networks,'' \emph{PLoS ONE}, vol.~8, no.~4, April 2013. [Online].
  Available: \url{doi: 10.1371/journal.pone.0059613}
\BIBentrySTDinterwordspacing

\bibitem{attack}
P.~Holme, B.~J. Kim, C.~N. Yoon, and S.~K. Han, ``Attack vulnerability of
  complex networks,'' \emph{Phys. Rev. E}, vol.~65, no.~5, May 2002.

\bibitem{vulnerWeighted}
L.~Dall'Asta, A.~Barrat, M.~Barth\'elemy, and A.~Vespignani, ``Vulnerability of
  weighted networks,'' \emph{J. Stat. Mech.}, p. P04006, 2006.

\bibitem{complexityMCF}
S.~Even, A.~Itai, and A.~Shamir, ``{On the Complexity of Timetable and
  Multicommodity Flow Problems},'' \emph{SIAM Journal on Computing}, vol.~5,
  no.~4, pp. 691--703, 1976.

\bibitem{EdmondsKarp}
J.~Edmonds and R.~M. Karp, ``Theoretical improvements in algorithmic efficiency
  for network flow problems,'' \emph{J. ACM}, vol.~19, no.~2, Apr. 1972.

\bibitem{vulner_flow}
I.~Mishkovski \emph{et~al.}, ``Vulnerability assessment of complex networks
  based on optimal flow measurements under intentional node and edge attacks,''
  in \emph{Proc. of ICT Innovations 2009}.\hskip 1em plus 0.5em minus
  0.4em\relax Springer, pp. 167--176.

\bibitem{error}
R.~Albert, H.~Jeong, and A.-L. Barabasi, ``{Error and attack tolerance of
  complex networks},'' \emph{Nature}, vol. 406, no. 6794, pp. 378--382, Jul.
  2000.

\bibitem{envelopes}
\BIBentryALTinterwordspacing
S.~Trajanovski \emph{et~al.}, ``Robustness envelopes of networks,'' Journal of
  Complex Networks, 2013. [Online]. Available:
  \url{http://comnet.oxfordjournals.org/content/early/2013/03/26/comnet.cnt004}
\BIBentrySTDinterwordspacing

\end{thebibliography}
\vspace{2in}
\appendix
\section{Appendices}
\label{sec:appendix}

\subsection{Employed Internet router-level Topologies}

In Tables~\ref{tab:nets} and~\ref{tab:ZOOnets} we present basic information about the network topologies
employed for our experiments. The four datasets of the former table contain binary graphs whereas 
%used for our
%correlation study and connectivity-related assessment of node removals.
%The dataset of 
the later table contains capacitated graphs used mainly for %the correlation study)
%and 
the experimentation with the traffic serving capacity.

\begin{table}[ht]%[!htbp]
\centering \caption{Properties of the Intra-domain ISP topologies}
\label{tab:nets} {\ssmall
\begin{tabular}{c   c  c c c }
\hline\hline
 %\textbf{Ego (R=1)} & DataSet  & ISP (AS number)& Mean Clustering coefficient & Diameter& Number of nodes (GCC) &Mean degree  & Spearman correlation coefficient \\
%\textit{$\alpha$ of G nodes} & \textit{$s$=1} & \textit{$s$=1.5} & \textit{$s$=2} \\
\multicolumn{1}{c}{Dataset}&\multicolumn{1}{c}{ISP(AS no.)}&\multicolumn{1}{c}{<Clust. Coeff. >}&\multicolumn{1}{c}{Diameter}&\multicolumn{1}{c}{Size }\\%&\multicolumn{1}{c}{<degree>}\\%&\multicolumn{2}{c}{\textbf{CBC \vs ego-CBC} }\\
 %\cline{1-7}
%\multicolumn{1}{c}{}&\multicolumn{1}{c}{}&\multicolumn{1}{c}{}&\multicolumn{1}{c}{}&\multicolumn{1}{c}{}&\multicolumn{1}{c}{}&\multicolumn{1}{c}{}\\
%& & & & & & & ego-net. \textbf{r=1} & ego-net. \textbf{r=2}& ego-net. \textbf{r=1} &ego-net. \textbf{r=2}&ego-net. \textbf{r=1}& 95\% \\    %95\% Conf. Inter.\\
\hline \hline
            &Global Crossing(3549)   & 0.546  &10     &76    \\%&3.71  \\%&0.9648 &0.9806  & 0.6720 & 0.9197  &0.9568 &0.008        \\ %&0.9568 &0.008 \\  commented :spearman for CBC vs egoCBC (r=1)
Mrinfo      &-//-                    & 0.479  &9      &100   \\%&3.78  \\%&0.9690 &0.9853  & 0.7029 & 0.9255    &0.9489 &0.013      \\ %&0.9489 &0.013\\
(Tier-1)    &NTTC-Gin(2914)          & 0.307  &11     &180   \\%&3.53  \\%&0.9209 &0.9565  & 0.7479 & 0.8561   &0.9554 &0.003         \\ %&0.9554 &0.003  \\
            &Sprint(1239)            & 0.298  &12     &216   \\%&3.07  \\%&0.9718 &0.9812  & 0.7470 & 0.8557   &0.9824 &0.002         \\ %&0.9824 &0.002 \\
            &Level-3(3356)           & 0.169  &25     &378   \\%&4.49  \\%&0.2708 &0.9393  & -0.0918 & 0.7982    &0.7336 &0.007     \\ %&0.7336 &0.007\\
            &-//-                    & 0.149  &28     &436   \\%&4.98  \\%&0.2055 & 0.9381 & -0.1217 & 0.7392     &0.7035 &0.005       \\ %&0.7035 &0.005 \\
            &Sprint(1239)            & 0.287  &16     &528   \\%&3.13  \\%&0.9866 &0.9928  & 0.5805 & 0.8488     &0.9847 &0.003    \\ %&0.9847 &0.003 \\
            &-//-                    & 0.251  &13     &741   \\%&3.29  \\%&0.9901 & 0.9930 & 0.7149 & 0.8622       &0.9884 &0.002   \\ %&0.9884 &0.002 \\
\hline
%T & & & & & & & & & & \\
            &JanetUK(786)           &0.132  &14      &336   \\%&2.69  \\%&0.9714 &0.9825  & 0.8049 & 0.9180     &0.9819& 0.001     \\ %&0.9819& 0.001\\
Mrinfo      &Iunet(1267)            &0.246  &11      &598   \\%&3.88  \\%&0.8506 &0.9468  & 0.8887 & 0.9688    &0.7825&0.033       \\%&0.7825&0.033\\
(Transit)   &-//-                   &0.231  &12      &645   \\%&3.75  \\%&0.8790 &0.9516  & 0.9094 & 0.9568     &0.8062&0.022     \\%&0.8062&0.022  \\
            &-//-                   &0.038  &13      &711   \\%&3.45  \\%&0.9470 &0.9826  & 0.5354 & 0.9536     &0.9370 &0.016       \\%&0.9370 &0.016\\
            &Telecom Italia(3269)   &0.037  &13      &995   \\%&3.65  \\%&0.7950 &0.9828  & 0.3362 & 0.8699        &0.9902 &0.001     \\%&0.9902 &0.001\\
            &TeleDanmark(3292)      & 0.058 &15      &1240       \\
            %t       & & & & & & & & & &  \\
\hline
               &   VSNL(4755)       & 0.263   &    6    & 41      \\%& 3.32  \\%& 0.9909 & 0.9971  &0.7286 & 0.9603 &  0.8740    &0.7842  \\
Rocket         &   Ebone(1755)      & 0.115   &    13   & 295     \\%& 3.68  \\%& 0.9736 & 0.9860  &0.6856 & 0.8895 &  0.9443    &0.7457  \\
Fuel           &   Tiscali(3257)    & 0.028   &    14   & 411     \\%& 3.18  \\%& 0.9522 & 0.9659  &0.6073 & 0.9281 &  0.9464    &0.7103    \\
               &   Exodus(3967)     & 0.273   &    14   & 353     \\%& 4.65  \\%&0.9125  & 0.9792  &0.6100 &0.9061  & 0.8204    &0.6241 \\
               &   Telstra (1221)   & 0.015   &    15   & 2515    \\%& 2.42  \\%&0.9990  & 0.9990  &0.3336 &0.7565  & 0.9783    &0.5172   \\
%           &        &  Abovenet(6461)    &          &         &  empty! &       &        &         &       &        &  &  \\
               &  Sprint(1239)      & 0.022   &    13   & 7303     \\%& 2.71  \\%&0.9980  &0.9990   &0.4770 &0.7977  & 0.9562   & 0.6537 \\
               &  Level-3(3356)      & 0.097   &    10   & 1620    \\%& 8.32  \\%&0.9841  &0.9923   &0.6346 &0.9075  & 0.9655   & 0.7045  \\
               &  AT\&T(7018)       & 0.005   &    14   & 9418     \\%& 2.48  \\%&0.9988  &0.9994   &0.3388 &0.5302  &  0.9882   & 0.4483   \\
               &  Verio (2914)      & 0.071   &    15   & 4607     \\%& 3.28  \\%& 0.9904 &0.9969   &0.4729 &0.8044  & 0.9315   & 0.6718   \\
\hline
               &  UUNet (701)         & 0.012  &  15     & 18281    \\%& 2.77  \\%&0.9841  &0.9886  &0.5430  &0.8752   &  0.9694 & 0.7544  \\
CAIDA          &  COGENT/PSI(174)     & 0.062  &  32     & 14413    \\%& 3.09  \\%&0.9638  &0.9599  &0.7272  &0.9354   &  0.8940 & 0.8791  \\
               &   LDComNet(15557)    & 0.021  &  40     &  6598    \\%& 2.47  \\%& 0.9674 &0.9245  &0.3782  &0.7676   &  0.9479 & 0.6634   \\
               &  TeliaNet(1299)      & 0.037  &  13     &  3820    \\%& 3.08  \\%&0.9593  &0.9764  &0.9176  &0.9628   &  0.9047 & 0.9594  \\
               &  ChinaTelecom(4134)  & 0.083  &  19     & 81121    \\%& 3.97  \\%&0.8324  &0.8986  &0.7861  &0.9714   &  0.7370 & 0.8795  \\
               &  FUSE-NET(6181)      & 0.018  &  10     & 1831     \\%& 2.38  \\%& 0.9903 & 0.9763 &0.6205  &0.8574   &  0.9536 & 0.7445\\
               &  JanetUK(786)        & 0.031  &  24     & 2259     \\%& 2.26  \\%& 0.9819 & 0.9834 &0.4444  &0.8506   &  0.9450 & 0.5765 \\
\hline\hline
\end{tabular}}
\end{table}

%\vspace{-0.1in}
\begin{table}[htp]%[!htbp]
\centering \caption{Properties of the capacitated IP-level Zoo topologies}
\label{tab:ZOOnets} {\ssmall
\begin{tabular}{c   c  c c c }
\hline\hline
 %\textbf{Ego (R=1)} & DataSet  & ISP (AS number)& Mean Clustering coefficient & Diameter& Number of nodes (GCC) &Mean degree  & Spearman correlation coefficient \\
%\textit{$\alpha$ of G nodes} & \textit{$s$=1} & \textit{$s$=1.5} & \textit{$s$=2} \\
\multicolumn{1}{c}{Network}&\multicolumn{1}{c}{Geo Location}&\multicolumn{1}{c}{Date of snapshot}&\multicolumn{1}{c}{Diameter}&\multicolumn{1}{c}{Size }\\%&\multicolumn{1}{c}{<degree>}\\%&\multicolumn{2}{c}{\textbf{CBC \vs ego-CBC} }\\
 %\cline{1-7}
%\multicolumn{1}{c}{}&\multicolumn{1}{c}{}&\multicolumn{1}{c}{}&\multicolumn{1}{c}{}&\multicolumn{1}{c}{}&\multicolumn{1}{c}{}&\multicolumn{1}{c}{}\\
%& & & & & & & ego-net. \textbf{r=1} & ego-net. \textbf{r=2}& ego-net. \textbf{r=1} &ego-net. \textbf{r=2}&ego-net. \textbf{r=1}& 95\% \\    %95\% Conf. Inter.\\
\hline \hline
Janet Lense      & UK                 &1/2011     &4   &20    \\%&3.71  \\%&0.9648 &0.9806  & 0.6720 & 0.9197  &0.9568 &0.008        \\ %&0.9568 &0.008 \\  commented :spearman for CBC vs egoCBC (r=1)
Belnet I         &Belgium             &2003       &3   &23   \\%&3.78  \\%&0.9690 &0.9853  & 0.7029 & 0.9255    &0.9489 &0.013      \\ %&0.9489 &0.013\\
Belnet II        &Belgium             &2006       &3    &23   \\%&3.53  \\%&0.9209 &0.9565  & 0.7479 & 0.8561   &0.9554 &0.003         \\ %&0.9554 &0.003  \\
Geant            &cross-Europe        &2009       &7    &34    \\%&3.07  \\%&0.9718 &0.9812  & 0.7470 & 0.8557   &0.9824 &0.002         \\ %&0.9824 &0.002 \\
Niif             &Hungary             &5/2009     &7     &36   \\%&4.49  \\%&0.2708 &0.9393  & -0.0918 & 0.7982    &0.7336 &0.007     \\ %&0.7336 &0.007\\
Bren             &Bulgaria            &10/2010    &8     &37   \\%&4.98  \\%&0.2055 & 0.9381 & -0.1217 & 0.7392     &0.7035 &0.005       \\ %&0.7035 &0.005 \\
Myren            &Malaysia            &3/2011     &4     &37   \\%&3.13  \\%&0.9866 &0.9928  & 0.5805 & 0.8488     &0.9847 &0.003    \\ %&0.9847 &0.003 \\
Kentman          &Kent, UK            &1/2011     &6     &38   \\%&3.29  \\%&0.9901 & 0.9930 & 0.7149 & 0.8622       &0.9884 &0.002   \\ %&0.9884 &0.002 \\
%T & & & & & & & & & & \\
Switch L3        &Switzerland         &2011       &6     & 42  \\%&2.69  \\%&0.9714 &0.9825  & 0.8049 & 0.9180     &0.9819& 0.001     \\ %&0.9819& 0.001\\
Renater          &France              &2010       &9     & 43  \\%&2.69  \\%&0.9714 &0.9825  & 0.8049 & 0.9180     &0.9819& 0.001     \\ %&0.9819& 0.001\\
Sanet            &Slovakia            &2008       &13     & 43  \\%&2.69  \\%&0.9714 &0.9825  & 0.8049 & 0.9180     &0.9819& 0.001     \\ %&0.9819& 0.001\\
Carnet           &Croatia             &8/2010     &6     &44   \\%&3.88  \\%&0.8506 &0.9468  & 0.8887 & 0.9688    &0.7825&0.033       \\%&0.7825&0.033\\
\hline
Uninett I\_min  &Norway                &2011   &9     &69   \\%&3.75  \\%&0.8790 &0.9516  & 0.9094 & 0.9568     &0.8062&0.022     \\%&0.8062&0.022  \\
Uninett I\_max  &-//-                  &-//-   &-//-     &69   \\%&3.45  \\%&0.9470 &0.9826  & 0.5354 & 0.9536     &0.9370 &0.016       \\%&0.9370 &0.016\\
Uninett I\_mean &                      &-/-   &-//-     &69  \\%&3.65  \\%&0.7950 &0.9828  & 0.3362 & 0.8699        &0.9902 &0.001     \\%&0.9902 &0.001\\
%t       & & & & & & & & & &  \\
\hline
Uninett II\_min  &-//-            &2010   &9     &74   \\%&3.75  \\%&0.8790 &0.9516  & 0.9094 & 0.9568     &0.8062&0.022     \\%&0.8062&0.022  \\
Uninett II\_max  &-//-            &-//-   &-//-     &74   \\%&3.45  \\%&0.9470 &0.9826  & 0.5354 & 0.9536     &0.9370 &0.016       \\%&0.9370 &0.016\\
Uninett II\_mean &                &-//-   &-//-     &74  \\%&3.65  \\%&0.7950 &0.9828  & 0.3362 & 0.8699        &0.9902 &0.001     \\%&0.9902 &0.001\\
\hline\hline
\end{tabular}}
\end{table}
%

%\vspace{3in}

\subsection{Averages of Kendall coefficients per dataset}
\label{subsec:Spearman_results}

%As far as
%The %is concerned which
As the graph representation shows (in Figure~\ref{fig:graphs}.b),
the Kendall correlation values appear to be roughly similar to the Spearman ones; we have identified 
as dominant the same centrality pairs with the ones captured
by the Spearman coefficient. %previous statistical study.
However, a closer inspection on the absolute Kendall values of Table~\ref{kendal_correl_results} shows a weaker
dependence among the metrics;
%The difference concerns the absolute values
%revealing
a somewhat looser relationship is therefore revealed among the centrality indices than what the Spearman values may suggest.
% Finally, the linear correlation captured by Pearson coefficient
% has also been found high for the highly rank-correlated centrality pairs~\cite{thesis}.
%\textbf{CONSIDER REMOVING THE PARA}
%\subsection{Rank-correlation averages}
%\label{tbl:Kendall_results}
%In the following table we provide the full set of our results regarding the Kendall coefficient averages.
%Note that
%(As before, each row value (\ie top to bottom) in every box of the table corresponds to averages measured over the CAIDA, Rocketfuel,
%MrInfo-Tier1 and MrInfo-Transit dataset, respectively.)

%\vspace{-28.1in}
\begin{table}[ht]
\centering \caption{Averages of Kendall coefficients for all datasets }
\label{kendal_correl_results} {\ssmall
\begin{tabular}{|c| c c c c c c c l|}
\hline
\multicolumn{1}{|c|}{}&\multicolumn{1}{c}{CC}&\multicolumn{1}{c}{HC}&\multicolumn{1}{c}{EC}&\multicolumn{1}{c}{ECC}&\multicolumn{1}{c}{DC}&\multicolumn{1}{c}{BC}&\multicolumn{1}{c|}{PG}&\multicolumn{1}{c|}{dataset}\\
% & & &
 %\cline{1-7}
%  & & ego-network \textbf{(r=1)} & ego-network \textbf{(r=2)}\\
\hline
CC   & 1   &        &      &      &     &      &  &  CAIDA\\
     & 1   &       &      &      &     &      &   &  Rocketfuel\\
     & 1   &      &      &      &     &      &    & Mrinfo-Tier1\\
     & 1   &      &      &      &     &      &    & Mrinfo-Transit\\
\hline
HC & 0.96 &     1 &  &      &     &      &    & -//-\\
   & 0.91 &     1 &   &      &     &      &   & -//-\\
   & 0.83 &     1 &   &      &     &      &   & -//-\\
   & 0.92 &     1 &   &      &     &      &   & -//-\\
\hline
EC & 0.82 &   0.84 & 1 &  &     &      &   & -//- \\
   & 0.66 &   0.68 & 1 &  &     &      &   & -//- \\
   & 0.54 &   0.55 & 1 &  &     &      &    & -//-\\
   & 0.72 &   0.74 & 1 &  &     &      &    & -//-\\
\hline
ECC & 0.73& 0.72 & 0.71 & 1 &  &      &    & -//-\\
    & 0.60& 0.56 & 0.48 & 1 &  &      &   & -//- \\
    & 0.66& 0.55 & 0.42 & 1 &  &      &    & -//-\\
    & 0.77& 0.75 & 0.62 & 1 &  &      &    & -//-\\
\hline
DC &0.22& 0.22 & 0.23 & 0.22 &1  &      &    & -//-\\
   &0.38& 0.43 & 0.35 & 0.32 &1  &      &    & -//-\\
   &0.33& 0.47 & 0.36 & 0.25 &1  &      &    & -//-\\
   &0.38& 0.42 & 0.38 & 0.38 &1  &      &    & -//-\\
\hline
BC &0.22 & 0.22 & 0.21 & 0.22 & 0.83&1& & -//-\\
   &0.35 & 0.39 & 0.30 & 0.30 & 0.87&1& & -//-\\
   &0.37 & 0.47 & 0.22 & 0.30 & 0.60&1& & -//-\\
   &0.40 & 0.44 & 0.35 & 0.38 & 0.77&1& & -//-\\
\hline
PG &0.01& 0.02 & 0.03 & -0.01 & 0.73 & 0.64 & 1 & -//-\\
   &0.16& 0.20 & 0.09 &  0.14 & 0.72 & 0.66 & 1 & -//-\\
   &0.22& 0.33 & 0.20 &  0.18 & 0.80 & 0.60 &1 & -//-\\
   &0.27& 0.30 & 0.24 &  0.27 & 0.82 & 0.74 &1 & -//-\\
\hline
\end{tabular}}
%\vspace{-0.12 in}
\end{table}

\subsection{Averages of Top-$k$ overlap values per dataset}
\label{subsec:topk_per_dataset}

%\vspace{-0.1in}

Next we present the top-$5\%$ overlap value between each index pair, averaged 
%values 
%for the Spearman  correlation 
%coefficient 
%as computed 
over all snapshots of the CAIDA, Rocketfuel, mrinfo-Tier-1 and -Transit datasets.

\begin{table}[ht]
\centering \caption{averages of top-$5\%$ overlap (\%) for all datasets }
\label{tbl:topk_av} {\ssmall
\begin{tabular}{|c| c c c c c c c l|}
\hline
\multicolumn{1}{|c|}{}&\multicolumn{1}{c}{CC}&\multicolumn{1}{c}{HC}&\multicolumn{1}{c}{EC}&\multicolumn{1}{c}{ECC}&\multicolumn{1}{c}{DC}&\multicolumn{1}{c}{BC}&\multicolumn{1}{c|}{PG}&\multicolumn{1}{c|}{dataset}\\
% & & &
 %\cline{1-7}
%  & & ego-network \textbf{(r=1)} & ego-network \textbf{(r=2)}\\
\hline
CC   & 1   &        &      &      &     &      &     &  CAIDA\\
     & 1   &        &      &      &     &      &     &  Rocketfuel\\
     & 1   &        &      &      &     &      &     & Mrinfo-Tier1\\
     & 1   &        &      &      &     &      &     & Mrinfo-Transit\\
\hline
HC & 90.1  &     1 &   &      &     &      &    & -//-\\
   & 90.7  &     1 &   &      &     &      &    & -//-\\
   & 79.8  &     1 &   &      &     &      &    & -//-\\
   & 82.9  &     1 &   &      &     &      &    & -//-\\
\hline
EC & 53.2  &   58.6 & 1 &  &     &      &    & -//- \\
   & 48.4  &   51.9 & 1 &  &     &      &    & -//- \\
   & 42.5  &   45.9 & 1 &  &     &      &    & -//-\\
   & 52.8  &   55.8 & 1 &  &     &      &    & -//-\\
\hline
ECC & 28.5 & 42.6 & 28.5 & 1 &  &      &    & -//-\\
    & 26.9 & 40.9 & 27.0 & 1 &  &      &   & -//- \\
    & 33.9 & 27.2 & 26.9 & 1 &  &      &    & -//-\\
    & 28.8 & 53.9 & 32.0 & 1 &  &      &    & -//-\\
\hline
DC &28.9   & 30.8 & 24.1 & 28.7 &1  &      &    & -//-\\
   &57.4   & 62.1 & 46.6 & 29.9 &1  &      &    & -//-\\
   &35.4   & 44.4 & 50.8 & 10.5 &1  &      &    & -//-\\
   &42.0   & 53.9 & 39.9 & 32.6 &1  &      &    & -//-\\
\hline
BC &27.8 & 29.4 & 22.5 & 25.8 & 70.9&1& & -//-\\
   &60.1 & 61.4 & 32.3 & 34.7 & 65.8&1& & -//-\\
   &67.8 & 63.1 & 30.1 & 34.7 & 33.8&1& & -//-\\
   &59.8 & 70.1 & 40.9 & 44.8 & 67.3&1& & -//-\\
\hline
PG &26.4& 27.9 & 22.9 &  25.6 & 89.6 & 72.1 & 1 & -//-\\
   &46.6& 50.8 & 35.2 &  20.7 & 80.7 & 64.9 & 1 & -//-\\
   &29.5& 39.7 & 39.3 &  10.8 & 75.0 & 30.4 &1 & -//-\\
   &33.0& 44.8 & 38.3 &  25.6 & 86.7 & 57.8 &1 & -//-\\
\hline
\end{tabular}}
%\vspace{-0.12 in}
\end{table}

\end{document}